\documentclass[prx,reprint,twocolumn,amsmath,superscriptaddress,longbibliography]{revtex4-2}
\usepackage{amsmath,amssymb,bm,mathrsfs,graphicx, braket, times,amsthm,enumerate, scrextend}
\usepackage[colorlinks=true,citecolor=blue,linkcolor=blue]{hyperref}
\usepackage{longtable}
\usepackage{multirow}
\usepackage{array}
\usepackage{bigstrut}
\usepackage[all,cmtip]{xy}
\usepackage[normalem]{ulem}
\usepackage[usenames,dvipsnames]{color}
\usepackage{makecell}
\usepackage{xcolor}
\usepackage{adjustbox}
\usepackage{booktabs}
\usepackage{tikz}
\usetikzlibrary{positioning}
\usepackage{placeins}
\usepackage{enumerate}
\usepackage{pifont}
\usepackage{soul}
\usepackage{diagbox}
\usepackage{appendix}

\makeatletter
\newcommand{\sectionnotoc}[1]{%
	\begingroup
	\let\addcontentsline\@gobblethree
	\section{#1}%
	\endgroup
}
\newcommand{\subsectionnotoc}[1]{%
	\begingroup
	\let\addcontentsline\@gobblethree
	\subsection{#1}%
	\endgroup
}
\makeatother

\makeatletter
\newcommand{\sectionnotocc}[1]{%
	\begingroup
	\let\addcontentsline\@gobblethree
	\section*{\MakeUppercase{#1}}%
	\endgroup
}
\newcommand{\subsectionnotocc}[1]{%
	\begingroup
	\let\addcontentsline\@gobblethree
	\subsection*{\MakeUppercase{#1}}%
	\endgroup
}
\makeatother

\begin{document}

	\title{Prediction of Magnetic Topological Materials Combining Spin and Magnetic Space Groups}
\author{Liangliang Huang}
\affiliation{National Laboratory of Solid State Microstructures and School of Physics, Nanjing University, Nanjing 210093, China}
\affiliation{Collaborative Innovation Center of Advanced Microstructures, Nanjing University, Nanjing 210093, China}
\author{Yuanze Song}
\affiliation{National Laboratory of Solid State Microstructures and School of Physics, Nanjing University, Nanjing 210093, China}
\affiliation{Collaborative Innovation Center of Advanced Microstructures, Nanjing University, Nanjing 210093, China}
\author{Houhao Wang}
\affiliation{National Laboratory of Solid State Microstructures and School of Physics, Nanjing University, Nanjing 210093, China}
\affiliation{Collaborative Innovation Center of Advanced Microstructures, Nanjing University, Nanjing 210093, China}
\author{Zhixian Yu}
\affiliation{National Laboratory of Solid State Microstructures and School of Physics, Nanjing University, Nanjing 210093, China}
\affiliation{Collaborative Innovation Center of Advanced Microstructures, Nanjing University, Nanjing 210093, China}
\author{Xiangang Wan}
\email{xgwan@nju.edu.cn}
\affiliation{National Laboratory of Solid State Microstructures and School of Physics, Nanjing University, Nanjing 210093, China}
\affiliation{Collaborative Innovation Center of Advanced Microstructures, Nanjing University, Nanjing 210093, China}
\affiliation{Hefei National Laboratory, Hefei 230088, China}
\affiliation{Jiangsu Physical Science Research Center, Nanjing, China}
\author{Feng Tang}
\email{fengtang@nju.edu.cn}
\affiliation{National Laboratory of Solid State Microstructures and School of Physics, Nanjing University, Nanjing 210093, China}
\affiliation{Collaborative Innovation Center of Advanced Microstructures, Nanjing University, Nanjing 210093, China}
\date{\today}	
\begin{abstract}	  
	Recent developments in spin space groups (SSGs) have led to a new classification of magnetic order and, with it, the emergence of altermagnetism. 
	However, an efficient and routine approach that predicts electronic band topology in magnetic materials by SSGs, and especially the evolution of band topology from negligible spin-orbit coupling (SOC) to finite SOC with the symmetry lowering from SSGs to magnetic space groups (MSGs) determined by magnetic-moment directions, has not yet been established.
	Here, we propose a scheme combining SSG and its MSG subgroups to diagnose band topology in collinear magnets using symmetry indicators of electronic band topology, established for all the 1,421 SSGs and 3,420 MSG subgroups.
	Specifically, the compatibility relations from SSG to the MSG subgroups can enable a complete topological classification for all possible magnetic-moment directions from the first-principles calculated numbers of occurrences of irreducible (co-)representations at high-symmetry points without SOC with respect to SSG.
	This scheme can be directly applied to collinear magnets whose magnetic structures have already been determined experimentally or theoretically. 
	Applying it to 488 collinear magnets from MAGNDATA with experimentally determined magnetic structures, we identify 26 materials that are trivial by MSG but are expected to inherit topology protected by SSG symmetry. 
	We showcase Dirac nodal lines and boundary states predicted by SSG but invisible to MSG in FePSe$_3$.
	On the other hand, we predict 224 collinear magnetic ground-state structures from 1,483 stoichiometric paramagnetic structures in Inorganic Crystal Structure Database belonging to space group No.~166 or No.~221 utilizing the symmetry-reduction hierarchy from a space group for the paramagnetic phase to SSG and then from SSG to MSG. 
	24 materials are identified as magnetic topological materials, including the experimentally confirmed magnetic Weyl semimetal Co$_3$Sn$_2$S$_2$, magnetic topological insulator MnBi$_2$Te$_4$, and magnetic Weyl nodal-line semimetal Fe$_3$Sn$_2$, whose computed magnetic structures and magnetic moments are in good agreement with experimental results. 
	Our work demonstrates the predictive power of the approach combining SSG and MSG over the approaches based on either group alone. 
	The results of high-throughput calculations are expected to extend the magnetic topological materials pool, facilitating future experimental realizations of more magnetic topological materials.	
\end{abstract}
\maketitle
\sectionnotoc{Introduction}

Magnetic topological materials have attracted considerable interest~\cite{mtm1,mtm2} due to their fascinating physical properties, such as Weyl semimetal phase~\cite{ahc1,ahc2,ahc9,css1,css2,css3,armitage2018weyl,lv2021experimental}, quantum anomalous Hall effect~\cite{ahc3,ahc4} and quantized magnetoelectric effect~\cite{qxl,ahc5}. Although the method based on symmetry indicators~\cite{si4,si6} or topological quantum chemistry~\cite{si1,si2} within the framework of space groups (SGs) or magnetic space groups (MSGs)~\cite{BC} is efficient and suitable for large-scale searches for topological materials in material databases~\cite{sg1,sg2,sg3,sg5,sg6,msg1,msg4}, only a very limited number of theoretically predicted magnetic topological materials have been experimentally confirmed~\cite{css4,ahc7,css1,css2,css3}. The scarcity of experimentally verified magnetic topological materials hinders their further optimization and design for device applications.

Indeed, for about seven decades, MSGs have served as the standard symmetry framework for characterizing materials with commensurate magnetic order~\cite{BC}, providing the basis for studying diverse properties of magnetic materials, including magnetic structures and domains~\cite{aizu1970,ms}, symmetry-allowed tensor responses such as magnetoelectric and transport coefficients~\cite{gallego2019tensor,etxebarria2025}, symmetry-breaking phase transitions~\cite{BC,aizu1970}, selection rules in magnetic crystals~\cite{cracknell1967}, topological quasiparticles~\cite{Yu2022Encyclopedia,212,msgcorep2023,bradlyn2016beyond,cano2019multifold,hua2018dirac}, and so on. 
However, a complete tabulation of irreducible (co-)representations [(co-)irreps] for little groups of all $\bm{k}$ points on the 1651 MSGs, which is essential for constructing symmetry indicators~\cite{si4,si6} and topological quantum chemistry~\cite{si1,si2}, had not been fully established until 2021~\cite{si2, 212}. 

\begin{figure*}[htbp]
	\includegraphics[width=\textwidth]{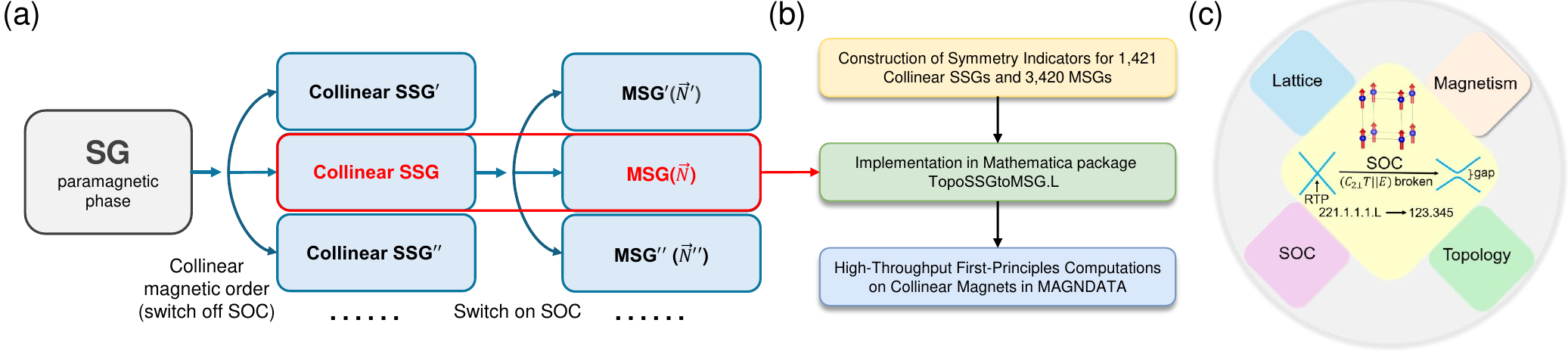}\\
	\caption{(a) Starting from the paramagnetic phase whose symmetry is described by an SG with time-reversal symmetry, the formation of a collinear magnetic order reduces the SG to an SSG first, which further reduces to an MSG. In such a symmetry-reduction sequence, SOC is hypothetically switched off first and then switched on. (b) Symmetry indicators are first constructed for the 1,421 collinear SSGs and 3,420 MSGs, then implemented in the Mathematica package TopoSSGtoMSG.L. It can be applied to high-throughput first-principles calculations of the electronic band topology of collinear magnetic materials in MAGNDATA with experimentally determined magnetic structures. (c) SOC reduces the symmetry group from an SSG to an MSG and thus intertwines lattice symmetry, magnetism, SOC, and topology. An example is shown in the center: a ferromagnetic cubic system described by SSG 221.1.1.L hosts a real triple point (RTP) at $\bm{k}=(0,0,0)$, which becomes gapped since the symmetry is reduced to MSG 123.345 in the Belov-Neronova-Smirnova (BNS) notation~\cite{BNS} with tetrahedral symmetry.
	}
	\label{fig:rtp}
\end{figure*}

\begin{figure}[htbp]
	\includegraphics[width=0.9\columnwidth]{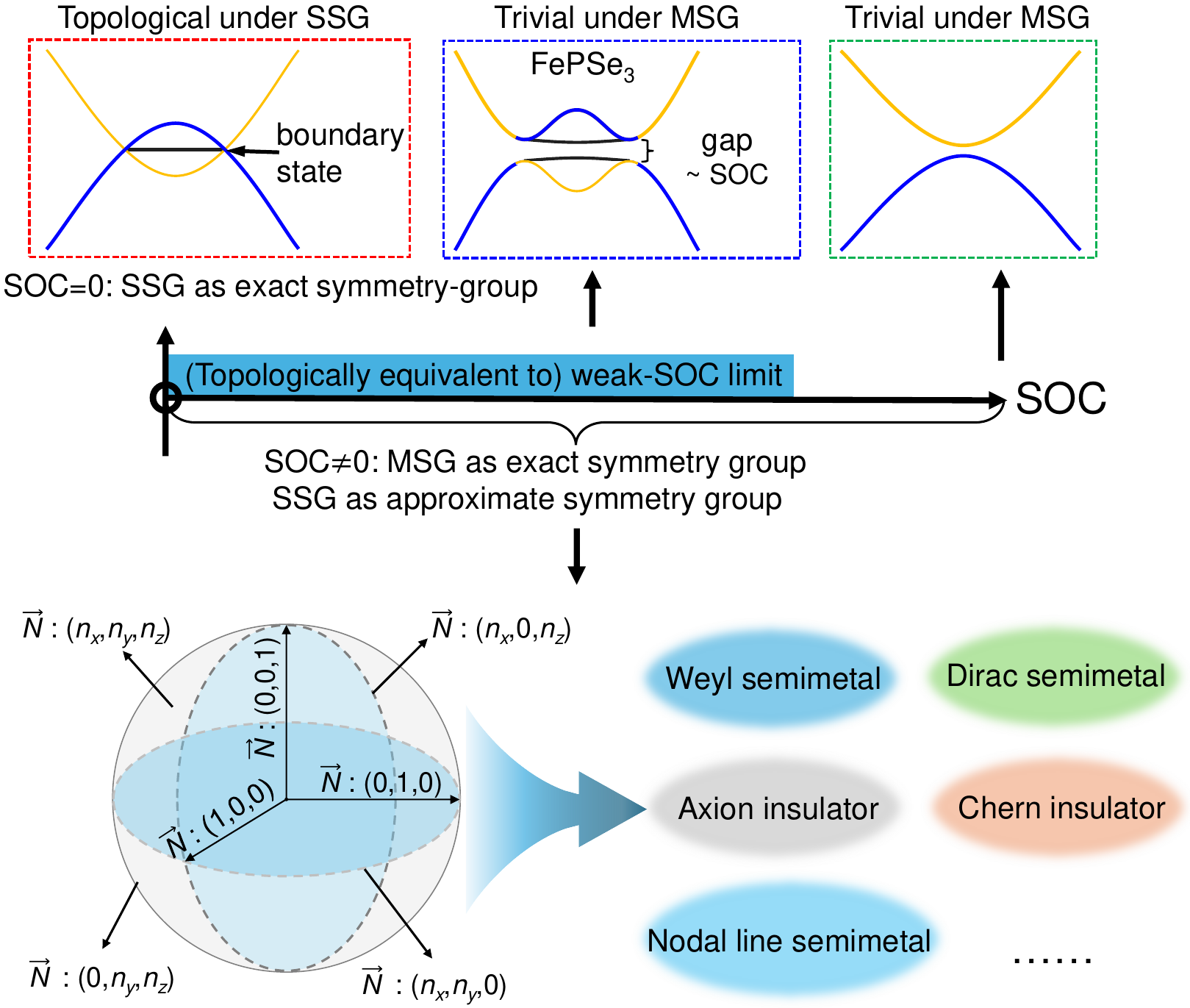}\\
	\caption{With SOC increased from vanishing to finite, the symmetry group for the electronic system of a magnet is reduced immediately from an SSG to an MSG. The resulting MSG is uniquely determined by the MM direction $(\vec{N})$ for a collinear magnet. As shown in the lower panel, one can obtain a specific electronic band topology for each MM direction. Though there are an infinite number of MM directions, they can be divided into several classes based on the SSG, with the MM directions in each class corresponding to one MSG. The compatibility relations from SSG to MSG for all classes of MM directions can be applied to quantitatively characterize whether the SOC is weak and the SSG is a reasonably approximate symmetry, and if so, one can exhaustively map from all MM directions to the topological classifications with only the HSP symmetry data obtained by first-principles calculations without SOC (based on SSG). In the upper panel, we demonstrate a possible evolution of band structure: when SOC is absent, SSG symmetry can protect some band nodes from band inversion with boundary states which can be diagnosed by SSG. Once SOC can be considered to be weak, the band inversion persists with the boundary states remaining only with a tiny gap opened by SOC, which cannot be diagnosed by MSG (we take FePSe$_3$ predicted in this work as a demonstration). When SOC cannot be regarded as weak, one can directly diagnose the band structure to be trivial by MSG.
	}
	\label{fig:change}
\end{figure}

Although an MSG is the exact symmetry group of a magnet, it may fail to capture certain electronic band degeneracies and unconventional magnon band degeneracies~\cite{nsoc3,magnon,nsoc7}. Very recently, the spin space group (SSG)~\cite{ssg1,ssg2,ssg3}, which accurately describes the symmetry of magnetic materials in the absence of spin–orbit coupling (SOC), has attracted renewed worldwide interest~\cite{differentFacets}. Notably, for nonmagnetic materials, the 230 SGs are sufficient regardless of whether SOC is included. In contrast, there are infinitely many SSGs~\cite{nsoc4,nsoc5,nsoc6}, among which 1,421 describe collinear magnets, i.e., magnets with parallel or antiparallel local magnetic moments (MMs). A complete representation theory based on the 1,421 SSGs has been applied in discovering unconventional magnons~\cite{magnon}. However, naturally nonvanishing SOC reduces the symmetry of the electronic system to MSG, and establishing a new framework considering such symmetry reduction is necessary for a complete characterization of electronic band topology in magnetic materials. 
In the following, we first provide a brief review of SSGs for collinear magnets and then propose a scheme of diagnosing electronic topology combining SSG and its MSG subgroups.

\sectionnotoc{A brief review of symmetry groups for collinear magnets}

\subsectionnotoc{SSG for collinear magnets}
The SSG for a collinear magnet can be written in the form of $G^S \times G^N$~\cite{ssg3}, where $G^S$ and $G^N$ are the spin-only and nontrivial parts, respectively. 
The operations of the nontrivial part $G^N$ can be written in the form of $(R_s||R_l)$, where $R_s$ and $R_l$ denote operations in spin space and real space, respectively, and $R_s$ is either the identity $E$ or time reversal $T$, which preserves or reverses spins, respectively.
Specifically, $R_l$ relates atomic positions in real space and can be written in Seitz notation as $R_l=\lbrace\alpha|\bm{t}\rbrace$, where $\alpha$ denotes a point group operation and $\bm{t}$ denotes a translation.
Hence, $(R_s||R_l)$ requires the spins at sites $\bm{r}$ and $R_l\bm{r}=\alpha\bm{r}+\bm{t}$ to be identical or opposite in sign for $R_s=E$ or $T$, respectively.
If $R_s$ takes only $E$, there exist 230 SSGs describing ferromagnets (FMs). If $R_s$ can take both $E$ and $T$, the remaining SSGs describe antiferromagnets (AFMs). When $G^N$ contains $(T||P)$ with $P$ denoting the spatial inversion, or $(T||\tau)$ with $\tau$ denoting a fractional lattice translation, the nonrelativistic band spin splitting is forced to vanish throughout the Brillouin zone, which corresponds to conventional AFMs (252/517 SSGs for $PT$/$\tau T$ AFMs, respectively). 
When $G^N$ contains neither $(T||P)$ nor $(T||\tau)$, the point group operation $\alpha$ in $R_l$ can only be a rotation or rotoinversion when $R_s=T$. There are 422 such SSGs, and the corresponding AFMs are altermagnets (AMs)~\cite{am2,am1,am12}, allowing nonvanishing band spin splitting~\cite{momentumSpinSplitting,am4,am5,am6,am8,am9,am14,am16,am11,am18} at generic $\bm{k}$ points (GPs). To summarize, if we further consider whether the SSG is centrosymmetric (namely, containing $P$ as a sole symmetry), there are 230 FM SSGs (92 contain $P$), 422 AM SSGs (160 contain $P$), 252 $PT$ AFM SSGs, and 517 $\tau T$ AFM SSGs (228 contain $P$).

\begin{figure*}[htbp]
	\includegraphics[width=\textwidth]{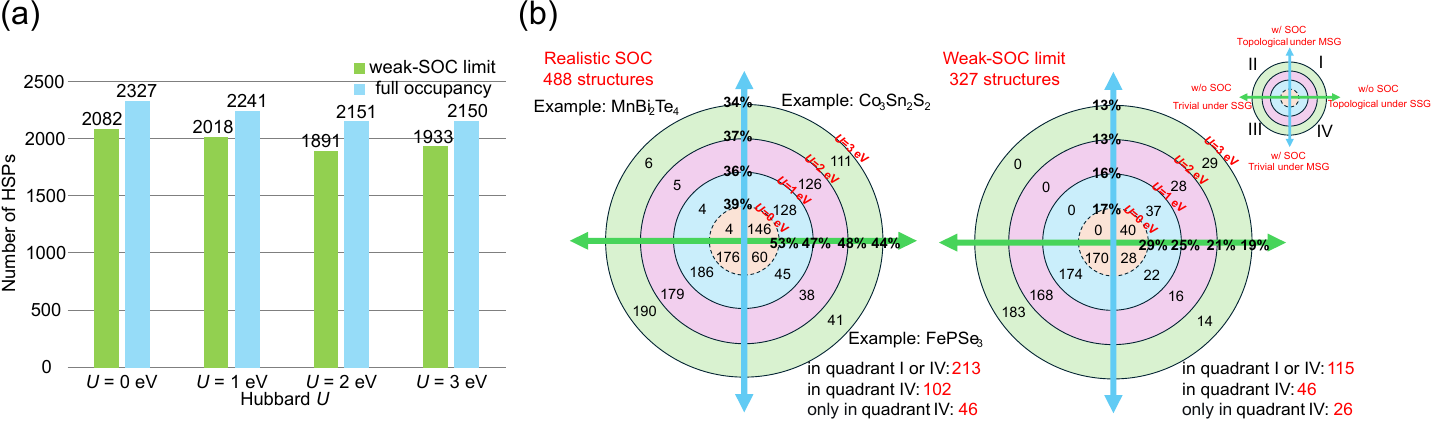}\\
	\caption{Statistics of HSPs and materials based on the high-throughput investigation of MAGNDATA using our scheme.
		(a) Statistics of HSPs satisfying the weak-SOC limit, where all $n_{\bm{k}}^{\mathrm{SSG},j}$ are required to be integers for full occupancy. (b) Left panel: Statistics of 488 materials classified as topological or trivial under SSG or MSG by symmetry indicators. The integers within each sector (corresponding to each value of $U$) represent the number of materials in four quadrants.  Right panel: The same statistics restricted to materials in the weak-SOC limit. Here, ``only in quadrant IV'' in panel (b) denotes materials that are classified as topological under SSG but trivial under MSG in all converged self-consistent first-principles calculations among $U=0$, 1, 2, and 3~eV.
	}
	\label{fig:statistics}
\end{figure*}

Besides, $G^S = \mathrm{SO}(2) + (C_{2\perp}T \,\|\, E)\mathrm{SO}(2)$ denotes the spin-only part. Here, $\mathrm{SO}(2)$ represents arbitrary spin rotations about the MM, while $C_{2\perp}$ denotes a twofold rotation about an axis perpendicular to the MM. $G^S$ constrains the nonrelativistic electronic Hamiltonian of a collinear magnet to take the block-diagonal form $H(\bm{k})=\begin{pmatrix}H_{\uparrow}(\bm{k})&\mathbf{0}\\ \mathbf{0}&H_{\downarrow}(\bm{k})\end{pmatrix}$, where $H_{\uparrow}(\bm{k})$ and $H_{\downarrow}(\bm{k})$ denote the spin-up and spin-down blocks, respectively.
In the absence of SOC, the topological classifications for $H_{\uparrow}(\bm{k})$ and $H_{\downarrow}(\bm{k})$ can be regarded as equivalent to those for the SG constituted by $(E||R_l)$ with spinless time-reversal symmetry, as completed in Ref.~\cite{nsocsg2}. 
However, this method treating $H_{\uparrow}(\bm{k})$ and $H_{\downarrow}(\bm{k})$ separately cannot capture spin-degenerate degeneracies such as the 12-fold nodal-point degeneracy~\cite{magnon}, and requires the assumption that there are no symmetry-enforced band crossings at $\bm{k}$ points other than GPs.

It is worth noting that the MM direction can be specified arbitrarily maintaining the group structure, while SOC reduces the symmetry of the electronic system to an MSG determined by the MM direction.
In the following, we clarify that the SSG can be viewed as an intermediate symmetry during the transition from a paramagnetic phase to a magnetic phase with collinear magnetic order~\cite{nsoc3,am2,am1}, as illustrated in Fig.~\hyperref[fig:rtp]{\ref*{fig:rtp}(a)}.

\begin{table}[htbp]
	\centering
	\caption{46 magnetic materials that are diagnosed to be trivial (in case I) by MSG when realistic SOC and MM directions are considered, but topological (in case II/III) by SSG neglecting SOC. These materials are categorized according to the type of collinear magnet, and are specified by chemical formula (ID in MAGNDATA). The 26 materials which can be thought to take the weak-SOC limit are printed in bold.}
	\label{tab:46}
	\resizebox{\columnwidth}{!}{%
		\begin{tabular}{clll}
			\hline
			\hline
			\noalign{\vskip 3pt}
			\textbf{Type} & \multicolumn{3}{c}{\textbf{Chemical formula (ID)}} \\
			\noalign{\vskip 3pt}
			\hline
			\noalign{\vskip 3pt}
			
			FM & \textbf{SrMnSb$_2$ (0.767)} & \textbf{Ca$_2$NiOsO$_6$ (0.796)} &  \\
			
			\noalign{\vskip 3pt}\hline
			\noalign{\vskip 3pt}
			
			$\tau T$ AFM & AlCr$_2$ (0.598) & Gd$_2$CuO$_4$ (1.104) & GdBiPt (1.111) \\
			& \textbf{NiWO$_4$ (1.194)} & \textbf{FePSe$_3$ (1.210)} & FeCl$_2$ (1.241) \\
			& FeBr$_2$ (1.242) & \textbf{YBaCuFeO$_5$ (1.281)} & \textbf{ErAuGe (1.33)} \\
			& LiErF$_4$ (1.35) & \textbf{VOCl (1.37)} & \textbf{CeNiGe$_3$ (1.414)} \\
			& U$_2$N$_2$As (1.465) & \textbf{UCr$_2$Si$_2$ (1.470)} & \textbf{GdAgSn (1.505)} \\
			& KErSe$_2$ (1.578) & DyVO$_4$ (1.664) & \textbf{CoO (1.69)} \\
			& YMn$_2$ (1.746) & & \\
			
			\noalign{\vskip 3pt}\hline
			\noalign{\vskip 3pt}
			
			$PT$ AFM & \textbf{EuTiO$_3$ (0.16)} & CeMnAsO (0.186) & \textbf{DyB$_4$ (0.22)} \\
			& \textbf{CuMnAs (0.222)} & \textbf{ErGe$_3$ (0.330)} & \textbf{TbGe$_2$ (0.343)} \\
			& DyCrO$_4$ (0.372) & \textbf{Sr$_4$Fe$_4$O$_{11}$ (0.401)} & \textbf{DyRuAsO (0.451)} \\
			& \textbf{TbRuAsO (0.452)} & \textbf{ErB$_4$ (0.468)} & U$_2$N$_2$Se (0.485) \\
			& CaMn$_2$Ge$_2$ (0.604) & \textbf{CeMnSbO (0.666)} & CaMnBi$_2$ (0.72) \\
			& \textbf{MnPd$_2$ (0.798)} & \textbf{CuMnAs (0.881)} &  \\
			
			\noalign{\vskip 3pt}\hline
			\noalign{\vskip 3pt}
			
			AM & FeCO$_3$ (0.116) & Ca$_3$Co$_{2-x}$Mn$_x$O$_6$ (0.13) & CoF$_3$ (0.334) \\
			& Ce$_4$Ge$_3$ (0.448) & CrSb (0.528) & \textbf{RuO$_2$ (0.607)} \\
			& \textbf{VNb$_3$S$_6$ (0.712)} & \textbf{CaIrO$_3$ (0.79)} &  \\
			
			\noalign{\vskip 3pt}
			\hline
			\hline
		\end{tabular}%
	}
\end{table}

\subsectionnotoc{Symmetry-reduction from SSG to MSG}
For a crystal structure in the paramagnetic phase, its symmetry is characterized by an SG whose elements can be specified by the real-space operation with $T$ as a sole symmetry~\cite{BC}.
When a paramagnetic phase undergoes a transition to a magnetic phase, it has been widely accepted that the symmetry is reduced to an MSG whose elements can be a pure real-space operation or a real-space operation multiplied by $T$~\cite{BC}. 
However, such symmetry reduction is rather drastic. As a matter of fact, one can hypothetically switch off SOC first, for which the symmetry is an SSG as shown in Fig.~\hyperref[fig:rtp]{\ref*{fig:rtp}(a)}. 
When SOC is then switched on, not all operations of the SSG leave the electronic system invariant: only those operations whose spin-space operation is identical to the rotational part of the real-space operation or the rotational part of the real-space operation multiplied by $T$ are maintained as symmetry operations and constitute an MSG. 
Note that the collinear SSG contains arbitrary rotation symmetry around the MM direction and $C_{2\perp}T$ symmetry in the spin space. Hence, the MM direction should be specified to determine the resulting MSG, which can be enumerated exhaustively by considering all rotation axes of the real-space operations in the collinear SSG and the directions perpendicular to these rotation axes as possible MM directions.
Following this procedure, we obtain a total of 3,420 MSGs from the 1,421 collinear SSGs by considering all possible MM directions, which are listed in Sec.~S4 of the Supplemental Material (SM)~\cite{SM}.

Our exhaustive list of all MM directions and the resulting MSG subgroups for all the 1,421 collinear SSGs can be applied widely in various properties rather than electronic topology in this work.
For example, a nonpolar SSG can be reduced to a polar MSG, indicating that electronic polarization is induced by SOC, which is tunable by the MM directions.
With respect to the band topology, SOC could induce a gap to a topological degeneracy, as illustrated in Fig.~\hyperref[fig:rtp]{\ref*{fig:rtp}(c)}: a ferromagnetic cubic system hosts a real RTP~\cite{rtp} at $\bm{k}=(0,0,0)$ in the absence of SOC, which becomes gapped once SOC is included. 
Rotation of the MM direction could alter the resulting MSG while the position and the degeneracy of the RTP, as protected by the original SSG symmetry, could change slightly by SOC following the MM rotation. To generalize to other symmetry settings, a complete and systematic framework of studying electronic topology based on SSG and MSG is needed as we described in the following.

\sectionnotoc{Scheme of topological classification combining SSGs and MSGs}

\subsectionnotoc{Symmetry indicators and topological classification}

As complete theories of symmetry-based classification of band topology, symmetry indicators~\cite{si4,si6} and topological quantum chemistry~\cite{si1,si2} have proven highly efficient for diagnosing topological materials. 
However, all previous frameworks have involved a specific symmetry group such as an SG or an MSG that remains unchanged in the classification or diagnosis. 
In order to study the evolution of electronic band topology with SOC in collinear magnets, either SSG or MSG alone is not sufficient. The vital role in relating SSG and MSG is played by the compatibility relations, which we will discuss in the following. We first complete the symmetry indicator theory for the 1,421 collinear SSGs and their MSG subgroups for electronic systems.
This involves computing the little-group (co-)irreps at all high-symmetry points (HSPs), the compatibility relations between $\bm{k}$ points, and the (co-)irreps of site-symmetry groups at all Wyckoff positions together with their induced atomic insulators~\cite{si4,case} for all the 1,421 SSGs and the 3,420 MSG subgroups. By expanding the first-principles HSP symmetry data [namely, $n_{\bm{k}}^{\mathrm{SSG},j}$ and $n_{\bm{k}}^{\mathrm{MSG},j}$ denoting the numbers of occurrences of (co-)irrep $j$ at HSP $\bm{k}$ for the SSG and MSG, respectively] in the basis of atomic insulators~\cite{case}, the electronic band topology can be quickly classified into three cases~\cite{case}:
\begin{description}
	\item[Case I] No topology is detected, and the material is possibly trivial.
	\item[Case II] For SSG, band nodes can occur at GPs, corresponding to Weyl points (WPs), Dirac points (DPs), Weyl nodal lines, or Dirac nodal lines (DNLs) [see Appendix~\ref{A3} for a complete tabulation of the (co-)irreps of the little groups at GPs for the 1,421 collinear SSGs and the types of topologically protected band nodes at GPs]; for MSG, the material is classified as a magnetic topological insulator or a Weyl semimetal~\cite{si2,tc}.
	\item[Case III] For both SSG and MSG, symmetry-enforced band nodes can occur at HSPs, high-symmetry lines (HSLs), and high-symmetry planes (HSPLs). The positions and topological characters of these band nodes can be determined from the compatibility relations between $\bm{k}$ points. Note that for SSG, symmetry-enforced band nodes can also occur at GPs since different (co-)irreps are allowed as shown in Appendix~\ref{A3}.
\end{description}
Based on the group-theoretical data for all 1,421 collinear SSGs and 3,420 MSGs, we then complete a subroutine, TopoSSGtoMSG.L~\cite{topo}, which can routinely diagnose the electronic band topology of any collinear magnet by SSG and MSG combined with first-principles calculated wavefunctions. It can also output the coordinates, little groups, and character tables of the $\bm{k}$ points for the SSG, as well as those for the MSG determined by each class of MM directions, and one can see Appendix~\ref{A6} for a user guide to the package.

\subsectionnotoc{Compatibility relations from SSGs to MSGs}\label{IIB}

\begin{figure*}[htbp]
	\includegraphics[width=0.9\textwidth]{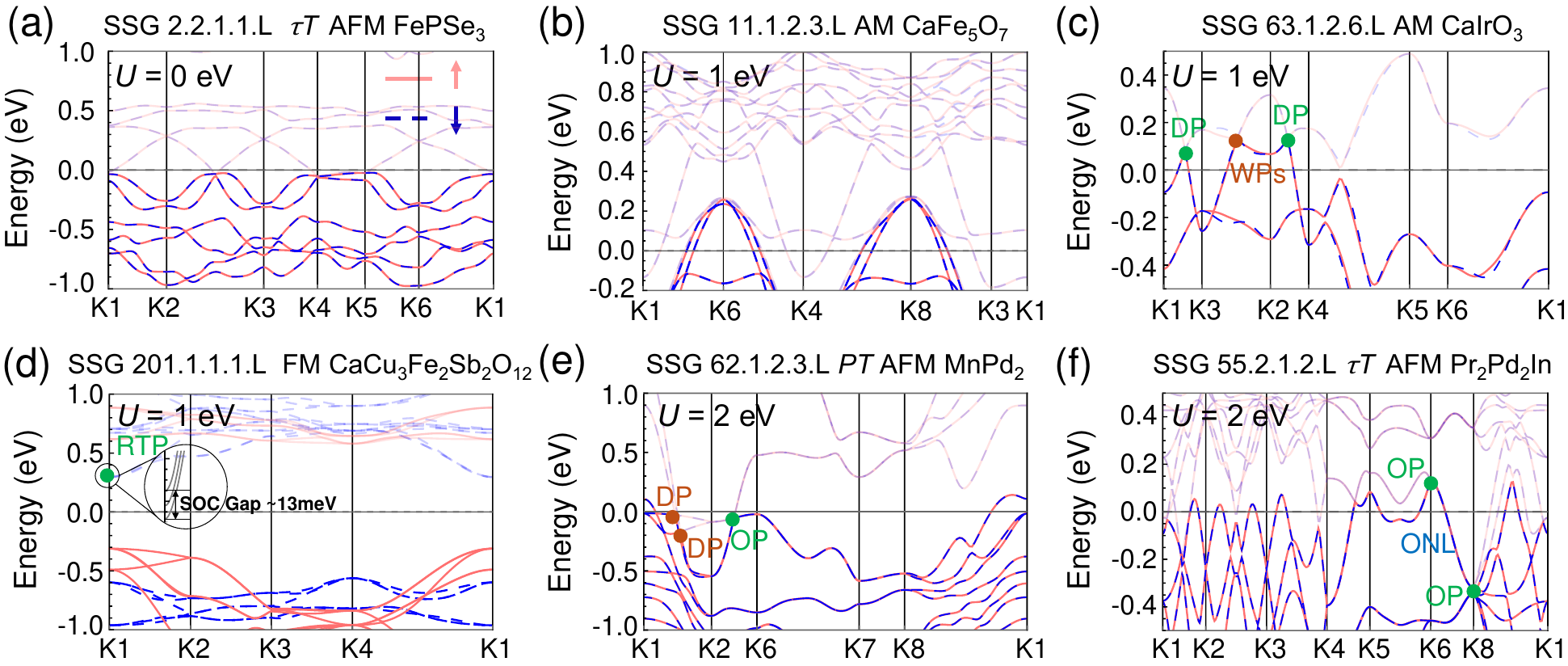}\\
	\caption{Electronic band structures of representative collinear magnets. (a) FePSe$_3$ (ID: 1.210), hosting DNLs at GPs. (b) CaFe$_5$O$_7$ (ID: 0.358), hosting Weyl nodal lines at GPs. (c) CaIrO$_3$ (ID: 0.79), hosting two DPs (green dots) along HSLs K1--K3 and K2--K4 and two WPs (brown dots) along HSL K3--K2. (d) CaCu$_3$Fe$_2$Sb$_2$O$_{12}$ (ID: 0.672), with one RTP (green dot) at HSP K1 formed by three spin-down bands. (e) MnPd$_2$ (ID: 0.798), hosting one OP (green dot) along HSL K2--K6 and two DPs (brown dots) along HSL K1--K2. (f) Pr$_2$Pd$_2$In (ID: 1.334), showing enforced OPs (green dots) at HSPs K6 and K8 and ONLs along HSL K6--K8. The coordinates of the HSPs shown in panels (a)--(f) can be obtained using TopoSSGtoMSG.L~\cite{topo}. In each band-structure panel, we also provide the name of the SSG, following the notation adopted in Ref.~\cite{iop}, and indicate the corresponding type of magnetic structure.}
	\label{fig:band}
\end{figure*}

For SGs or MSGs, compatibility relations have been utilized to study band connectivity leading to a comprehensive classification of band structures for band insulators~\cite{si4,si6,si1,si2}. For the 1,421 collinear SSGs focused on here, we computed all compatibility relations and found that the dimension of band structure space is exactly equal to that of atomic insulator space~\cite{si4,si6} for each collinear SSG. On the other hand, for the symmetry lowering from SSG to MSG, a complete calculation of the compatibility relations could provide a quantitative characterization of the strength of SOC. The compatibility relations from the little-group (co-)irrep $j$ of SSG at HSP $\bm{k}$ to the little-group (co-)irrep $j'$ of MSG at HSP $\bm{k}$ are given by
\begin{equation}\label{eq:1}
	D_{\bm{k}}^{\mathrm{SSG},j} \rightarrow \bigoplus_{j'} c_{j'j} D_{\bm{k}}^{\mathrm{MSG},j'}, 
\end{equation}
where $c_{j'j}$ are non-negative integers, $D_{\bm{k}}^{\mathrm{SSG},j}$ and $D_{\bm{k}}^{\mathrm{MSG},j'}$ denote the little-group (co-)irreps $j$ and $j'$ of SSG and MSG at HSP $\bm{k}$, respectively, and $\bigoplus$ represents the direct sum of (co-)irreps. From Eq.~\eqref{eq:1}, once we have $n_{\bm{k}}^{\mathrm{SSG},j'}$ (assumed to be integers, namely, the valence electrons fully occupy the energy levels at $\bm{k}$), we can obtain $n_{\bm{k},\text{weak-SOC}}^{\mathrm{MSG},j}$ as follows, assuming that SOC is negligibly weak:
\begin{equation}\label{eq:2}
	n_{\bm{k},\text{weak-SOC}}^{\mathrm{MSG},j}= \sum_{j'} c_{jj'}\, n_{\bm{k}}^{\mathrm{SSG},j'}.
\end{equation}
Then we check whether $n_{\bm{k},\text{weak-SOC}}^{\mathrm{MSG},j}$ are equal to $n_{\bm{k}}^{\mathrm{MSG},j}$ computed for the specified MM direction with realistic SOC, i.e., SOC included at its physical strength in the self-consistent first-principles calculation. If this equality holds, the system is considered to satisfy the weak-SOC limit at $\bm{k}$. In Sec.~S4 of the SM~\cite{SM}, we provide the explicit formulas in the form of Eq.~\eqref{eq:2} for all possible MM directions of the 1,421 collinear SSGs. It is worth mentioning that these results are not only applicable to the topological classification of collinear magnets for all MM directions, but are also expected to be applied to the complete calculation of other SOC-related properties (e.g. electronic polarization) for all MM directions~\cite{socani,rigidbody,stm,nsoc1,AHNeelTextures}.

By combining SSGs and MSGs for topological diagnosis, this scheme offers two advantages.
The first advantage is illustrated in the upper panel of Fig.~\ref{fig:change}, which shows a system that is trivial under MSG but inherits topology protected by SSG. When SOC is absent, the system hosts topologically protected boundary states. 
Upon turning on SOC, these states persist but acquire a small gap proportional to the SOC strength, as long as the band inversion remains intact in the weak-SOC limit, as exemplified by FePSe$_3$. These states persist until SOC becomes sufficiently strong to eliminate the band inversion. 
In contrast, the MSG framework alone may be insufficient to fully capture the band inversion and the associated boundary states in the weak-SOC limit.  
Another advantage is illustrated in the lower panel of Fig.~\ref{fig:change}, the topological phases for all MM directions in the weak-SOC limit can be obtained from the compatibility relations from SSG to MSG without explicit first-principles calculations with SOC. 
It is also noteworthy that a system may be trivial under SSG and remain trivial under MSG in the weak-SOC limit, while sufficiently strong SOC may induce a band inversion and drive it into a topological phase, as exemplified by MnBi$_2$Te$_4$~\cite{ahc5, ahc6}.

\begin{figure*}[htbp]
	\includegraphics[width=0.98\textwidth]{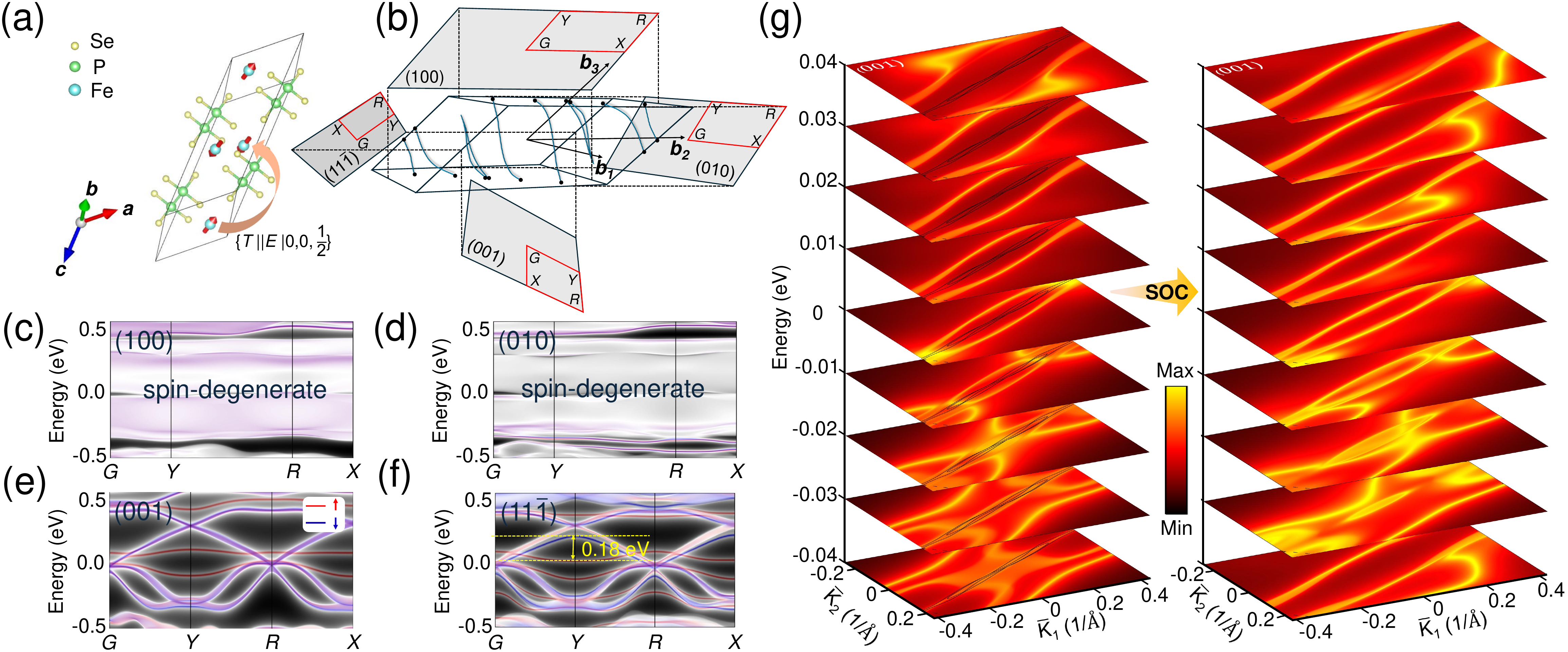}\\
	\caption{DNLs and surface states in FePSe$_3$ with $U=0~\mathrm{eV}$.
		(a) Crystal structure of FePSe$_3$, which crystallizes in SSG 2.2.1.1.L, where the orange arrow highlights two Fe atoms related by the SSG operation $\{T \,||\, E \,|\, 0,0,\tfrac{1}{2}\}$.
		(b) Bulk Brillouin zone, the $(100)$, $(010)$, $(001)$, and $(11\bar{1})$ surface Brillouin zones, and the DNLs at GPs.
		Blue lines denote the DNLs, and black dots mark their intersections with the Brillouin zone boundaries.
		(c--f) Surface states calculated without SOC along the red paths in panel (b) for the $(100)$, $(010)$, $(001)$, and $(11\bar{1})$ surface Brillouin zones, respectively.
		Panels (c) and (d) are spin-degenerate, whereas panels (e) and (f) exhibit spin splitting; the splitting in panel (f) reaches 0.18~eV.
		(g) Left: stacked electronic isoenergy contours on the $(001)$ surface Brillouin zone without SOC, where the central black lines denote the projection of the DNLs onto the $(001)$ surface Brillouin zone, forming one nodal ring and two open nodal lines.
		Right: stacked electronic isoenergy contours on the $(001)$ surface Brillouin zone with SOC.
	}
	\label{fig:jx}
\end{figure*}

\begin{figure}[htbp]
	\centering
	\includegraphics[width=0.85\columnwidth]{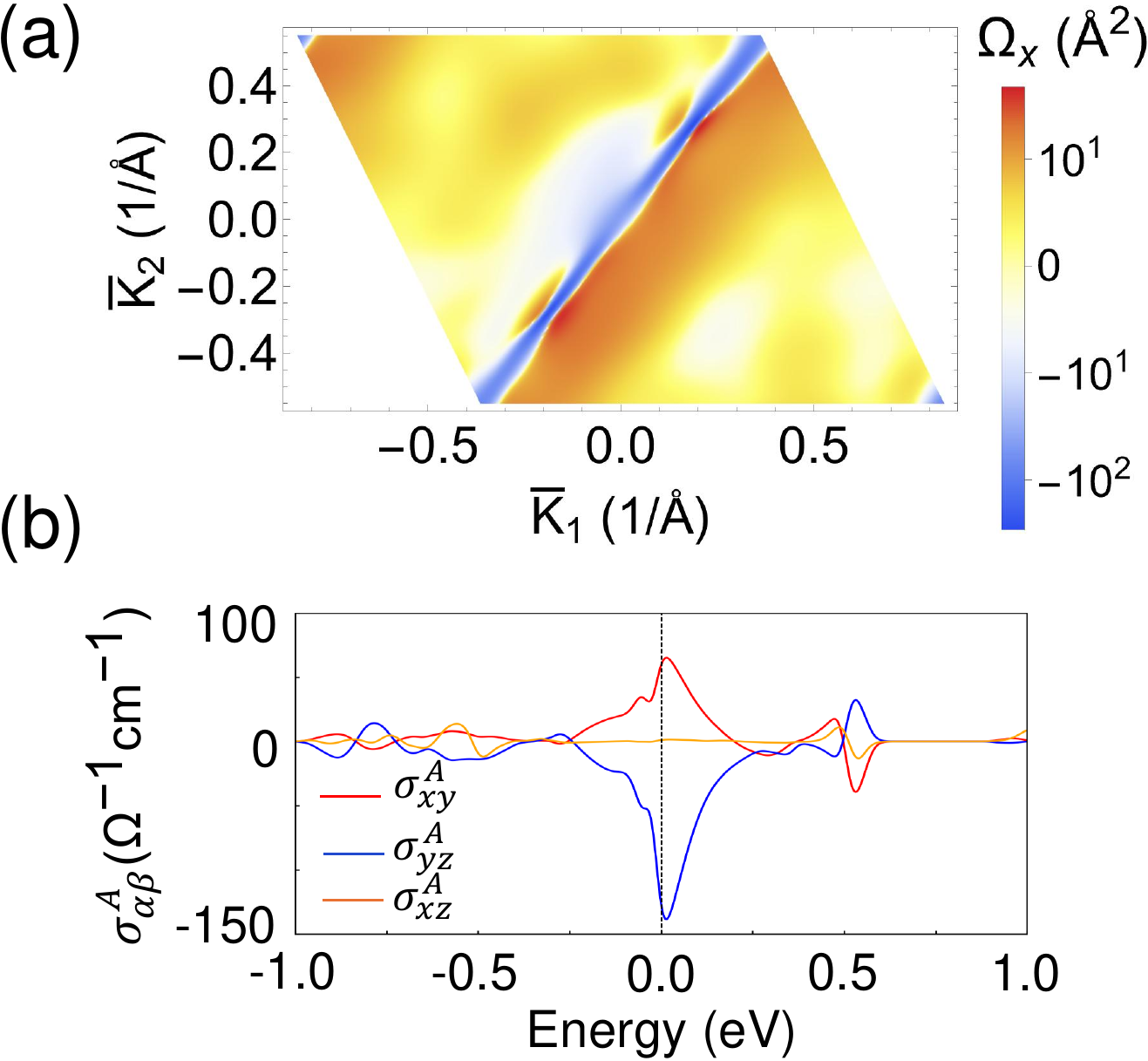}
	\caption{Berry curvature and anomalous Hall conductivity of FePSe$_3$ including SOC with $U=0~\mathrm{eV}$. 
		(a) Distributions of the Berry curvature component $\Omega_x$ for states below the Fermi level on the plane spanned by the reciprocal lattice basis vectors $\bm{b}_1$ and $\bm{b}_2$, as shown in the Brillouin zone in Fig.~\hyperref[fig:jx]{\ref*{fig:jx}(b)}.
		(b) Energy-dependent anomalous Hall conductivity $\sigma_{\alpha\beta}^{A}$, with the Fermi level set to zero.}
	\label{fig:ahc}
\end{figure}

\begin{figure*}[t]
	\includegraphics[width=0.98\textwidth]{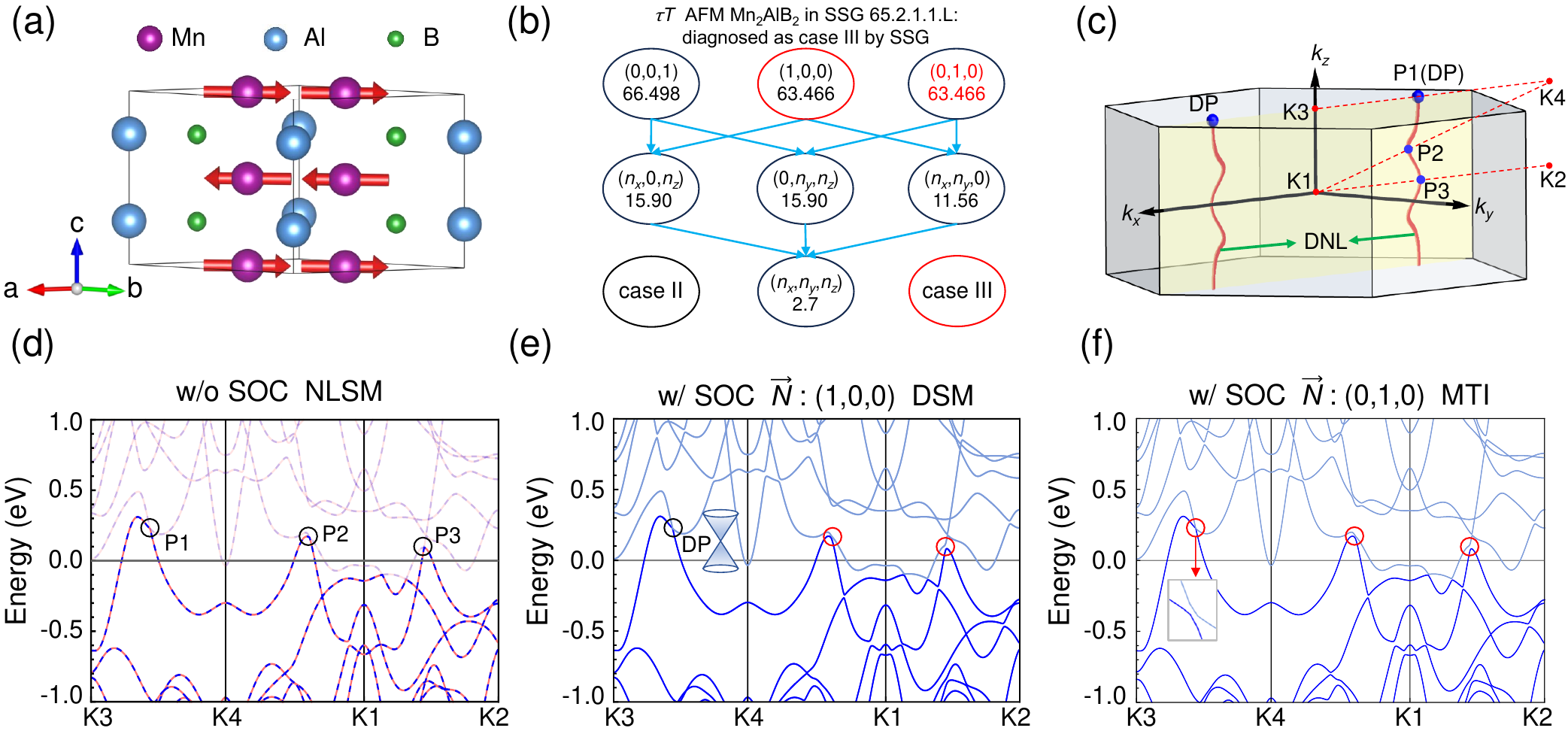}\\
	\caption{Electronic band topology of Mn$_2$AlB$_2$ tuned by the MM direction with $U=0~\mathrm{eV}$.
		(a) Crystal structure with the MM aligned along the experimentally determined direction $(0,1,0)$, where the magnetic order formed at Mn atoms is of a $\tau T$ AFM type.
		(b) Topological classification by MSG for all MM directions. The relationships between MM directions are illustrated by a tree graph, where each node (circle) represents a class of MM directions and is labeled with its corresponding MSG in the BNS notation~\cite{BNS} and the identified topology. Cyan arrows indicate inclusion relations between MM directions, where the arrow points to the direction that contains the one at its tail. For example, the node in the right-upper corner denoting the experimental MM direction (0,1,0) with MSG 63.466 can be tuned to rotate in the $(0,n_y,n_z)$ or $(n_x,n_y,0)$ plane, reducing the symmetry to MSG 15.90 or 11.56, respectively.
		(c) Evolution of nodal structures in the Brillouin zone. Without SOC, Mn$_2$AlB$_2$ is a nodal-line semimetal (NLSM), hosting a pair of inversion-related DNLs in the $k_x$–$k_z$ plane. The band structure along the red HSLs is shown in (d), where P1, P2, and P3 denote the intersections between the DNLs and the corresponding HSLs. 
		When SOC is included and the MM is aligned along $(1,0,0)$, the DNLs become gapped, and Mn$_2$AlB$_2$ turns into a Dirac semimetal (DSM), with only a pair of inversion-related DPs remaining (P1 and $-$P1). The corresponding band structure is shown in (e). 
		For other MM directions, all DNLs are fully gapped, and the system becomes a magnetic topological insulator (MTI). Panel (f) shows the band structure for the MM along $(0,1,0)$. In panels (d)--(f), the two bands enclosed by the black circles exhibit band crossings, whereas those enclosed by the red circles do not touch and are separated by a gap.
		The coordinates of HSPs K1, K2, K3, and K4 in panels (d)--(f) are shown in Table~\ref{h1} of Appendix~\ref{A6}.
	}
	\label{fig:mmd}
\end{figure*}

\sectionnotoc{High-throughput calculations on collinear magnets in MAGNDATA}

\subsectionnotoc{Statistics of the results}

\begin{figure}[htbp]
	\includegraphics[width=\columnwidth]{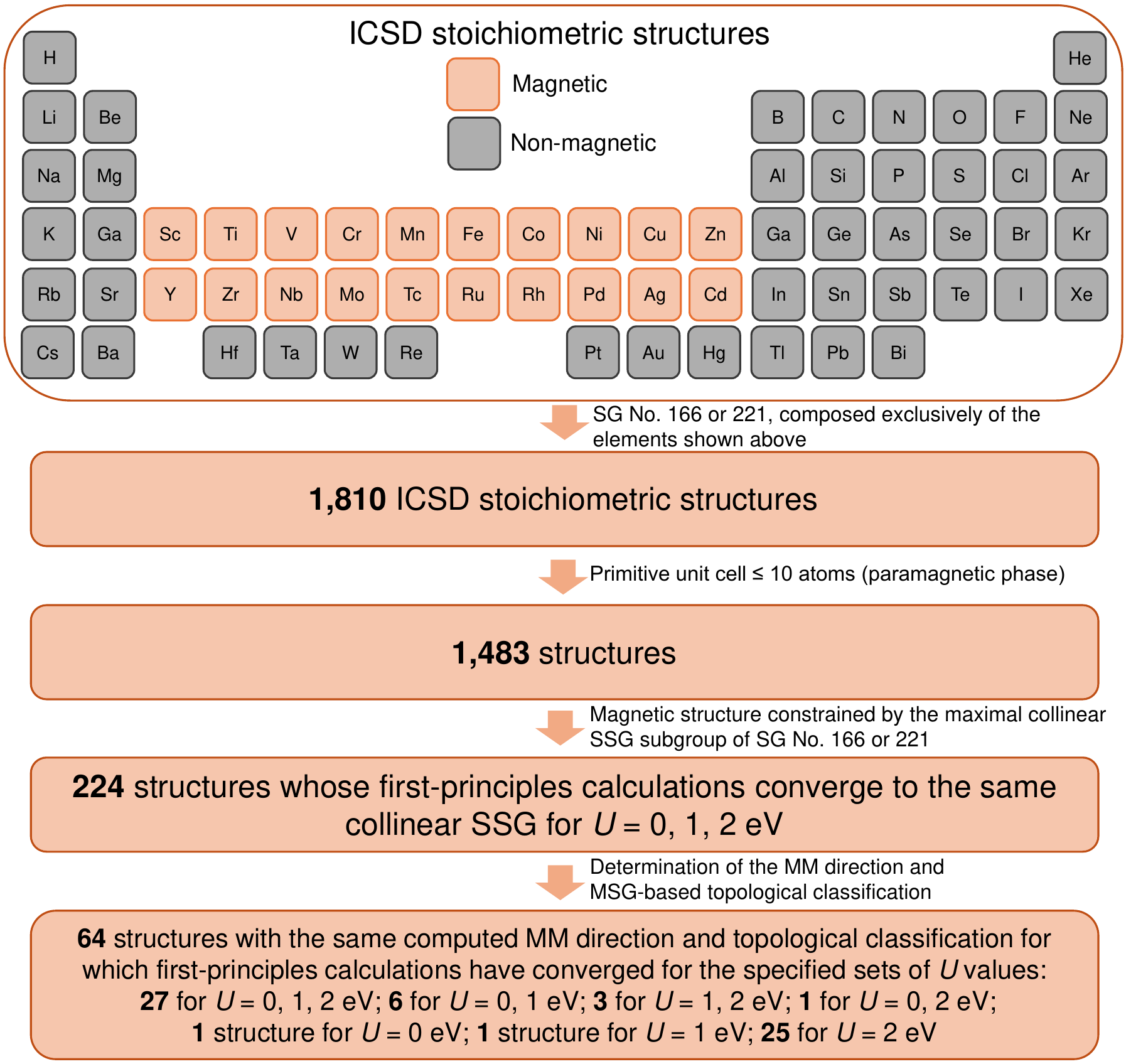}\\
	\caption{Workflow for determining ground-state collinear magnetic structures and selecting magnetic topological materials from paramagnetic crystal structures in ICSD. After distinguishing magnetic from nonmagnetic elements, we select 1,810 stoichiometric ICSD structures belonging to SG 166 or 221 and further restrict to those whose paramagnetic primitive cells contain no more than 10 atoms, leaving 1,483 structures. We then construct all candidate magnetic structures consistent with the SSG operations of the maximal collinear SSG subgroup of the parent SG (see Sec.~S4 of the SM~\cite{SM}), compute their total energies without SOC, and retain only materials whose ground-state magnetic structure remains the same for $U=0$, $1$, and $2~\mathrm{eV}$, with all magnetic structures converged and at least one magnetic atom carrying a moment $\ge 0.01\,\mu_\mathrm{B}$, resulting in 224 structures. We next compute the total energies of all specific MM directions of the ground-state magnetic structure with SOC, identify the lowest-energy MM direction satisfying the same convergence criterion, and determine the electronic band topology under the MSG dictated by that direction. This procedure yields 64 structures with the same computed MM direction and topological classification, for which first-principles calculations have converged at the specified sets of \(U\) values, corresponding to 24 magnetic topological materials with distinct chemical formulas. Specifically, among these 64 structures, 27 yield identical results for \(U=0, 1, 2~\mathrm{eV}\); 6 yield identical results for \(U=0, 1~\mathrm{eV}\); 3 yield identical results for \(U=1, 2~\mathrm{eV}\); 1 yields identical results for \(U=0, 2~\mathrm{eV}\); 1 yields identical results for \(U=0~\mathrm{eV}\); 1 yields identical results for \(U=1~\mathrm{eV}\); and the remaining 25 are determined at \(U=2~\mathrm{eV}\). Materials that are diagnosed as trivial at some \(U\) values for which the first-principles calculations have converged are not discarded, provided that all nontrivial diagnoses correspond to the same topological phase. Details are summarized in Table~\ref{tab:materials}.}
	\label{fig:icsd1}
\end{figure}

We first perform high-throughput first-principles calculations on 877 experimentally synthesized collinear magnets from MAGNDATA~\cite{magndata}, among which 488 materials achieve convergence both without and with SOC (the computational details are provided in Appendix~\ref{A1}). For these 488 materials, we then compare the HSP symmetry data $n_{\bm{k}}^{\mathrm{MSG},j}$ obtained under realistic SOC with $n_{\bm{k},\text{weak-SOC}}^{\mathrm{MSG},j}$, which are derived from $n_{\bm{k}}^{\mathrm{SSG},j}$ obtained neglecting SOC through the compatibility relations in Eq.~\eqref{eq:2}. 
Whether realistic SOC induces any exchange between occupied and unoccupied bands carrying different (co-)irreps can thus be quantitatively predicted.
Our calculations show that, for the vast majority of HSPs in the computed collinear magnets, SOC cannot induce such exchange, and thus the weak-SOC limit is satisfied (see Fig.~\hyperref[fig:statistics]{\ref*{fig:statistics}(a)}).

To more clearly illustrate the effect of SOC on electronic band topology, we count the number of materials in each of the four quadrants (as shown in Fig.~\hyperref[fig:statistics]{\ref*{fig:statistics}(b)}): (I) topological under both SSG and MSG; (II) topological under MSG but trivial under SSG; (III) trivial under both SSG and MSG; and (IV) topological under SSG but trivial under MSG. 
Considering the realistic SOC and experimentally determined MM directions, the left panel of Fig.~\hyperref[fig:statistics]{\ref*{fig:statistics}(b)} is for all 488 collinear magnetic materials: 213 structures are diagnosed as topological by SSG. Among these, 102 are classified as trivial under MSG with realistic SOC (see Sec.~S1 of the SM~\cite{SM}), of which 46 are predicted to be topological under SSG but trivial under MSG, with their specific chemical formulas and material IDs in MAGNDATA listed in Table~\ref{tab:46}. 
On the other hand, as shown in the right panel of Fig.~\hyperref[fig:statistics]{\ref*{fig:statistics}(b)}, among the 327 collinear magnets that satisfy the weak-SOC limit at all HSPs, 115 structures are topological according to their SSGs.
Of these, 46 are classified as trivial by MSG, and 26 are predicted to be topological under SSG but trivial under MSG. 
It is interesting to note that the additional topological materials identified through SSG increase the fraction of materials classified as topological by approximately 10 percentage points, as indicated by the difference between the horizontal and vertical percentages in the left and right panels of Fig.~\hyperref[fig:statistics]{\ref*{fig:statistics}(b)}. 
Besides, the evolution of electronic band topology in Co$_3$Sn$_2$S$_2$~\cite{ahc9} and MnBi$_2$Te$_4$~\cite{ahc5, ahc6} is consistent with previous theoretical studies.

We note that in the weak-SOC limit, once a magnet is diagnosed to be trivial by SSG, it must also be diagnosed as trivial by MSG, irrespective of the MM direction.
We then perform a complete classification of electronic band topology considering realistic SOC for all MM directions for these 115 materials that are topological by SSG, based on the computed HSP symmetry data using Eq.~\eqref{eq:2}. The results are listed in Sec.~S2 of the SM~\cite{SM}.

\subsectionnotoc{Representative material examples}

Then we choose several material examples to showcase the necessity of utilizing the SSG symmetry in the prediction of electronic band topology of collinear magnets, which is invisible to previous schemes based on MSGs.  
They include $\tau T$ AFM FePSe$_3$ (ID: 1.210), AM CaFe$_5$O$_7$ (ID: 0.358), AM CaIrO$_3$ (ID: 0.79), FM CaCu$_3$Fe$_2$Sb$_2$O$_{12}$ (ID: 0.672), $PT$ AFM MnPd$_2$ (ID: 0.798), and $\tau T$ AFM Pr$_2$Pd$_2$In (ID: 1.334).
The electronic band structures without SOC for these materials are plotted in Fig.~\ref{fig:band}, with relatively clean bands near the Fermi level. By SSG, FePSe$_3$ and CaFe$_5$O$_7$ are classified to be in case II, while the rest are classified to be in case III.
As detailed in Appendix~\ref{A4}, FePSe$_3$ hosts DNLs at GPs, whereas CaFe$_5$O$_7$ hosts Weyl nodal lines at GPs.
CaIrO$_3$ exhibits DPs along HSLs K1--K3 and K2--K4, as well as WPs along HSL K3--K2, approximately $0.1$~eV away from the Fermi level. 
MnPd$_2$ exhibits DPs along HSL K1--K2 and octuple points (OPs) along HSL K2--K6, where the OPs cannot be protected by MSG symmetry. 
CaCu$_3$Fe$_2$Sb$_2$O$_{12}$ and Pr$_2$Pd$_2$In host an RTP at HSP K1 and OPs at HSPs K6 and K8, respectively. 
In addition, Pr$_2$Pd$_2$In hosts octuple nodal lines (ONLs) along HSL K6--K8, which is another high-dimensional band degeneracy that cannot be protected by MSG symmetry.
All of the above-mentioned topological band degeneracies become gapped after SOC is included. 
Therefore, if the topological diagnosis were performed directly on the first-principles calculation results including SOC based solely on MSGs, these topological degeneracies would be missed.

As discussed above, the nontrivial band topology in FePSe$_3$ without SOC cannot be predicted by MSG alone.
As shown in Fig.~\hyperref[fig:jx]{\ref*{fig:jx}(b)}, the DNLs in the bulk Brillouin zone of FePSe$_3$ consist of six open nodal lines and one nodal ring. For each of the $k_i = \frac{1}{2}$ planes $(i=1,2,3)$, the DNLs intersect the plane a number of times congruent to $2$ modulo $4$, consistent with the symmetry indicators computed from the spatial-inversion eigenvalues of the spin-up and spin-down bands at the eight time-reversal-invariant momenta~\cite{nsocsg2}. Here, $k_1$, $k_2$, and $k_3$ are the components of the wave vector $\bm{k}$ in the reciprocal-lattice basis defined by $\bm{b}_1$, $\bm{b}_2$, and $\bm{b}_3$. Upon inclusion of SOC, the DNLs become gapped [shown in Fig.~\hyperref[fig:jx]{\ref*{fig:jx}(g)}], rendering the material topologically trivial as diagnosed by symmetry indicators of MSG. Although the nodal lines are gapped out by SOC, the comparison with and without SOC in Fig.~\hyperref[fig:jx]{\ref*{fig:jx}(g)} shows that part of the nodal-line-derived surface states survives. Moreover, large Berry curvature remains concentrated near the former nodal-line regions [Fig.~\hyperref[fig:ahc]{\ref*{fig:ahc}(a)}], giving rise to a sizable anomalous Hall conductivity $\sigma_{yz}^{A}$ of nearly $-150~\Omega^{-1}\mathrm{cm}^{-1}$ [Fig.~\hyperref[fig:ahc]{\ref*{fig:ahc}(b)}].

Although conventional AFMs do not exhibit band spin splitting in the bulk Brillouin zone, band spin-split surface states may emerge when $PT$ or $\tau T$ symmetry is broken at specific surfaces. For example, in $\tau T$ AFM FePSe$_3$, this corresponds to breaking the SSG operation $\lbrace T || E | 0,0,\tfrac{1}{2}\rbrace$, as illustrated in Fig.~\hyperref[fig:jx]{\ref*{fig:jx}(a)}.
Figures~\hyperref[fig:jx]{\ref*{fig:jx}(c)--\ref*{fig:jx}(d)} show spin-degenerate surface states for the (100) and (010) surface Brillouin zones which preserve the symmetry, whereas Figs.~\hyperref[fig:jx]{\ref*{fig:jx}(e)--\ref*{fig:jx}(f)} display topologically protected spin-split surface states for the (001) and $(11\bar{1})$ surface Brillouin zones, where the symmetry is broken. The spin splitting in the $(11\bar{1})$ surface Brillouin zone reaches 0.18 eV. Extending this analysis beyond FePSe$_3$, we cross-reference the 115 materials listed in Sec.~S2 of the SM~\cite{SM} with Ref.~\cite{surfam1} and identify five materials that can host topologically protected spin-split altermagnetic surface states on specific surfaces. Specifically, EuTiO$_3$ (ID: 0.16) exhibits $d$-wave spin splitting on the $(001)$, $(100)$, and $(010)$ surfaces; U$_2$Ni$_2$Sn (ID: 1.479) exhibits $d$-wave spin splitting on the $(001)$ surface; BaMnSb$_2$ (ID: 0.611) exhibits $d$-wave spin splitting on the $(001)$, $(1\bar{1}0)$, and $(110)$ surfaces; and FeSn$_2$ (ID: 1.556) and FeGe$_2$ (ID: 1.557) exhibit $g$-wave spin splitting on the $(001)$ surface.

Then we take $\tau T$ AFM Mn$_2$AlB$_2$ (ID: 1.508) to show that our scheme can be applied to an efficient classification of electronic band topology for all MM directions. The crystal structure is shown in Fig.~\hyperref[fig:mmd]{\ref*{fig:mmd}(a)}, and the topological classifications under MSG inferred from the compatibility relations from SSG to MSG for all MM directions are summarized as a tree graph in Fig.~\hyperref[fig:mmd]{\ref*{fig:mmd}(b)}. To verify these classifications, we consider 100 different MM directions (see Fig.~\ref{fig:sphere} in Appendix~\ref{A6} for details) and perform first-principles calculations including SOC for each direction. The computed values of $n_{\bm{k}}^{\mathrm{MSG},j}$ are exactly equal to those derived from $n_{\bm{k}}^{\mathrm{SSG},j}$ according to the compatibility relations from SSG to MSG. In the absence of SOC, Mn$_2$AlB$_2$ is a nodal-line semimetal hosting a pair of inversion-related DNLs in the $k_x$-$k_z$ plane (Fig.~\hyperref[fig:mmd]{\ref*{fig:mmd}(c)}). The band structure along the red HSLs in Fig.~\hyperref[fig:mmd]{\ref*{fig:mmd}(d)} reveals the intersections P1, P2, and P3 between the DNLs and the HSLs. When SOC is included and the MM is oriented along \((1,0,0)\), the DNLs become gapped except for a pair of inversion-related DPs, transforming the system into a Dirac semimetal (Fig.~\hyperref[fig:mmd]{\ref*{fig:mmd}(e)}). For all other MM directions, the DNLs are fully gapped and Mn$_2$AlB$_2$ becomes a magnetic topological insulator; the band structure for the MM direction along \((0,1,0)\) is shown in Fig.~\hyperref[fig:mmd]{\ref*{fig:mmd}(f)}, while the topological classifications for the remaining directions are provided in Sec.~S2 of the SM~\cite{SM}. 

We hope the above mentioned predicted topological conventional AFM materials with various spin-split topologically protected surface states will provide fertile platforms for exploring robust emergent surface magnetism, including surface FM and surface AM, and thereby promote the development of spintronics.

\begin{table*}[htbp]
	\caption{Magnetic topological materials identified using the workflow in Fig.~\ref{fig:icsd1}. 
		For each material, we list the chemical formula, collinear SSG, type of predicted magnetic structure, MM direction, MM magnitude, and topological phase. 
		The MM magnitude is averaged over the \(U\) values for which the first-principles calculations have converged, and reported as \(\mu_{\mathrm{B}}\) per specified magnetic atom. Specifically, the \(U\) values for which the first-principles calculations have converged are \(0\), \(1\), and \(2\) eV for Co$_3$Sn$_2$S$_2$, LiCoO$_2$, MnBi$_2$Te$_4$, MnPt$_3$, Fe$_3$Sn$_2$, LiNiO$_2$, BaFeO$_3$, and Pd$_3$N; \(0\) and \(1\) eV for Fe$_4$N, NNi$_4$, Ni$_3$Pt, and CoN; \(1\) and \(2\) eV for CAlFe$_3$, MgNi$_3$, and AuTi; \(0\) and \(2\) eV for CCo$_3$Mg; \(2\) eV for AlNi$_3$, InNi$_3$, Mn$_3$Sb, GaNi$_3$, Cr$_3$Fe, and CdN; \(1\) eV for MnN; and \(0\) eV for AuTi$_3$.}
	\label{tab:materials}
	\centering
	\small
	
	\resizebox{0.9\textwidth}{!}{
		\begin{tabular}{ccccccc}
			\hline\hline
			\textbf{Case} &
			\textbf{Formula} &
			\textbf{SSG} &
			\makecell{\textbf{Type of predicted}\\ \textbf{magnetic structure}} &
			\textbf{MM direction} &
			\makecell{\textbf{MM magnitude}\\ \textbf{($\mu_{\mathrm{B}}$/magnetic atom)}} &
			\textbf{Topological phase} \\
			\hline
			
			\multicolumn{7}{l}{\textbf{Case II}}\\
			& Co$_3$Sn$_2$S$_2$ & 166.1.1.1.L & FM & $(0,0,1)$ & $0.396/\mathrm{Co}$ & Weyl semimetal \\
			& LiCoO$_2$ & 166.2.1.1.L & $\tau T$ AFM & $(0,0,1)$ & $2.645/\mathrm{Co}$ & Axion insulator \\
			& MnBi$_2$Te$_4$ & 166.2.1.1.L & $\tau T$ AFM & $(0,0,1)$ & $4.236/\mathrm{Mn}$ & Axion insulator \\
			& MnPt$_3$ & 221.1.1.1.L & FM & $(1,-1,1)$ & $3.836/\mathrm{Mn}$ & Axion insulator \\
			
			\hline
			\multicolumn{7}{l}{\textbf{Case III}}\\
			& Fe$_3$Sn$_2$ & 166.1.1.1.L & FM & $(1,-\sqrt{3},0)$ & $2.423/\mathrm{Fe}$ & Weyl nodal-line semimetal \\
			& LiNiO$_2$ & 166.1.1.1.L & FM & $(1,-\sqrt{3},0)$ & $0.851/\mathrm{Ni}$ & Weyl nodal-line semimetal \\
			& BaFeO$_3$ & 221.1.1.1.L & FM & $(1,0,1)$ & $3.255/\mathrm{Fe}$ & Weyl nodal-line semimetal \\
			& Fe$_4$N & 221.1.1.1.L & FM & $(1,-1,1)$ & $2.590/\mathrm{Fe}$ & Weyl semimetal \\
			& AlNi$_3$ & 221.1.1.1.L & FM & $(0,1,0)$ & $0.329/\mathrm{Ni}$ & Weyl semimetal \\
			& InNi$_3$ & 221.1.1.1.L & FM & $(0,1,0)$ & $0.333/\mathrm{Ni}$ & Weyl nodal-line semimetal \\
			& NNi$_4$ & 221.1.1.1.L & FM & $(0,1,0)$ & $0.526/\mathrm{Ni}$ & Weyl nodal-line semimetal \\
			& CAlFe$_3$ & 221.1.1.1.L & FM & $(0,1,0)$ & $1.880/\mathrm{Fe}$ & Weyl nodal-line semimetal \\
			& CCo$_3$Mg & 221.1.1.1.L & FM & $(0,1,0)$ & $0.770/\mathrm{Co}$ & Weyl nodal-line semimetal \\
			& Mn$_3$Sb & 221.1.1.1.L & FM & $(0,1,0)$ & $3.670/\mathrm{Mn}$ & Weyl nodal-line semimetal \\
			& GaNi$_3$ & 221.1.1.1.L & FM & $(0,1,0)$ & $0.351/\mathrm{Ni}$ & Weyl semimetal \\
			& Ni$_3$Pt & 221.1.1.1.L & FM & $(0,1,0)$ & $0.769/\mathrm{Ni}$ & Weyl semimetal \\
			& Cr$_3$Fe & 221.2.1.1.L & $\tau T$ AFM & $(1,-1,1)$ & $2.260/\mathrm{Fe}$ & Dirac semimetal \\
			& MgNi$_3$ & 221.1.1.1.L & FM & $(0,1,0)$ & $0.504/\mathrm{Ni}$ & Weyl nodal-line semimetal \\
			& MnN & 221.1.1.1.L & FM & $(0,1,0)$ & $1.540/\mathrm{Mn}$ & Weyl nodal-line semimetal \\
			& CoN & 221.1.1.1.L & FM & $(0,1,0)$ & $1.067/\mathrm{Co}$ & Weyl nodal-line semimetal \\
			& AuTi & 221.1.1.1.L & FM & $(0,1,0)$ & $0.899/\mathrm{Ti}$ & Weyl nodal-line semimetal \\
			& AuTi$_3$ & 221.1.1.1.L & FM & $(0,1,0)$ & $0.054/\mathrm{Ti}$ & Weyl nodal-line semimetal \\
			& CdN & 221.1.1.1.L & FM & $(1,-1,1)$ & $0.055/\mathrm{Cd}$ & Weyl semimetal \\
			& Pd$_3$N & 221.1.1.1.L & FM & $(0,1,0)$ & $0.595/\mathrm{Pd}$ & Weyl nodal-line semimetal \\
			
			\hline\hline
		\end{tabular}
	}
\end{table*}

\sectionnotoc{Predicting collinear topological magnets from paramagnetic structures}\label{V}

With experimentally determined magnetic structures, one can readily apply our subroutine TopoSSGtoMSG.L~\cite{topo} to study the evolution of electronic band topology from vanishing to finite SOC, where collinear SSGs play a vital role. As a matter of fact, utilizing SSG could also accelerate the prediction of collinear magnetic structure as we will describe in the following.
Indeed, the magnetic structures for only a very tiny portion of magnetic materials have been determined. Inorganic Crystal Structure Database (ICSD)~\cite{icsd} contains over one hundred thousand stoichiometric crystal structures, around 70\% of which contain elements that can host local MMs, such as Fe, Co, and Ni, and are therefore potentially magnetic materials.
Determining the magnetic structure is essential for identifying the corresponding SSG and MSG, which further enable topological classification using TopoSSGtoMSG.L~\cite{topo}.
However, experimentally determining magnetic structures on a large scale is clearly impractical. 
Previous studies~\cite{frey2020high,su2022high} first theoretically predicted collinear magnetic ground states and then diagnosed the band topology of the predicted structures based on the corresponding MSGs. Frey \textit{et al.}~\cite{frey2020high} generated symmetry-distinct candidate magnetic configurations for more than 3,000 transition-metal oxides, determined their ground states by comparing first-principles calculated total energies, and identified 18 candidate magnetic topological materials. Su \textit{et al.}~\cite{su2022high} applied a similar workflow to 1,049 transition-metal compounds without oxygen, determined the corresponding MSGs, and performed topological classifications based on magnetic topological quantum chemistry, identifying 117 and 99 candidate magnetic topological materials without and with the Hubbard $U$ correction, respectively.
To the best of our knowledge, a systematic determination of magnetic structures for candidate magnetic materials in ICSD based on SSGs has not yet been achieved.

\begin{figure*}[htbp]
	\includegraphics[width=0.95\textwidth]{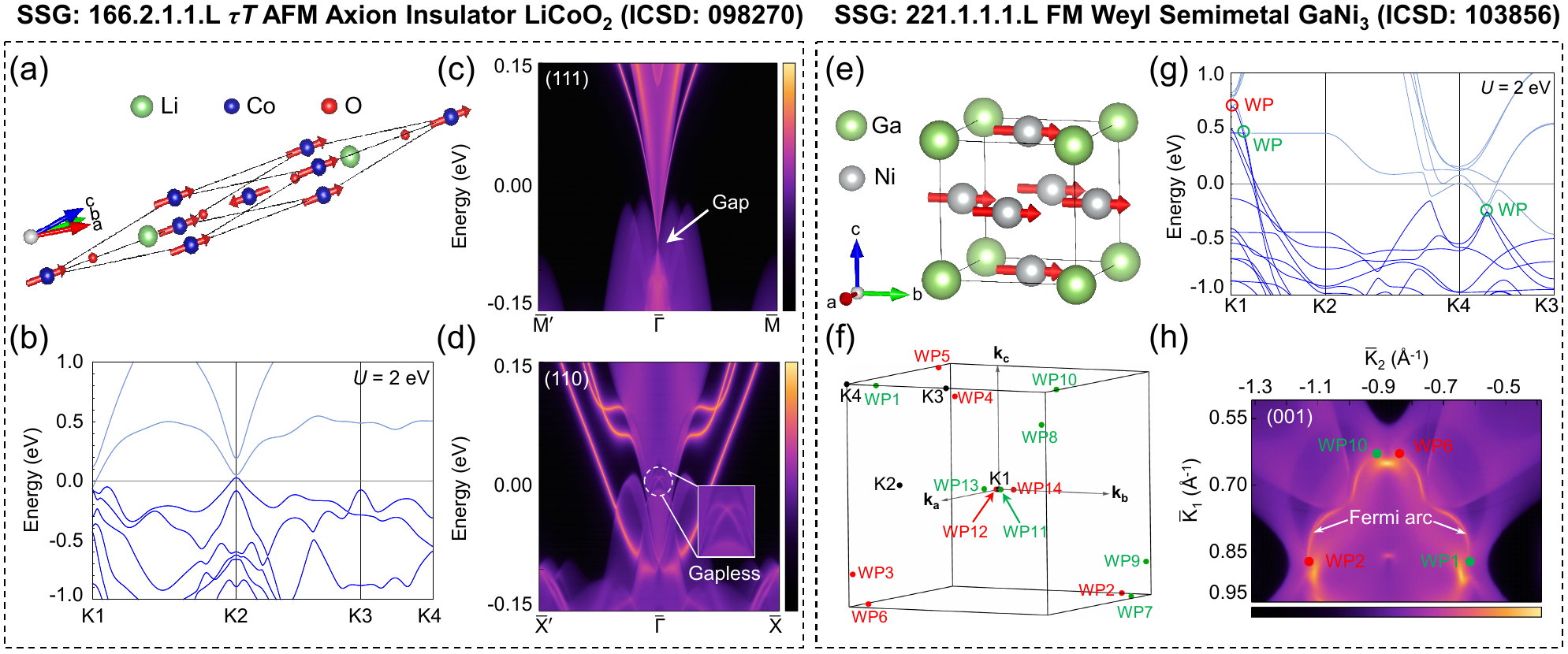}\\
	\caption{Electronic band topology of the axion insulator LiCoO$_2$ and the Weyl semimetal GaNi$_3$ calculated with SOC with $U=2~\mathrm{eV}$. (a) Crystal structure of LiCoO$_2$ (ICSD: 098270), each Co atom carries an MM of $2.716~\mu_{\mathrm{B}}$. (b) Electronic band structure of LiCoO$_2$. The coordinates of the HSPs shown in panel (b) can be obtained using TopoSSGtoMSG.L~\cite{topo}. (c) Gapped surface state of LiCoO$_2$ on the $(111)$ surface, which breaks the $\tau T$ symmetry. The surface high-symmetry path in the $(111)$ surface Brillouin zone is $\bar{\text{M}}'(\frac{1}{2},0)-\bar{\Gamma}(0,0)-\bar{\text{M}}(0,-\frac{1}{2})$. (d) Gapless surface state of LiCoO$_2$ on the $(110)$ surface, which preserves the $\tau T$ symmetry. The surface high-symmetry path in the $(110)$ surface Brillouin zone is $\bar{\text{X}}'(-\frac{1}{2},0)-\bar{\Gamma}(0,0)-\bar{\text{X}}(\frac{1}{2},0)$. (e) Crystal structure of GaNi$_3$ (ICSD: 103856). (f) WPs of GaNi$_3$ in the bulk Brillouin zone, where the red WPs carry topological charge $+1$ and the green WPs carry topological charge $-1$. (g) Electronic band structure of GaNi$_3$. (h) Fermi arcs of GaNi$_3$ in the (001) surface Brillouin zone at $E= -0.2365$ eV. The coordinates of HSPs K1, K2, K3, and K4 in panels (f) and (g) are $(0,0,0)$, $(0,-\frac{1}{2},0)$, $(\frac{1}{2},0,\frac{1}{2})$, and $(\frac{1}{2},-\frac{1}{2},\frac{1}{2})$, respectively, in the Brillouin zone shown in panel (g).}
	\label{fig:material}
\end{figure*}

In this work, we constrain the possible collinear magnetic structures using the maximal collinear SSG subgroups of the SG of a paramagnetic structure, as shown in Fig.~\hyperref[fig:rtp]{\ref*{fig:rtp}(a)}. 
The SO(2) spin-rotation symmetry in any collinear SSG constrains the MM to lie along an arbitrarily chosen spin axis and behave as a scalar suitable to be self-consistent field calculated in first-principles calculations without SOC.
And any element in the nontrivial part of a collinear SSG, $(R_s||R_l)$ requires that the magnitudes of MMs at any pair of atomic sites that are related by $R_l$ are identical or opposite in sign when $R_s=E$ or $T$, respectively.
On the other hand, the number of maximal collinear SSG subgroups of a given SG is finite, including FM/AM/$PT$ AFM and $\tau T$ AFM types. Note that for $\tau T$ AFM SSGs, the original translation symmetry of the paramagnetic phase is broken.
Consequently, there is a finite number of the corresponding collinear magnetic structures to be evaluated.
Then the magnetic structure with the least total energy is selected as the most promising one. In this step, we can determine the collinear SSG and the corresponding magnetic order with SOC switched off as shown in Fig.~\hyperref[fig:rtp]{\ref*{fig:rtp}(a)}. Next, we include SOC in the first-principles calculations for all possible MM directions: we only consider the concrete MM directions (namely, the MM directions that contain no free parameters such as $n_x$, $n_y$, $n_z$) as listed in Sec.~S4 of the SM~\cite{SM} for the identified collinear SSG in the first step, since these directions correspond to maximal MSG subgroups of the collinear SSG. Finally, the MM direction with the lowest total energy is selected as the most likely alignment direction.

\begin{table*}[htbp]
	\centering
	\caption{
		The reciprocal space positions in fractional coordinates, energies, and topological charges of the WPs in GaNi$_3$. The momentum coordinates are defined by
		$\bm{k}=k_1\bm{b}_1+k_2\bm{b}_2+k_3\bm{b}_3$,
		where $\bm{b}_i$ are the reciprocal lattice basis vectors.
	}
	\label{tab:weyl_points}
	
	\renewcommand{\arraystretch}{1.18}
	\setlength{\tabcolsep}{18pt}
	
	\begin{tabular}{cccc}
		\hline\hline
		WP
		& Position $(k_1,k_2,k_3)$
		& Energy (eV)
		& Charge \\
		\hline\hline
		
		WP1
		& $(0.5,\,-0.353,\,0.5)$
		& $-0.236$
		& $-1$ \\
		
		WP2
		& $(-0.5,\,0.353,\,-0.5)$
		& $-0.236$
		& $+1$ \\

		WP3
		& $( 0.5,\,-0.479,\,-0.353)$
		& $-0.206$
		& $+1$ \\
		
		WP4
		& $(-0.5,\,-0.479,\, 0.353)$
		& $-0.206$
		& $+1$ \\
		
		WP5
		& $(-0.353,\,-0.479,\, 0.5)$
		& $-0.206$
		& $+1$ \\
		
		WP6
		& $( 0.353,\,-0.479,\,-0.5)$
		& $-0.206$
		& $+1$ \\
		
		WP7
		& $(-0.353,\, 0.479,\,-0.5)$
		& $-0.206$
		& $-1$ \\
		
		WP8
		& $( 0.5,\, 0.479,\, 0.353)$
		& $-0.206$
		& $-1$ \\
		
		WP9
		& $(-0.5,\, 0.479,\,-0.353)$
		& $-0.206$
		& $-1$ \\
		
		WP10
		& $( 0.353,\, 0.479,\, 0.5)$
		& $-0.206$
		& $-1$ \\
		
		WP11
		& $(0,\,0.011,\,0)$
		& $0.695$
		& $-1$ \\
		
		WP12
		& $(0,\,-0.011,\,0)$
		& $0.695$
		& $+1$ \\
		
		WP13
		& $(0,\,-0.074,\,0)$
		& $0.459$
		& $-1$ \\
		
		WP14
		& $(0,\,0.074,\,0)$
		& $0.459$
		& $+1$ \\
		
		\hline\hline
	\end{tabular}
	
\end{table*}

We then apply the above strategy to stoichiometric structures listed in ICSD composed exclusively of the elements as shown in Fig.~\ref{fig:icsd1}, some of which are potentially magnetic. 
First, we concentrate on structures that crystallize in SG Nos.~166 and 221, leading to 1,810 structures. 
Second, we restrict the number of atoms per primitive unit cell to $\leq 10$, and we then obtain 1,483 structures. 
From these crystal structures, one can obtain concrete SG operations (see Appendix~\ref{A7} for the concrete SG operations of SG Nos.~166 and 221). 
Using these SG operations, we construct all possible maximal collinear SSG subgroups: 1. One FM SSG whose nontrivial part can be written as $\{(E||R_l)\}$ where $\{R_l\}$ constitute all the SG operations. 2. $PT$ AFM/AM/$\tau T$ AFM SSGs whose nontrivial parts can be written as $\{(E||R_l)\}+\{T||R_l'\}$  where $\{R_l\}$ and $\{R_l'\}$ constitute all the SG operations and $\{R_l\}$ is is a subgroup of the SG of 2nd order; in this case, $\{R_l\}$ are half of the SG operations, whose translation parts can be exactly equal to either the whole or half of the SG operations (for $PT$ AFM/AM SSGs and $\tau T$ AFM SSGs, respectively). See a complete tabulation of the maximal collinear SSG subgroups for SG Nos.~166 and 221 in Appendix~\ref{A7}. 
By these concrete SSG operations, we then construct candidate magnetic structures (see Appendix~\ref{A7} for examples) and perform first-principles calculations without SOC. 
Then, we obtain 224 structures (listed in Sec.~S3 of the SM~\cite{SM}), for which the first-principles calculations all converge to the same collinear SSG with the least total energy, for $U=0,1,2$ eV. 
Finally, we include SOC in the first-principles calculations, fixing the MM to take concrete directions.
By comparing the total energies for the calculations corresponding to these MM directions, we obtain the most promising MM direction and thus obtain the corresponding MSG and topological classification. Hence 64 structures are found to be topological, corresponding to 24 distinct chemical formulas, and their predicted magnetic structures, MM directions and magnitudes, and topological phases are summarized in Table~\ref{tab:materials}.
It is worth noting that none of the candidate magnetic structures generated for the 1,483 stoichiometric ICSD structures (see Fig.~\ref{fig:icsd1}) are AMs. Although $PT$ AFM structures are allowed for some candidates, they do not correspond to the lowest-energy converged magnetic structures. Consequently, Table~\ref{tab:materials} contains neither AMs nor $PT$ AFMs.

Notably, three experimentally verified magnetic topological materials, including magnetic Weyl semimetal Co$_3$Sn$_2$S$_2$~\cite{css2,css3}, magnetic topological insulator MnBi$_2$Te$_4$~\cite{css4,ahc7}, and magnetic Weyl nodal-line semimetal Fe$_3$Sn$_2$~\cite{fe3sn2,fe3sn22,fe3sn2_nodal}, are predicted by the above workflow. Besides, the calculated MM magnitudes also agree well with the experimentally measured values, $0.396~\mu_{\mathrm{B}}$/Co for Co$_3$Sn$_2$S$_2$, $4.236~\mu_{\mathrm{B}}$/Mn for MnBi$_2$Te$_4$, and $2.423~\mu_{\mathrm{B}}$/Fe for Fe$_3$Sn$_2$, are comparable to the experimentally reported values of approximately $0.29~\mu_{\mathrm{B}}$/Co~\cite{cosns}, $4.04~\mu_{\mathrm{B}}$/Mn~\cite{mnbite}, and $1.94$--$2.02~\mu_{\mathrm{B}}$/Fe~\cite{fe3sn2,fe3sn22}, respectively. This demonstrates that combining SSGs and MSGs provides a reliable and powerful framework for predicting collinear magnetic structures and determining their corresponding band topology based on paramagnetic crystal structures. The remaining 21 materials in Table~\ref{tab:materials} are newly predicted magnetic topological materials. We thus take the axion insulator LiCoO$_2$ and the Weyl semimetal GaNi$_3$ as examples (see Fig.~\ref{fig:material}) to illustrate details for determining their magnetic structures and the resulting topological properties.

Firstly, LiCoO$_2$ is crystallized in SG No.~166 (ICSD: 098270) as shown in Fig.~\hyperref[fig:material]{\ref*{fig:material}(a)}. It owns 8 maximal collinear SSG subgroups, as listed in Table~\ref{tab:LiCoO2_SSG} of Appendix~\ref{A7}. Only 2 of them constrain the Co atoms to have nonvanishing MM, namely, allowing the material to be magnetic. The corresponding two magnetic structures, with an FM SSG and a $\tau T$ AFM SSG, respectively, are schematically shown in Fig.~\ref{fig:LiCoO2_mag} of Appendix~\ref{A7}. For $U=0,1,2$ eV, the first-principles calculations without SOC all indicate that the magnetic structure for the $\tau T$ AFM SSG is the ground state. The $\tau T$ AFM SSG is further designated as SSG 166.2.1.1.L~\cite{iop}, for which the MM directions can be as listed in Table~S10409 of Sec.~S4 of the SM~\cite{SM}. We then choose the first 4 MM directions in the first-principles calculations with SOC, and find that for $U=0,1,2$ eV, the MM direction (0,0,1) corresponds to the ground state, with the MSG designated as MSG 167.108 in the BNS notation~\cite{BNS}. Furthermore, for this MM direction in these calculations, the material is diagnosed to be topological (case II) by MSG using TopoSSGtoMSG.L~\cite{topo}, implying that the magnetic material is either a topological insulator or a Weyl semimetal. The electronic band structure for $U=2$ eV with SOC is shown in Fig.~\hyperref[fig:material]{\ref*{fig:material}(b)}, with a relatively clear band structure and large direct gaps around the Fermi level.
Since the magnetic structure preserves $P$ symmetry (breaking $T$ symmetry), these two possibilities can be distinguished by the parity-based invariant $\mathbb{Z}_{4}=\sum_{\alpha=1}^{8}\sum_{n=1}^{n_{\mathrm{occ}}}[1+\xi_{n}(\boldsymbol{k}_{\alpha})]/2\pmod{4}$~\cite{si9,si6,si10}, where $\boldsymbol{k}_{\alpha}$ denotes the eight inversion-invariant momenta, $n_{\mathrm{occ}}$ is the number of occupied valence bands, and $\xi_{n}(\boldsymbol{k}_{\alpha})=\pm1$ is the inversion eigenvalue of the $n$-th band at $\boldsymbol{k}_{\alpha}$. Odd values, $\mathbb{Z}_{4}=1$ or $3$, indicate a Weyl semimetal, whereas $\mathbb{Z}_{4}=2$ identifies an axion insulator. For $\mathrm{LiCoO_2}$, we obtain $\mathbb{Z}_{4}=2$ and therefore identify it as an axion insulator. 
Accordingly, we observe gapped surface states in the surface Brillouin zone, for example, on the $(111)$ surface [Fig.~\hyperref[fig:material]{\ref*{fig:material}(c)}]. For a surface that preserves $\tau T$ symmetry, we further observe $\tau T$-protected gapless surface states at the center $\bar{\Gamma}(0,0)$ of the surface Brillouin zone~\cite{tc}, as shown in Fig.~\hyperref[fig:material]{\ref*{fig:material}(d)}.

For GaNi$_3$, it crystallizes in SG No.~221 (ICSD: 103856), as shown in Fig.~\hyperref[fig:material]{\ref*{fig:material}(e)}. It owns 8 maximal collinear SSG subgroups, as listed in Table~\ref{tab:GaNi3_SSG} of Appendix~\ref{A7}. Only one of them constrains the Ni atoms to have nonvanishing MMs. The corresponding magnetic structure, with an FM SSG, is schematically shown in Fig.~\ref{fig:GaNi3_mag} of Appendix~\ref{A7}. For $U=0,1,2$~eV, the first-principles calculations without SOC all reach self-consistency and identify the magnetic structure for the FM SSG as the ground state.
The FM SSG is further designated as SSG 221.1.1.1.L~\cite{iop}, for which the MM directions can be as listed in Table~S13519 of Sec.~S4 of the SM~\cite{SM}. We then choose the first 13 MM directions in the first-principles calculations with SOC, and find that for $U=0,1,2$ eV, the MM direction (0,0,1) corresponds to the ground state, with the MSG designated as MSG 123.345 in the BNS notation~\cite{BNS}.
Furthermore, for this MM direction, the material is diagnosed to be topological (case III) by MSG using TopoSSGtoMSG.L~\cite{topo}. The electronic band structure for $U=2$~eV with SOC is shown in Fig.~\hyperref[fig:material]{\ref*{fig:material}(g)}, exhibiting WPs along the HSLs K1--K2 and K4--K3. The WP along K4--K3 is identified by TopoSSGtoMSG.L~\cite{topo}. It carries a topological charge of $-1$ and is labeled WP1 in Fig.~\hyperref[fig:material]{\ref*{fig:material}(f)}. The other identified WP, labeled WP2 in Fig.~\hyperref[fig:material]{\ref*{fig:material}(f)}, carries a topological charge of $+1$ and is related to WP1 by spatial inversion,
$P:(k_1,k_2,k_3)\rightarrow(-k_1,-k_2,-k_3)$,
where $(k_1,k_2,k_3)$ are the fractional coordinates defined by
$\bm{k}=k_1\bm{b}_1+k_2\bm{b}_2+k_3\bm{b}_3$,
with $\bm{b}_1$, $\bm{b}_2$, and $\bm{b}_3$ denoting the reciprocal lattice basis vectors.
A further search based on the Wannier tight-binding model of GaNi$_3$ identifies 12 additional WPs, denoted as WP3--WP14, which are marked in Fig.~\hyperref[fig:material]{\ref*{fig:material}(f)}. 
The positions, energies, and topological charges of WP1--WP14 are listed in Table~\ref{tab:weyl_points}.
WP3--WP10 are related by the fourfold rotation $C^{+}_{4y}:(k_1,k_2,k_3)\rightarrow(k_3,k_2,-k_1)$ and spatial inversion. 
WP11 and WP12 are related by the mirror operation
$\sigma_y:(k_1,k_2,k_3)\rightarrow(k_1,-k_2,k_3)$. WP13 and WP14 are related by spatial inversion.
The constant-energy contour of the $(001)$ surface at $E=-0.2365~\mathrm{eV}$ displays two open Fermi arcs, each connecting the surface projections of two WPs with opposite chiral charges, as shown in Fig.~\hyperref[fig:material]{\ref*{fig:material}(h)}. Specifically, one Fermi arc connects the surface projections of WP1 and WP6, which carry chiral charges of $-1$ and $+1$, respectively, whereas the other connects those of WP2 and WP10, which carry chiral charges of $+1$ and $-1$, respectively.

These two materials possess relatively clean electronic band structures near the Fermi level, and their predicted topological properties and magnetic structures are worthy of future experimental verification.

\sectionnotoc{Conclusion and outlook}

To conclude, we first establish a framework suitable for studying the band topology of electronic band structures in collinear magnets that evolves with SOC through a symmetry-reduction hierarchy from SSGs to MSGs using symmetry indicators~\cite{case,si4,sg4}. 
We compute all compatibility relations between the (co-)irreps of the HSPs of the 1,421 collinear SSGs and their MSG subgroups (3,420 in total), which can be used to check whether any switching of (co-)irreps between the valence and conduction bands occurs when realistic SOC is considered. 
Once there is not any such kind of switching, the diagnosis result by MSG is thus equivalent to that obtained by considering negligibly weak SOC, a case dubbed the weak-SOC limit.
In the weak-SOC limit, an exhaustive topological classification for all possible MM directions by MSGs can be obtained since the HSP symmetry data for these MSGs can be computed by the HSP symmetry data computed neglecting SOC with respect to SSG and the compatibility relations without any further realistic first-principles calculations.

We perform high-throughput calculations with and without SOC on collinear magnets and identify 46 materials that are diagnosed as topological by SSGs but trivial by MSGs, 26 of which can satisfy the weak-SOC limit for at least one value of $U$.
As a representative example among these 26 materials, FePSe$_3$ hosts surface states inherited from its SSG topology when realistic SOC is taken into account, which cannot be predicted within the MSG framework alone. 
We also present the topological classifications for all possible MM directions by MSGs for 115 materials that are topological by SSG and satisfy the weak-SOC limit, as listed in detail in Sec.~S2 of the SM~\cite{SM}. 
Using Mn$_2$AlB$_2$ as an example, we illustrate the evolution of its electronic band topology for different MM directions. 
By varying the electron filling, we also investigate topological features within a broad energy window around the Fermi level for all computed collinear magnets from MAGNDATA. 
All computational data for the materials studied here, which span diverse chemical compositions and crystal structures, have been deposited in an online database at Ref.~\cite{magndata_results}, providing a platform facilitating future studies of MM-direction control of electronic topology in collinear magnets.

We also combine SSG and MSG with first-principles calculations to predict ground-state collinear magnetic structures from paramagnetic structures in ICSD~\cite{icsd} that crystallize in SG No.~166 or 221. This procedure identifies 24 magnetic topological materials, including three experimentally confirmed materials, magnetic Weyl semimetal Co$_3$Sn$_2$S$_2$~\cite{css2,css3}, magnetic
topological insulator MnBi$_2$Te$_4$~\cite{css4,ahc7}, and magnetic Weyl nodal-line semimetal Fe$_3$Sn$_2$~\cite{fe3sn2,fe3sn22,fe3sn2_nodal}, for which the calculated MM magnitudes also agree well with the experimental values. 
We expect the workflow shown in Fig.~\ref{fig:icsd1} can be generalized to all ICSD entries in future work.

\sectionnotocc{Acknowledgments}

This work was supported by the National Natural Science Foundation of China (NSFC) under Grants No. 12188101 and No. 12322404; the National Key R\&D Program of China (Grant No. 2025YFA1411301 and Grant No. 2022YFA1403601); Innovation Program for Quantum Science and Technology (Grant No. 2021ZD0301902), Quantum Science and Technology-National Science and Technology Major Project (Grant No. 2024ZD0300101); Fundamental and Interdisciplinary Disciplines Breakthrough Plan of the Ministry of Education of China (JYB2025XDXM411); the Fundamental Research Funds for the Central Universities (KG202501); the Natural Science Foundation of Jiangsu Province (Grants No. BK20233001 and No. BK20243011); and the Basic Research Program of Jiangsu (Grant No. BK20253054). X.W. also acknowledges support from the Tencent Foundation through the XPLORER PRIZE and the New Cornerstone Science Foundation. We acknowledge the e-Science Center of the Collaborative Innovation Center of Advanced Microstructures, Nanjing University, for providing computational resources and technical support.

\sectionnotocc{Author contributions} 

F.T. conceived and designed the project. F.T. and X.W. supervised the project. F.T. established the symmetry-indicator theory for the 1,421 collinear SSGs and the corresponding MSGs, as well as the compatibility relations between (co-)irreps at HSPs from SSGs to MSGs. L.H. performed the ab initio calculations and developed the Mathematica package. All authors contributed to the writing and editing of the manuscript.

\sectionnotocc{Data availability}
The computational data for the MAGNDATA materials that support the
findings of this study are openly available~\cite{magndata_results}.
Additional data supporting the figures and other findings of this
study are available from the corresponding author upon reasonable
request.

\bibliography{Ref}

\clearpage

\onecolumngrid
\begin{appendices}
\appendix
\makeatletter
\@addtoreset{table}{section}
\makeatother
\renewcommand{\thetable}{\Alph{section}\arabic{table}}
\makeatletter
\@addtoreset{figure}{section}
\makeatother
\renewcommand{\thefigure}{\Alph{section}\arabic{figure}}

\sectionnotoc{First-principles calculations}\label{A1}

The computational settings used in all first-principles calculations in this work, including those for materials from MAGNDATA~\cite{magndata} and structures from the ICSD~\cite{icsd}, are summarized below. For materials from MAGNDATA, the experimentally determined crystal and magnetic structures were adopted, whereas for structures from the ICSD, the experimentally determined crystal structures of the paramagnetic phases were used. For the collinear magnets from MAGNDATA, the structures are firstly adapted to the convention so that their SSG operations are the same as those listed in Sec.~S2 of SM~\cite{SM}. All calculations were performed using the Vienna \textit{Ab initio} Simulation Package (VASP)~\cite{ck1,ck2}, with exchange-correlation effects treated within the Perdew--Burke--Ernzerhof generalized-gradient approximation~\cite{ck4}.

Among the 1,612 magnetic materials in MAGNDATA, 962 exhibit collinear magnetic order. After restricting the primitive cell to no more than 50 atoms, 877 collinear magnets remain. Electronic-structure calculations both with and without SOC were performed for these 877 materials. For the localized orbitals of the magnetic atoms, the Hubbard parameter was set as $U=0$, $1$, $2$, and $3~\mathrm{eV}$. A plane-wave energy cutoff of 400~eV was used. $\Gamma$-centered Monkhorst--Pack $k$-point meshes were generated using VASPKIT~\cite{ck5}, with a uniform Brillouin-zone sampling resolution of $0.03\times 2\pi~\text{\AA}^{-1}$. The electronic self-consistency criterion was set to $10^{-5}$~eV. 

For structures from the ICSD, all calculations, both without and with SOC, employed a plane-wave energy cutoff of 520~eV and a $\Gamma$-centered $7\times7\times7$ $k$-point mesh, with the electronic self-consistency criterion set to $10^{-5}$~eV. For the localized orbitals of the magnetic atoms, the Hubbard parameter was set to $U=0$, $1$, and $2~\mathrm{eV}$. Calculations without SOC were first performed for the selected 1,483 stoichiometric paramagnetic structures. For each paramagnetic structure, candidate collinear magnetic structures were constructed, one for each maximal collinear SSG subgroup of the parent SG. In these calculations, each magnetic atom in every candidate structure was assigned an initial MM whose magnitude is $1~\mu_{\mathrm{B}}$, and the total energies of all candidate structures were evaluated. We retained only cases in which all candidate structures converged, the same lowest-energy magnetic structure was obtained for all three $U$ values, and at least one magnetically active atom in this structure had a converged MM of at least $0.01~\mu_{\mathrm{B}}$. These criteria yielded 224 structures with consistent collinear magnetic ground states. 
As shown in Fig.~\ref{fig:change}, at this step, the maximal collinear SSGs of the 224 structures have already been determined. Upon including SOC, the symmetry is reduced to an MSG. Here, we consider the MSG as a maximal subgroup of the corresponding collinear SSG. In this case, the MM direction takes a specific value and dose not contain the free variables $n_x$, $n_y$, and $n_z$. The possible MM directions for each collinear SSG are listed in Sec.~S4 of the SM~\cite{SM}.
For these 224 structures, calculations with SOC were subsequently performed for these MM directions of the corresponding ground-state magnetic structure. For each $U$ value for which these calculations converged, the lowest-energy MM direction and the corresponding band topology under its MSG were determined.

\sectionnotoc{Computation of (co-)irreps of HSPs for SSG by first-principles wavefunctions}\label{A2}
In the context of VASP, the Bloch wave function is expanded in a plane-wave (PW) basis. Since each band is spin polarized, the Bloch wave function of the $n$-th band can be written as
\begin{equation}
	\psi_{n\bm{k}}^{\sigma}(\mathbf{r})=\sum_{j}
	C^{\sigma}_{n\bm{k},j}
	e^{\mathrm{i}(\bm{k}+\bm{G}_j)\cdot\mathbf{r}},
	\tag{B1}
	\label{eq:pw}
\end{equation}
where $\bm{k}$ denotes a wave vector, $\sigma = \uparrow, \downarrow$ represents the spin index, $\mathrm{i}$ is the imaginary unit, while $C^{\sigma}_{n\bm{k},j}$ and $\bm{G}_j$ denote the $j$-th coefficient and the $j$-th reciprocal lattice vector for the $n$-th energy band and spin $\sigma$ at momentum $\bm{k}$, as obtained from VASP calculations. 

We first consider an SSG operation \( g = \{C_n(\phi) ||\alpha | \mathbf{t} \} \) of the little group at momentum \(\bm{k}\), which does not couple spin-up and spin-down bands. Here, $C_n(\phi)$ [$\phi \in [0,2\pi)$] is a rotation in spin space around the MM direction , while \(\{\alpha | \mathbf{t} \}\) acts on real space, with \(\alpha\) being the rotational part and \(\mathbf{t}\) the translational part. Then, its action on the Bloch state reads
\begin{align}
	\{C_n(\phi)\,||\,\alpha\,|\,\mathbf{t}\}\psi_{n\bm{k}}^{\sigma}(\mathbf{r})
	&= e^{-\mathrm{i}\bm{k}\cdot\mathbf{t}} \sum_{j} C^{\sigma}_{n\bm{k},j} U^{\sigma\sigma}_{\mathrm{spin}}(C_n(\phi)) e^{-\mathrm{i}\bm{G}_{j'}\cdot\mathbf{t}} e^{\mathrm{i}(\bm{k}+\bm{G}_{j'})\cdot\mathbf{r}},
	\tag{B2}
\end{align}
where $\bm{G}_{j'}=\alpha(\bm{k}+\bm{G}_j)-\bm{k}$, \( U_{\text{spin}}(C_n(\phi)) \) is the rotation matrix associated with \( g \) acting on spin space. The operation matrix acting on the combined basis can thus be expressed as
$U(g)=U_{\mathrm{spin}}(C_n(\phi))\otimes U_{\mathrm{space}}(\lbrace \alpha|\mathbf{t}\rbrace)$, with $U^{j'j}_{\text{space}}(\lbrace \alpha|\mathbf{t}\rbrace)=e^{-\mathrm{i}\bm{k}\cdot\mathbf{t}}e^{-\mathrm{i}\bm{G}_{j'}\cdot\mathbf{t}}$,
and $U_{\mathrm{spin}}(C_n(\phi))=\begin{pmatrix}e^{-\mathrm{i}\phi/2} & 0 \\0 & e^{\mathrm{i}\phi/2}\end{pmatrix}$. Next, we consider the case where the SSG operation \( g = \{C_{2\perp}C_n(\phi) ||\alpha | \mathbf{t} \} \) of the little group at momentum \(\bm{k}\) connects spin-up and spin-down bands. Here, $C_{2\perp}$ is a two-fold rotation in spin space perpendicular to the MM direction. The action of $g$ on the Bloch state then becomes
\begin{align}
	\{C_{2\perp}C_n(\phi)\,||\,\alpha\,|\,\mathbf{t}\}\psi_{n\bm{k}}^{\sigma}(\mathbf{r})
	&= e^{-\mathrm{i}\bm{k}\cdot\mathbf{t}} \sum_{j,\sigma'} C^{\sigma}_{n\bm{k},j}
	U^{\sigma\sigma'}_{\mathrm{spin}}(C_{2\perp}C_n(\phi))  e^{-\mathrm{i}\bm{G}_{j'}\cdot\mathbf{t}}
	e^{\mathrm{i}(\bm{k}+\bm{G}_{j'})\cdot\mathbf{r}} .
	\tag{B3}
\end{align}
In this case, $U_{\mathrm{spin}}(C_{2\perp}C_n(\phi))	=\begin{pmatrix}0 & -\mathrm{i}e^{\mathrm{i}\phi/2} \\-\mathrm{i}e^{-\mathrm{i}\phi/2} & 0\end{pmatrix}$. The action of $g$ on the coefficient vectors can be written as
\begin{equation}
	U(g) C_n = \sum_m \rho_{mn}(g)C_m ,
	\tag{B4}
\end{equation}
where $\rho(g)$ is the required representation matrix of $g$, and $C_{m}$ is the coefficient vector corresponding to the $m$-th energy band. Defining $C=(C_1,\ldots,C_N)$, with $N$ the total number of bands, $\rho(g)$ can be obtained as
\begin{equation}
	\rho(g)=\left(C^\dagger C\right)^{-1}C^\dagger U_g C.
	\tag{B5}
\end{equation}
Here, $C$ is a $2N_{G}\times N$ matrix, $\rho(g)$ is an $N\times N$ matrix and $N_G$ is the total number of reciprocal lattice vectors. The matrix $\rho(g)$ is block diagonal, with each block corresponding to a (non-)degenerate energy level. The character of each level is obtained by taking the trace of the corresponding block. By comparing these characters with the character table of the little group at $\bm{k}$, the (co-)irrep of each (non-)degenerate band can be uniquely determined.

\sectionnotoc{\texorpdfstring{(Co-)irreps of the little groups at GPs for four types of collinear magnets}{(Co-)irreps of the little groups at generic k points for four types of collinear magnets}}\label{A3}
When a collinear magnets is diagnosed as case II by the symmetry indicators of its collinear SSG, the corresponding topological feature is the presence of band nodes at GPs. Here, we list the (co-)irreps of the little groups and the types of band nodes at GPs for four types of collinear magnets. Based on Tables~\ref{g-1}--\ref{g-5}, we conclude that the band nodes at GPs correspond to WPs in noncentrosymmetric FMs and AMs, Weyl nodal lines in centrosymmetric FMs and AMs, DPs in $PT$ AFMs and noncentrosymmetric $\tau T$ AFMs, and DNLs in centrosymmetric $\tau T$ AFMs.

\begin{table}[htbp] 
	\caption{Irreps of the little groups at GPs for FMs or AMs without an inversion center (noncentrosymmetric), where $D_{\uparrow}$ ($D_{\downarrow}$) is the irrep for spin-up (spin-down) states, and $C_n(\phi)$ [$\phi \in [0,2\pi)$] is a rotation around the MM direction. A total of 138 (262) collinear SSGs describing noncentrosymmetric FMs (AMs) fall into this case.}
	\centering
	\renewcommand{\arraystretch}{1.1}
	\resizebox{0.3\textwidth}{!}{%
		\begin{tabular}{c|c}
			\hline
			\hline
			\diagbox{irrep}{little group} & $(C_n(\phi)||E)$ \\
			\hline
			$D_{\uparrow}$ & $(e^{-\mathrm{i} \phi/2})$ \\
			$D_{\downarrow}$ & $(e^{\mathrm{i} \phi/2})$ \\
			\hline
			\hline
		\end{tabular}%
	}\label{g-1}
\end{table}

\begin{table}[htbp] 
	\caption{Co-irreps of the little groups at GPs for FMs or AMs with an inversion center (centrosymmetric), where $D_{\uparrow}$ ($D_{\downarrow}$) is the co-irrep for spin-up (spin-down) states,  $C_n(\phi)$ [$\phi \in [0,2\pi)$] is a rotation around the MM direction,  \(T\) represents time-reversal symmetry,  \(P\) denotes spatial inversion, and \(K\) stands for the complex conjugation operator. A total of 92 (160) collinear SSGs describing centrosymmetric FMs (AMs) fall into this case.} 
	\centering
	\renewcommand{\arraystretch}{1.1}
	\resizebox{0.4\textwidth}{!}{%
		\begin{tabular}{c|cc}
			\hline
			\hline
			\diagbox{co-irrep}{little group} & $(C_n(\phi)||E)$&$(C_{2\perp}T||P)$ \\
			\hline
			$D_{\uparrow}$ & $(e^{-\mathrm{i} \phi/2})$&$(1)K$ \\
			$D_{\downarrow}$ & $(e^{\mathrm{i} \phi/2})$ &$(1)K$\\
			\hline
			\hline
		\end{tabular}%
	}\label{g-2}
\end{table}

\begin{table}[htbp] 
	\caption{Co-irreps of the little groups at GPs for $PT$ AFMs, where $D_{\uparrow+\downarrow}$ denotes the co-irrep for spin-degenerate states, $C_n(\phi)$ [$\phi \in [0,2\pi)$] is a rotation around the MM direction,  \(T\) represents time-reversal symmetry,  \(P\) denotes spatial inversion, and \(K\) stands for the complex conjugation operator. A total of 252 collinear SSGs describing $PT$ AFMs fall into this case.} 
	\centering
	\renewcommand{\arraystretch}{1.1}
	\resizebox{0.4\textwidth}{!}{%
		\begin{tabular}{c|cc}
			\hline
			\hline
			\diagbox{co-irrep}{little group} & $(C_n(\phi)||E)$&$(T||P)$ \\
			\hline
			$D_{\uparrow+\downarrow}$ & $\begin{pmatrix}
				e^{-\mathrm{i} \phi/2} & 0 \\
				0 & e^{\mathrm{i} \phi/2}
			\end{pmatrix}$& $\begin{pmatrix}
				0 & 1 \\
				-1 & 0
			\end{pmatrix}K$ \\
			\hline
			\hline
		\end{tabular}%
	}\label{g-3}
\end{table}

\begin{table}[htbp] 
	\caption{Irreps of the little groups at GPs for $\tau T$ AFMs without an inversion center (noncentrosymmetric), where $D_{\uparrow+\downarrow}$ denotes the irrep for spin-degenerate states, $C_{2\perp}$ is a two-fold rotation perpendicular to the MM direction, $C_n(\phi)$ [$\phi \in [0,2\pi)$] is a rotation around the MM direction, and $\tau$ represents a fractional translation in real space. A total of 289 collinear SSGs describing noncentrosymmetric $\tau T$ AFMs fall into this case.} 
	\centering
	\renewcommand{\arraystretch}{1.1}
	\resizebox{0.4\textwidth}{!}{%
		\begin{tabular}{c|cc}
			\hline
			\hline
			\diagbox{irrep}{little group} & $(C_n(\phi)||E)$&$(C_{2\perp}C_n(\phi)||\tau)$ \\
			\hline
			$D_{\uparrow+\downarrow}$ & $\begin{pmatrix}
				e^{-\mathrm{i} \phi/2} & 0 \\
				0 & e^{\mathrm{i} \phi/2}
			\end{pmatrix}$& $\begin{pmatrix}
				0 & -e^{\mathrm{i} \phi/2} \\
				e^{-\mathrm{i} \phi/2} & 0
			\end{pmatrix}$ \\
			\hline
			\hline
		\end{tabular}%
	}\label{g-4}
\end{table}

\begin{table}[htbp] 
	\caption{Co-irreps of the little groups at GPs for $\tau T$ AFMs with an inversion center (centrosymmetric), where $D_{\uparrow+\downarrow}$ denotes the co-irrep for spin-degenerate states, $C_{2\perp}$ is a two-fold rotation perpendicular to the MM direction, $C_n(\phi)$ [$\phi \in [0,2\pi)$] is a rotation around the MM direction, $\tau$ represents a fractional translation in real space, \(T\) represents time-reversal symmetry,  \(P\) denotes spatial inversion, and \(K\) stands for the complex conjugation operator. A total of 228 collinear SSGs describing centrosymmetric $\tau T$ AFMs fall into this case.} 
	\centering
	\renewcommand{\arraystretch}{1.1}
	\resizebox{0.4\textwidth}{!}{%
		\begin{tabular}{c|ccc}
			\hline
			\hline
			\diagbox{co-irrep}{little group} & $(C_n(\phi)||E)$&$(C_{2\perp}C_n(\phi)||\tau)$ &$(C_{2\perp}T||P)$\\
			\hline
			$D_{\uparrow+\downarrow}$ & $\begin{pmatrix}
				e^{-\mathrm{i} \phi/2} & 0 \\
				0 & e^{\mathrm{i} \phi/2}
			\end{pmatrix}$& $\begin{pmatrix}
				0 & -e^{\mathrm{i} \phi/2} \\
				e^{-\mathrm{i} \phi/2} & 0
			\end{pmatrix}$ &$\begin{pmatrix}
				0 & 1 \\
				-1 & 0
			\end{pmatrix}K$ \\
			\hline
			\hline
		\end{tabular}%
	}\label{g-5}
\end{table}

\FloatBarrier
\sectionnotoc{Parity-based symmetry indicators for FePSe$_3$ and CaFe$_5$O$_7$}\label{A4}
Following Ref.~\cite{nsocsg2}, we calculate the numbers of bands with positive and negative parities, denoted as $N_{+}(\bm{k})_{\uparrow},N_{-}(\bm{k})_{\uparrow}\ (N_{+}(\bm{k})_{\downarrow},N_{-}(\bm{k})_{\downarrow})$, for the spin-up (spin-down) states at the eight time-reversal-invariant momenta (TRIMs) $\bm{k}$: $(0,0,0)$, $(\frac{1}{2},0,0)$, $(0,\frac{1}{2},0)$, $(0,0,\frac{1}{2})$, $(\frac{1}{2},\frac{1}{2},0)$, $(\frac{1}{2}, 0, \frac{1}{2})$, $(0,\frac{1}{2},\frac{1}{2})$, and $(\frac{1}{2},\frac{1}{2},\frac{1}{2})$, expressed in the reciprocal-lattice basis $\bm{b}_1,\bm{b}_2,\bm{b}_3$. The results for FePSe$_3$ and CaFe$_5$O$_7$ are listed in Table~\ref{s1} and Table~\ref{s2}, respectively. Following the formulation in Ref.~\cite{nsocsg2}, we list the corresponding symmetry indicators $(z_{2,1},, z_{2,2},, z_{2,3}, z_{4})_{\uparrow(\downarrow)}$ of the SI group $\mathbb{Z}_2 \times \mathbb{Z}_2 \times \mathbb{Z}_2 \times \mathbb{Z}_4$ for SG 2 for spin-up and spin-down bands as follows:
\begin{equation}
	\small
	\begin{aligned}
		z_{2,i}&\equiv\sum_{\substack{\bm{k} \in \text{TRIM}\\ \text{at } k_i=\tfrac{1}{2}}}\frac{N_{-}(\bm{k})_{\uparrow(\downarrow)}-N_{+}(\bm{k})_{\uparrow(\downarrow)}}{2} \mod 2,\ i=1,2,3,\\
		z_{4}&\equiv\sum_{\bm{k}\in \text{TRIM}}\frac{N_{-}(\bm{k})_{\uparrow(\downarrow)}-N_{+}(\bm{k})_{\uparrow(\downarrow)}}{2} \mod 4.
	\end{aligned}
	\tag{D1}
	\label{eq:a1}
\end{equation}
Using Eq.~(\ref{eq:a1}) and the data in Table~\ref{s1}, we obtain $(z_{2,1}, z_{2,2}, z_{2,3}, z_4)_{\uparrow} = (1,1,1,1)$ for the spin-up bands and $(z_{2,1}, z_{2,2}, z_{2,3}, z_4)_{\downarrow} = (1,1,1,1)$ for the spin-down bands. Similarly, for CaFe$_5$O$_7$, we obtain $(z_{2,1}, z_{2,2}, z_{2,3}, z_4)_{\uparrow} = (0,0,1,0)$ for the spin-up bands and $(z_{2,1}, z_{2,2}, z_{2,3}, z_4)_{\downarrow} =  (0,0,1,0)$ for the spin-down bands.

\begin{table}[htbp]
	\centering
	\caption{Numbers of spin-up and spin-down bands with positive and negative parity eigenvalues at the eight TRIMs of FePSe$_3$.}
	\resizebox{0.7\columnwidth}{!}{%
		\begin{tabular}{c|cccc}
			\hline
			\diagbox{TRIM}{\scalebox{0.7}{Number of occurrences of even/odd parity}} & $N_{+}(\bm{k})_{\uparrow}$ & $N_{-}(\bm{k})_{\uparrow}$ & $N_{+}(\bm{k})_{\downarrow}$ & $N_{-}(\bm{k})_{\downarrow}$ \\
			\hline
			$(0,0,0)$         & 31 & 31 & 31 & 31 \\[3pt]
			$(\frac{1}{2},0,0)$ & 31 & 31 & 31 & 31 \\[3pt]
			$(0,\frac{1}{2},0)$ & 31 & 31 & 31 & 31 \\[3pt]
			$(0,0,\frac{1}{2})$ & 31 & 31 & 31 & 31 \\[3pt]
			$(\frac{1}{2},\frac{1}{2},0)$ & 31 & 31 & 31 & 31 \\[3pt]
			$(0,\frac{1}{2},\frac{1}{2})$ & 31 & 31 & 31 & 31 \\[3pt]
			$(\frac{1}{2},0,\frac{1}{2})$ & 31 & 31 & 31 & 31 \\[3pt]
			$(\frac{1}{2},\frac{1}{2},\frac{1}{2})$ & 30 & 32 & 32 & 30 \\[3pt]
			\hline
	\end{tabular}}
	\label{s1}
\end{table}

\begin{table}[htbp]
	\centering
	\caption{Numbers of bands with positive and negative parities for the spin-up (spin-down) states of CaFe$_5$O$_7$ at the eight TRIMs.}
	\resizebox{0.7\columnwidth}{!}{%
		\begin{tabular}{c|cccc}
			\hline
			\diagbox{TRIM}{\scalebox{0.7}{Number of occurrences of even/odd parity}} & $N_{+}(\bm{k})_{\uparrow}$ & $N_{-}(\bm{k})_{\uparrow}$ & $N_{+}(\bm{k})_{\downarrow}$ & $N_{-}(\bm{k})_{\downarrow}$ \\
			\hline
			$(0,0,0)$ & 49 & 43 & 49 & 43 \\[3pt]
			$(\frac{1}{2},0,0)$ & 48 & 44 & 48 & 44 \\[3pt]
			$(0,\frac{1}{2},0)$ & 48 & 44 & 44 & 48 \\[3pt]
			$(0,0,\frac{1}{2})$ & 43 & 49 & 43 & 49 \\[3pt]
			$(\frac{1}{2},\frac{1}{2},0)$ & 48 & 44 & 44 & 48 \\[3pt]
			$(0,\frac{1}{2},\frac{1}{2})$ & 44 & 48 & 48 & 44 \\[3pt]
			$(\frac{1}{2},0,\frac{1}{2})$ & 48 & 44 & 48 & 44 \\[3pt]
			$(\frac{1}{2},\frac{1}{2},\frac{1}{2})$ & 44 & 48 & 48 & 44 \\[3pt]
			\hline
	\end{tabular}}
	\label{s2}
\end{table}

\FloatBarrier

\begin{figure*}[htbp]
	\centering
	\includegraphics[width=1\textwidth]{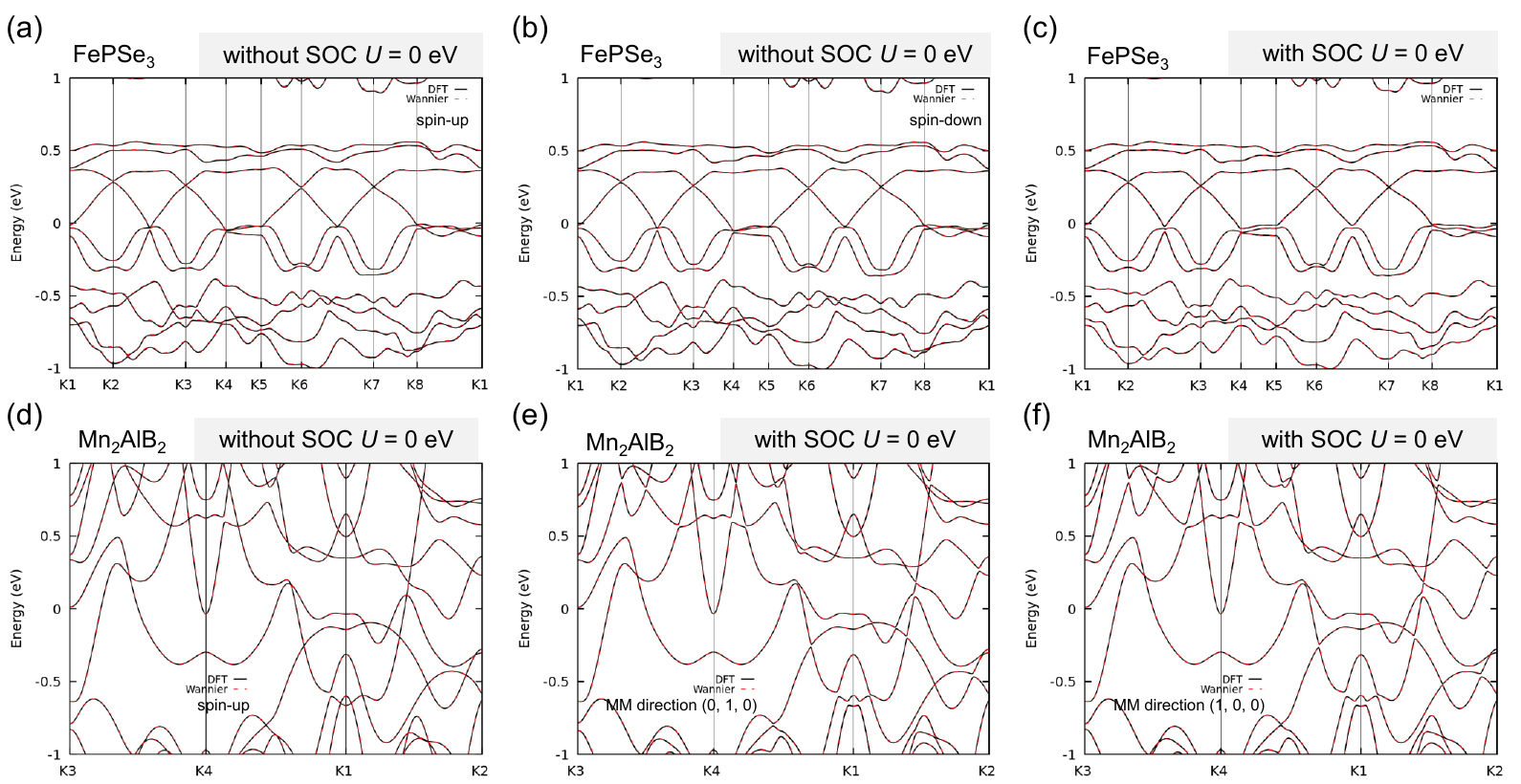}
	\caption{Comparison of Wannier-interpolated and first-principles band structures at $U = 0$~eV.
		(a) Spin-up and (b) spin-down bands of FePSe$_3$ without SOC.
		(c) FePSe$_3$ with SOC.
		For Mn$_2$AlB$_2$:
		(d) spin-up bands without SOC,
		(e)--(f) bands with SOC for the MM directions along $(0,1,0)$ and $(1,0,0)$, respectively.}
	\label{fig:wa1}
\end{figure*}

\sectionnotoc{Comparison of Wannier-interpolated and first-principles band structures for FePSe$_3$ and Mn$_2$AlB$_2$}\label{A5}
For FePSe$_3$, we built two tight-binding models (with and without SOC) using maximally localized Wannier functions from Fe 3$d$ and Se 4$p$ orbitals. Figures~\hyperref[fig:wa1]{\ref*{fig:wa1}(a)} and~\hyperref[fig:wa1]{\ref*{fig:wa1}(b)} respectively compare the spin-up and spin-down bands from Wannier interpolation and first-principles calculations without SOC at $U = 0$ eV, while Fig.~\hyperref[fig:wa1]{\ref*{fig:wa1}(c)} compares the band structures from Wannier interpolation and first-principles calculations with SOC. For Mn$_2$AlB$_2$, analogous models were constructed from Mn 3$d$ and 4$s$, B 2$s$ and 2$p$, and Al 3$s$ and 3$p$ orbitals. Since spin-up and spin-down bands are degenerate without SOC, only the spin-up comparison is shown in Fig.~\hyperref[fig:wa1]{\ref*{fig:wa1}(d)}.
Figures~\hyperref[fig:wa1]{\ref*{fig:wa1}(e)} and \hyperref[fig:wa1]{\ref*{fig:wa1}(f)} compare the band structures from Wannier interpolation and first-principles calculations with SOC at $U = 0$ eV for the MM directions along $(0,1,0)$ and $(1,0,0)$,
respectively. All tight-binding models used for Wannier interpolation were constructed using the Wannier90 package~\cite{ck6}. Based on these models, the surface states and anomalous Hall conductivity were calculated, and the nodal structures were analyzed, using the WannierTools package~\cite{ck7}.

\begin{figure}[htbp]
	\centering
	\includegraphics[width=0.8\columnwidth]{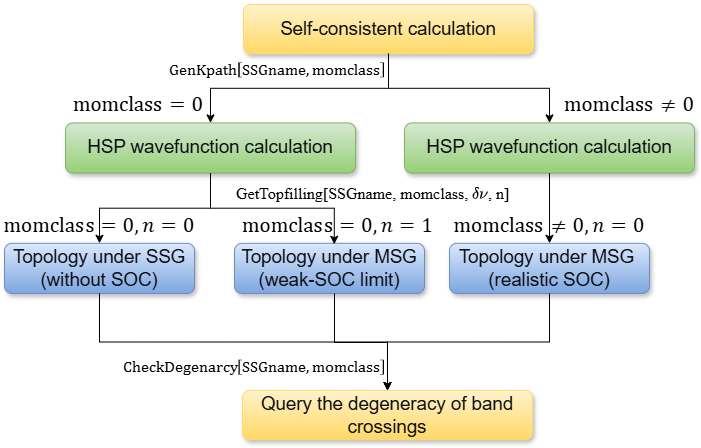}
	\caption{Computational workflow of \textbf{TopoSSGtoMSG.L} for diagnosing the topology of collinear magnetic materials.}
	\label{fig:sm2}
\end{figure}

\sectionnotoc{User guide to the Mathematica package TopoSSGtoMSG.L}\label{A6}
In this section, we present a useguide to the Mathematica package TopoSSGtoMSG.L~\cite{topo}. The TopoSSGtoMSG.L package enables the computation of topological properties of arbitrary collinear magnets using HSP wavefunctions obtained from first-principles calculations with and without SOC. It includes the following main modules: \textbf{CheckDegeneracy} module, \textbf{GenKpath} module, and \textbf{GetTopfilling} module. In the following, we first introduce how to use TopoSSGtoMSG.L to compute the topology of collinear magnetic materials under different SOC-strength regimes (see the central axis shown in Fig.~\ref{fig:rtp}), namely: without SOC, the weak-SOC limit, and the realistic SOC. For the latter two cases, the topological classification for all possible MM directions can be explored. The computational workflow is illustrated in the flowchart (Fig.~\ref{fig:sm2}). Specifically, the workflow consists of three steps: a self-consistent VASP calculation, a non-self-consistent calculation to obtain wavefunctions at HSPs with the contents of the KPOINTS file generated by \textbf{GenKpath}, and a topological calculation using \textbf{GetTopfilling}, which requires the WAVECAR and OUTCAR files from the second step. 
Here, \textbf{GetTopfilling[SSGname, momclass, $\delta\nu$, n]} computes in parallel the electronic band topology at a relative electronic filling $\delta\nu$, where $\delta\nu$ denotes the deviation from the valence-electron filling. The combination \textbf{momclass = 0, n = 0} corresponds to the SSG-based topological diagnosis using HSP wavefunctions obtained from first-principles calculations without SOC. The combination \textbf{momclass $\neq$ 0, n = 1} corresponds to the MSG-based topological diagnosis for the MM direction specified by \textbf{momclass} in the weak-SOC limit, which is obtained from HSP wavefunctions calculated without SOC together with the compatibility relations from SSG to MSG. The combination \textbf{momclass $\neq$ 0, n = 0} corresponds to the MSG-based topological diagnosis for the MM direction specified by \textbf{momclass}, using HSP wavefunctions obtained from first-principles calculations with realistic SOC.

Using the material Mn$_2$AlB$_2$ as an example, we introduce how to use the various modules of \textbf{TopoSSGtoMSG.L}. The basic information of Mn$_2$AlB$_2$ can be obtained via \textbf{CheckSSGinfo["65.2.1.1.L"]}\textbf{(*CheckSSGinfo[SSGname]*)} (Table~\ref{fig:sm1}), where \textbf{SSGname} denotes the collinear SSG name, following the notation adopted in Ref.~\cite{iop}.  

\begin{table}[htbp]
	\centering
	\caption{Primitive lattice vectors and the nontrivial part of SSG operations for Mn$_2$AlB$_2$. The quantities $n_x$, $n_y$, and $n_z$ are the components of the unit vector of the MM direction along the $x$-, $y$-, and $z$-axes in the Cartesian coordinate system.}
	\label{fig:sm1}
	\small
	\begin{tabular}{|cp{0.43\textwidth}|}
		\hline
		\multicolumn{2}{|c|}{\textbf{Spin space group 65.2.1.1.L}} \\
		\hline
		\multicolumn{2}{|c|}{\textbf{Primitive lattice vectors}} \\
		\hline
		\multicolumn{2}{|c|}{
			$\bm{a}_1=(a,-b,0),\quad
			\bm{a}_2=(a,b,0),\quad
			\bm{a}_3=(0,0,c)$
		} \\
		\hline
		\multicolumn{2}{|c|}{\textbf{Spin space group operations}} \\
		\hline
		1  & $(E\,||\,x,y,z)$ \\
		2  & $(E\,||\,-x,-y,z)$ \\
		3  & $(E\,||\,-y,-x,-z)$ \\
		4  & $(E\,||\,y,x,-z)$ \\
		5  & $(E\,||\,-x,-y,-z)$ \\
		6  & $(E\,||\,x,y,-z)$ \\
		7  & $(E\,||\,y,x,z)$ \\
		8  & $(E\,||\,-y,-x,z)$ \\
		9  & $(T\,||\,x,y,\frac{1}{2}+z)$ \\
		10 & $(T\,||\,-x,-y,\frac{1}{2}+z)$ \\
		11 & $(T\,||\,-y,-x,\frac{1}{2}-z)$ \\
		12 & $(T\,||\,y,x,\frac{1}{2}-z)$ \\
		13 & $(T\,||\,-x,-y,\frac{1}{2}-z)$ \\
		14 & $(T\,||\,x,y,\frac{1}{2}-z)$ \\
		15 & $(T\,||\,y,x,\frac{1}{2}+z)$ \\
		16 & $(T\,||\,-y,-x,\frac{1}{2}+z)$ \\
		\hline
		\multicolumn{2}{|c|}{\textbf{Magnetic moment directions}} \\
		\hline
		Class 1 & $(n_x,n_y,n_z)$ \\
		Class 2 & $(0,1,0)$ \\
		Class 3 & $(n_x,0,n_z)$ \\
		Class 4 & $(1,0,0)$ \\
		Class 5 & $(0,n_y,n_z)$ \\
		Class 6 & $(0,0,1)$ \\
		Class 7 & $(n_x,n_y,0)$ \\
		\hline
	\end{tabular}
\end{table}

The \textbf{CheckDegeneracy} module displays the labels and degeneracies of the (co-)irreps at HSPs, HSLs, HSPLs, or GP for either SSG or MSG determined by a specific class of MM directions. This module is useful for understanding the degeneracies of symmetry-enforced band nodes in case~III for both SSG and MSG. First, we present the results for SSG $G = 65.2.1.1.\mathrm{L}$ (Table~\ref{fig:ssg}). We then present the results for MSG~63.466 (BNS notation~\cite{BNS}, type~IV), determined by selecting class~2 $(0,1,0)$ of MM directions in SSG $G = 65.2.1.1.\mathrm{L}$ (Table~\ref{fig:msg}) and class~4 $(1,0,0)$ (Table~\ref{fig:msg2}).

\begin{table}
	\centering
	\caption{Labels and degeneracies of the (co-)irreps at HSPs, HSLs, HSPLs, or GP $(u,v,w)$ for SSG $G = 65.2.1.1.\mathrm{L}$, obtained using \textbf{CheckDegeneracy["65.2.1.1.L", 0](*CheckDegeneracy[SSGname, momclass]*)}, where \textbf{momclass = 0} corresponds to calculations without SOC.}
	\label{fig:ssg}
	\scriptsize
	\renewcommand{\arraystretch}{1.25}
	\begin{adjustbox}{width=0.8\textwidth}
		\begin{tabular}{|c|c|c|c|}
			\hline
			\multicolumn{4}{|c|}{\textbf{Spin space group 65.2.1.1.L}} \\
			\hline
			HSP/HSL/HSPL/GP label
			& HSP/HSL/HSPL/GP coordinate
			& (Co-)irrep labels 
			& (Co-)irrep degeneracy \\
			\hline
			
			K1 & $(0,0,0)$ 
			& $\{D_{33},D_{34},D_{35},D_{36},D_{37},D_{38},D_{39},D_{40}\}$ 
			& $\{2,2,2,2,2,2,2,2\}$ \\
			\hline
			
			K2 & $(-\frac{1}{2},\frac{1}{2},0)$ 
			& $\{D_{33},D_{34},D_{35},D_{36},D_{37},D_{38},D_{39},D_{40}\}$ 
			& $\{2,2,2,2,2,2,2,2\}$ \\
			\hline
			
			K3 & $(0,0,\frac{1}{2})$ 
			& $\{D_{17},D_{18},D_{19},D_{20},D_{21},D_{22},D_{23},D_{24}\}$ 
			& $\{2,2,2,2,2,2,2,2\}$ \\
			\hline
			
			K4 & $(-\frac{1}{2},\frac{1}{2},\frac{1}{2})$ 
			& $\{D_{17},D_{18},D_{19},D_{20},D_{21},D_{22},D_{23},D_{24}\}$ 
			& $\{2,2,2,2,2,2,2,2\}$ \\
			\hline
			
			K5 & $(-\frac{1}{2},0,0)$ 
			& $\{D_{17},D_{18},D_{19},D_{20}\}$ 
			& $\{2,2,2,2\}$ \\
			\hline
			
			K6 & $(-\frac{1}{2},0,\frac{1}{2})$ 
			& $\{D_{9},D_{10},D_{11},D_{12}\}$ 
			& $\{2,2,2,2\}$ \\
			\hline
			
			L1 & $(0,0,w)$ 
			& $\{D_{17},D_{18},D_{19},D_{20}\}$ 
			& $\{2,2,2,2\}$ \\
			\hline
			
			L2 & $(-\frac{1}{2},\frac{1}{2},w)$ 
			& $\{D_{17},D_{18},D_{19},D_{20}\}$ 
			& $\{2,2,2,2\}$ \\
			\hline
			
			L3 & $(-\frac{1}{2},0,w)$ 
			& $\{D_{9},D_{10}\}$ 
			& $\{2,2\}$ \\
			\hline
			
			L4 & $(-w,w,\frac{1}{2})$ 
			& $\{D_{9},D_{10},D_{11},D_{12}\}$ 
			& $\{2,2,2,2\}$ \\
			\hline
			
			L5 & $(-w,w,0)$ 
			& $\{D_{17},D_{18},D_{19},D_{20}\}$ 
			& $\{2,2,2,2\}$ \\
			\hline
			
			L6 & $(-w,-w,0)$ 
			& $\{D_{17},D_{18},D_{19},D_{20}\}$ 
			& $\{2,2,2,2\}$ \\
			\hline
			
			L7 & $(-w,-w,\frac{1}{2})$ 
			& $\{D_{9},D_{10},D_{11},D_{12}\}$ 
			& $\{2,2,2,2\}$ \\
			\hline
			
			P1 & $(-u,u,v)$ 
			& $\{D_{9},D_{10}\}$ 
			& $\{2,2\}$ \\
			\hline
			
			P2 & $(u,u,v)$ 
			& $\{D_{9},D_{10}\}$ 
			& $\{2,2\}$ \\
			\hline
			
			P3 & $(-v,u,0)$ 
			& $\{D_{9},D_{10}\}$ 
			& $\{2,2\}$ \\
			\hline
			
			P4 & $(-v,u,\frac{1}{2})$ 
			& $\{D_{5},D_{6}\}$ 
			& $\{2,2\}$ \\
			\hline
			
			GP & $(u,v,w)$ 
			& $\{D_{5}\}$ 
			& $\{2\}$ \\
			\hline
			
		\end{tabular}
	\end{adjustbox}
\end{table}

\begin{table}
	\centering
	\caption{Labels and degeneracies of the (co-)irreps at HSPs, HSLs, HSPLs, or GP $(u,v,w)$ for MSG 63.466, obtained using \textbf{CheckDegeneracy["65.2.1.1.L", 2](*CheckDegeneracy[SSGname, momclass]*)}, where \textbf{momclass $\neq$ 0} corresponds to calculations with SOC, and the specific nonzero value of \textbf{momclass} labels the corresponding class of MM directions in the given SSG.}
	\label{fig:msg}
	\scriptsize
	\renewcommand{\arraystretch}{1.25}
	\begin{adjustbox}{width=0.8\textwidth}
		\begin{tabular}{|c|c|c|c|}
			\hline
			\multicolumn{4}{|c|}{\textbf{SSG 65.2.1.1.L, MM direction: (0, 1, 0) MSG 63.466 (BNS notation, MSG type IV)} } \\
			\hline
			HSP/HSL/HSPL/GP label
			& HSP/HSL/HSPL/GP coordinate
			& (Co-)irrep labels
			& (Co-)irrep degeneracy \\
			\hline
			
			K1 & $(0,0,0)$
			& $\{D^{d}_{1},D^{d}_{2}\}$
			& $\{2,2\}$ \\
			\hline
			
			K2 & $(-\frac{1}{2},\frac{1}{2},0)$
			& $\{D^{d}_{1},D^{d}_{2}\}$
			& $\{2,2\}$ \\
			\hline
			
			K3 & $(0,0,\frac{1}{2})$
			& $\{D^{d}_{1},D^{d}_{2}\}$
			& $\{2,2\}$ \\
			\hline
			
			K4 & $(-\frac{1}{2},\frac{1}{2},\frac{1}{2})$
			& $\{D^{d}_{1},D^{d}_{2}\}$
			& $\{2,2\}$ \\
			\hline
			
			K5 & $(-\frac{1}{2},0,0)$
			& $\{D^{d}_{1},D^{d}_{2}\}$
			& $\{2,2\}$ \\
			\hline
			
			K6 & $(0,\frac{1}{2},0)$
			& $\{D^{d}_{1},D^{d}_{2}\}$
			& $\{2,2\}$ \\
			\hline
			
			K7 & $(-\frac{1}{2},0,\frac{1}{2})$
			& $\{D^{d}_{1}\}$
			& $\{2\}$ \\
			\hline
			
			K8 & $(0,\frac{1}{2},-\frac{1}{2})$
			& $\{D^{d}_{1}\}$
			& $\{2\}$ \\
			\hline
			
			L1 & $(0,0,w)$
			& $\{D^{d}_{1}\}$
			& $\{2\}$ \\
			\hline
			
			L2 & $(-\frac{1}{2},\frac{1}{2},w)$
			& $\{D^{d}_{1}\}$
			& $\{2\}$ \\
			\hline
			
			L3 & $(-\frac{1}{2},0,w)$
			& $\{D^{d}_{1}\}$
			& $\{2\}$ \\
			\hline
			
			L4 & $(0,\frac{1}{2},-w)$
			& $\{D^{d}_{1}\}$
			& $\{2\}$ \\
			\hline
			
			L5 & $(-w,w,\frac{1}{2})$
			& $\{D^{d}_{1},D^{d}_{2}\}$
			& $\{2,2\}$ \\
			\hline
			
			L6 & $(-w,w,0)$
			& $\{D^{d}_{1}\}$
			& $\{2\}$ \\
			\hline
			
			L7 & $(-w,-w,0)$
			& $\{D^{d}_{1}\}$
			& $\{2\}$ \\
			\hline
			
			L8 & $(-w,-w,\frac{1}{2})$
			& $\{D^{d}_{1}\}$
			& $\{2\}$ \\
			\hline
			
			P1 & $(-u,u,v)$
			& $\{D^{d}_{1}\}$
			& $\{2\}$ \\
			\hline
			
			P2 & $(u,u,v)$
			& $\{D^{d}_{1}\}$
			& $\{2\}$ \\
			\hline
			
			P3 & $(-v,u,0)$
			& $\{D^{d}_{1}\}$
			& $\{2\}$ \\
			\hline
			
			P4 & $(-v,u,\frac{1}{2})$
			& $\{D^{d}_{1}\}$
			& $\{2\}$ \\
			\hline
			
		\end{tabular}
	\end{adjustbox}
\end{table}

\begin{table}
	\centering
	\caption{Labels and degeneracies of the (co-)irreps at HSPs, HSLs, HSPLs, or GP $(u,v,w)$ for MSG 63.466, obtained using \textbf{CheckDegeneracy["65.2.1.1.L", 4](*CheckDegeneracy[SSGname, momclass]*)}, where \textbf{momclass $\neq$ 0} corresponds to calculations with SOC, and the specific nonzero value of \textbf{momclass} labels the corresponding class of MM directions in the given SSG.}
	\label{fig:msg2}
	\scriptsize
	\renewcommand{\arraystretch}{1.25}
	\begin{adjustbox}{width=0.8\textwidth}
		\begin{tabular}{|c|c|c|c|}
			\hline
			\multicolumn{4}{|c|}{\textbf{SSG 65.2.1.1.L, MM direction: (1, 0, 0) MSG 63.466 (BNS notation, MSG type IV)} } \\
			\hline
			HSP/HSL/HSPL/GP label
			& HSP/HSL/HSPL/GP coordinate
			& (Co-)irrep labels
			& (Co-)irrep degeneracy \\
			\hline
			
			K1 & $(0,0,0)$
			& $\{D^{d}_{1},D^{d}_{2}\}$
			& $\{2,2\}$ \\
			\hline
			
			K2 & $(-\frac{1}{2},\frac{1}{2},0)$
			& $\{D^{d}_{1},D^{d}_{2}\}$
			& $\{2,2\}$ \\
			\hline
			
			K3 & $(0,0,\frac{1}{2})$
			& $\{D^{d}_{1},D^{d}_{2}\}$
			& $\{2,2\}$ \\
			\hline
			
			K4 & $(-\frac{1}{2},\frac{1}{2},\frac{1}{2})$
			& $\{D^{d}_{1},D^{d}_{2}\}$
			& $\{2,2\}$ \\
			\hline
			
			K5 & $(-\frac{1}{2},0,0)$
			& $\{D^{d}_{1},D^{d}_{2}\}$
			& $\{2,2\}$ \\
			\hline
			
			K6 & $(0,\frac{1}{2},0)$
			& $\{D^{d}_{1},D^{d}_{2}\}$
			& $\{2,2\}$ \\
			\hline
			
			K7 & $(-\frac{1}{2},0,\frac{1}{2})$
			& $\{D^{d}_{1}\}$
			& $\{2\}$ \\
			\hline
			
			K8 & $(0,\frac{1}{2},-\frac{1}{2})$
			& $\{D^{d}_{1}\}$
			& $\{2\}$ \\
			\hline
			
			L1 & $(0,0,w)$
			& $\{D^{d}_{1}\}$
			& $\{2\}$ \\
			\hline
			
			L2 & $(-\frac{1}{2},\frac{1}{2},w)$
			& $\{D^{d}_{1}\}$
			& $\{2\}$ \\
			\hline
			
			L3 & $(-\frac{1}{2},0,w)$
			& $\{D^{d}_{1}\}$
			& $\{2\}$ \\
			\hline
			
			L4 & $(0,\frac{1}{2},-w)$
			& $\{D^{d}_{1}\}$
			& $\{2\}$ \\
			\hline
			
			L5 & $(-w,-w,\frac{1}{2})$
			& $\{D^{d}_{1}\}$
			& $\{2\}$ \\
			\hline
			
			L6 & $(-w,w,0)$
			& $\{D^{d}_{1}\}$
			& $\{2\}$ \\
			\hline
			
			L7 & $(-w,-w,0)$
			& $\{D^{d}_{1}\}$
			& $\{2\}$ \\
			\hline
			
			L8 & $(-w,w,\frac{1}{2})$
			& $\{D^{d}_{1},D^{d}_{2}\}$
			& $\{2,2\}$ \\
			\hline
			
			P1 & $(-u,u,v)$
			& $\{D^{d}_{1}\}$
			& $\{2\}$ \\
			\hline
			
			P2 & $(u,u,v)$
			& $\{D^{d}_{1}\}$
			& $\{2\}$ \\
			\hline
			
			P3 & $(-v,u,0)$
			& $\{D^{d}_{1}\}$
			& $\{2\}$ \\
			\hline
			
			P4 & $(-v,u,\frac{1}{2})$
			& $\{D^{d}_{1}\}$
			& $\{2\}$ \\
			\hline
			
		\end{tabular}
	\end{adjustbox}
\end{table}

Below, we illustrate the computational procedure using the altermagnetic material Mn$_2$AlB$_2$ ($U$ = 0 eV) as an example. We first introduce how to compute the topology without SOC, which is diagnosed by SSG. First, perform the self-consistent calculation. After it is completed, use \textbf{GenKpath["65.2.1.1.L", 0](GenKpath[SSGname, momclass])} to generate the contents of the KPOINTS file required for the non-self-consistent calculation. Next, place the WAVECAR and OUTCAR files obtained from the non-self-consistent calculation into a folder (e.g., \textbf{dir1}) and run \textbf{GetTopfilling["65.2.1.1.L", 0, 0, 0]} to compute the band topology at relative electronic filling $\delta\nu = 0$. For the weak-SOC limit, the topology is diagnosed by MSG. Setting \textbf{momclass = 4} (MM direction along class 4, i.e., (1,0,0)), the topology can be computed using \textbf{GetTopfilling["65.2.1.1.L", 4, 0, 1]}. For calculations with realistic SOC, first perform a self-consistent calculation and generate the contents of the KPOINTS file using \textbf{GenKpath["65.2.1.1.L", 4]}, where \textbf{momclass $\neq$ 0} labels the specific class of MM directions. Then, using the WAVECAR and OUTCAR files from the non-self-consistent calculation in \textbf{dir2}, run \textbf{GetTopfilling["65.2.1.1.L", 4, 0, 0]}.

\begin{figure}[bp]
	\centering
	\includegraphics[width=0.4\columnwidth]{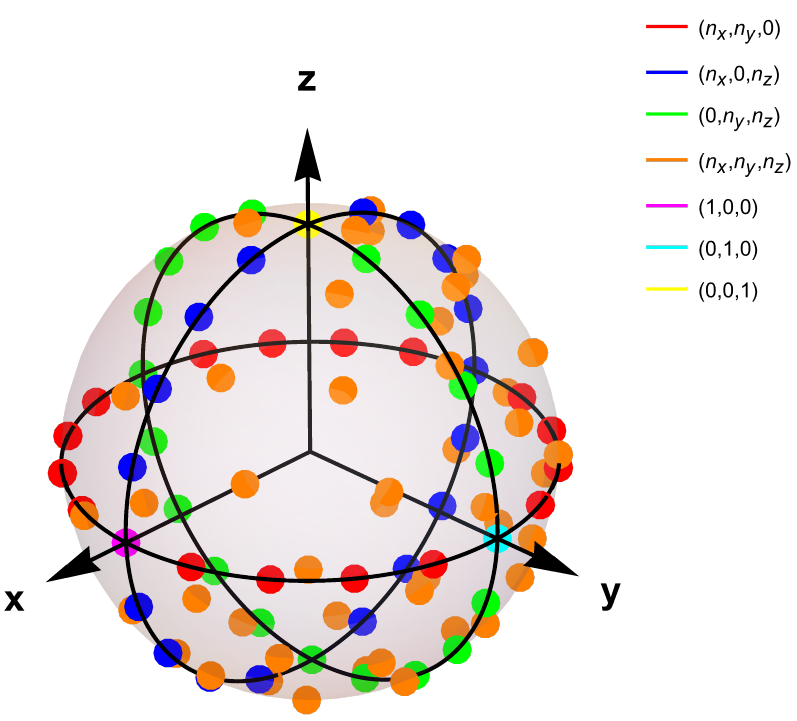}
	\caption{The 100 selected MM directions used for first-principles calculations of Mn$_2$AlB$_2$ with SOC included. The colored dots represent the sampled MM directions on the unit sphere in the Cartesian coordinate system $(x,y,z)$, and the curves denote three classes of MM directions: $(n_x,n_y,0)$, $(n_x,0,n_z)$, and $(0,n_y,n_z)$.}
	\label{fig:sphere}
\end{figure}

In the following, we show that Mn$_2$AlB$_2$ satisfies the weak-SOC limit at all HSPs and we can obtain topological phases under MSG for all MM directions without performing realistic calculations. First, we compare the HSP symmetry data for all MM directions derived from compatibility relations with those obtained from first-principles calculations for Mn$_2$AlB$_2$. Table~\ref{c1} lists the HSP symmetry data for the SSG, while the little-group information of the corresponding HSPs is given in Table~\ref{h1}. Using the compatibility relations from SSG 65.2.1.1.L to MSG given in Sec.~S4 of the SM~\cite{SM} and Eq.~\eqref{eq:2}, we obtain the HSP symmetry data in the weak-SOC limit for different MM directions without performing explicit SOC calculations, as presented in Table~\ref{c2}. On the other hand, we carried out first-principles calculations with SOC for 100 different MM directions (the selected directions are shown on a unit sphere in Fig.~\ref{fig:sphere}). By comparing the resulting $n_{\bm{k}}^{\mathrm{MSG},j}$, we find that $n_{\bm{k},\text{weak-SOC}}^{\mathrm{MSG},j}=n_{\bm{k}}^{\mathrm{MSG},j}$ holds for all HSPs across all 100 directions. This demonstrates that Mn$_2$AlB$_2$ satisfies the weak-SOC limit at all HSPs.

\begin{table*}[htbp]
	\caption{HSP symmetry data for Mn$_2$AlB$_2$ without SOC, where $n_{\bm{k}}^{\mathrm{SSG},j}$ is the number of occurrences of the (co-)irrep $j$ at $\bm{k}$ based on SSG.}
	\label{tab:Mn2AlB2_HSP}
	\begin{tabular}{lp{12cm}}
		\hline
		\hline
		HSP & \multicolumn{1}{c}{HSP symmetry data} \\
		\hline
		\noalign{\vskip 5pt}
		K1:$(0,0,0)$ &
		$n_{(0,0,0)}^{\text{SSG},D_{33}} = 7$,
		$n_{(0,0,0)}^{\text{SSG},D_{34}} = 3$,
		$n_{(0,0,0)}^{\text{SSG},D_{35}} = 2$,
		$n_{(0,0,0)}^{\text{SSG},D_{36}} = 2$,
		$n_{(0,0,0)}^{\text{SSG},D_{37}} = 2$,
		$n_{(0,0,0)}^{\text{SSG},D_{38}} = 1$,
		$n_{(0,0,0)}^{\text{SSG},D_{39}} = 3$,
		$n_{(0,0,0)}^{\text{SSG},D_{40}} = 3$ \\
		\noalign{\vskip 5pt}
		\hline
		\noalign{\vskip 5pt}
		K2:$(-\frac{1}{2}, \frac{1}{2}, 0)$ &
		$n_{(-\frac{1}{2},\frac{1}{2},0)}^{\text{SSG},D_{33}} = 5$,
		$n_{(-\frac{1}{2},\frac{1}{2},0)}^{\text{SSG},D_{34}} = 3$,
		$n_{(-\frac{1}{2},\frac{1}{2},0)}^{\text{SSG},D_{35}} = 1$,
		$n_{(-\frac{1}{2},\frac{1}{2},0)}^{\text{SSG},D_{36}} = 2$,
		$n_{(-\frac{1}{2},\frac{1}{2},0)}^{\text{SSG},D_{37}} = 2$,
		$n_{(-\frac{1}{2},\frac{1}{2},0)}^{\text{SSG},D_{38}} = 2$,
		$n_{(-\frac{1}{2},\frac{1}{2},0)}^{\text{SSG},D_{39}} = 3$,
		$n_{(-\frac{1}{2},\frac{1}{2},0)}^{\text{SSG},D_{40}} = 5$ \\
		\noalign{\vskip 5pt}
		\hline
		\noalign{\vskip 5pt}
		K3:$(0,0,\frac{1}{2})$ &
		$n_{(0,0,\frac{1}{2})}^{\text{SSG},D_{17}} = 5$,
		$n_{(0,0,\frac{1}{2})}^{\text{SSG},D_{18}} = 3$,
		$n_{(0,0,\frac{1}{2})}^{\text{SSG},D_{19}} = 2$,
		$n_{(0,0,\frac{1}{2})}^{\text{SSG},D_{20}} = 2$,
		$n_{(0,0,\frac{1}{2})}^{\text{SSG},D_{21}} = 2$,
		$n_{(0,0,\frac{1}{2})}^{\text{SSG},D_{22}} = 1$,
		$n_{(0,0,\frac{1}{2})}^{\text{SSG},D_{23}} = 5$,
		$n_{(0,0,\frac{1}{2})}^{\text{SSG},D_{24}} = 3$ \\
		\noalign{\vskip 5pt}
		\hline
		\noalign{\vskip 5pt}
		K4:$(-\frac{1}{2}, \frac{1}{2}, \frac{1}{2})$ &
		$n_{(-\frac{1}{2},\frac{1}{2},\frac{1}{2})}^{\text{SSG},D_{17}} = 4$,
		$n_{(-\frac{1}{2},\frac{1}{2},\frac{1}{2})}^{\text{SSG},D_{18}} = 4$,
		$n_{(-\frac{1}{2},\frac{1}{2},\frac{1}{2})}^{\text{SSG},D_{19}} = 1$,
		$n_{(-\frac{1}{2},\frac{1}{2},\frac{1}{2})}^{\text{SSG},D_{20}} = 2$,
		$n_{(-\frac{1}{2},\frac{1}{2},\frac{1}{2})}^{\text{SSG},D_{21}} = 2$,
		$n_{(-\frac{1}{2},\frac{1}{2},\frac{1}{2})}^{\text{SSG},D_{22}} = 2$,
		$n_{(-\frac{1}{2},\frac{1}{2},\frac{1}{2})}^{\text{SSG},D_{23}} = 4$,
		$n_{(-\frac{1}{2},\frac{1}{2},\frac{1}{2})}^{\text{SSG},D_{24}} = 4$ \\
		\noalign{\vskip 5pt}
		\hline
		\noalign{\vskip 5pt}
		K5:$(-\frac{1}{2}, 0, 0)$ &
		$n_{(-\frac{1}{2},0,0)}^{\text{SSG},D_{17}} = 6$,
		$n_{(-\frac{1}{2},0,0)}^{\text{SSG},D_{18}} = 5$,
		$n_{(-\frac{1}{2},0,0)}^{\text{SSG},D_{19}} = 5$,
		$n_{(-\frac{1}{2},0,0)}^{\text{SSG},D_{20}} = 7$ \\
		\noalign{\vskip 5pt}
		\hline
		\noalign{\vskip 5pt}
		K6:$(-\frac{1}{2}, 0, \frac{1}{2})$ &
		$n_{(-\frac{1}{2},0,\frac{1}{2})}^{\text{SSG},D_{9}} = 5$,
		$n_{(-\frac{1}{2},0,\frac{1}{2})}^{\text{SSG},D_{10}} = 6$,
		$n_{(-\frac{1}{2},0,\frac{1}{2})}^{\text{SSG},D_{11}} = 6$,		
		$n_{(-\frac{1}{2},0,\frac{1}{2})}^{\text{SSG},D_{12}} = 6$ \\
		\noalign{\vskip 5pt}
		\hline
		\hline
	\end{tabular}
	\label{c1}
\end{table*}

\clearpage

\renewcommand\arraystretch{1.15}\setlength{\LTcapwidth}{\textwidth}
\begin{longtable*} [htbp]
	{c p{12.9cm}}
	\caption{Symmetry data at the HSPs for Mn$_2$AlB$_2$ using compatibility relations from SSG to MSG listed in Sec.~S4 of the SM~\cite{SM} in the weak-SOC limit for different MM directions, where $n_{\bm{k},\text{weak-SOC}}^{\mathrm{MSG},j}$ is the number of occurrences of the (co-)irrep $j$ at $\bm{k}$ based on MSG.}
	\label{tab:Mn2AlB2_MM_HSP}\\
	\hline
	\hline
	\noalign{\vskip 3pt}
	\multicolumn{1}{c}{MM direction} & \multicolumn{1}{c}{HSP symmetry data} \\
	\noalign{\vskip 5pt}
	\hline
	\noalign{\vskip 5pt}
	\endfirsthead
	
	\multicolumn{2}{c}{}\\
	\hline
	\hline
	\noalign{\vskip 5pt}
	\multicolumn{1}{c}{MM direction} & \multicolumn{1}{c}{HSP symmetry data} \\
	\noalign{\vskip 5pt}
	\endhead
	
	\multicolumn{2}{r}{}\\
	\endfoot
	
	\endlastfoot
	$(n_x, n_y, n_z)$ &
	$n_{(0,0,0),\text{weak-SOC}}^{\text{MSG},D^d_1}=14$, $n_{(0,0,0),\text{weak-SOC}}^{\text{MSG},D^d_2}=9$; \newline
	$n_{(-\frac{1}{2},\frac{1}{2},0),\text{weak-SOC}}^{\text{MSG},D^d_1}=11$, $n_{(-\frac{1}{2},\frac{1}{2},0),\text{weak-SOC}}^{\text{MSG},D^d_2}=12$; \newline
	$n_{(0,0,\frac{1}{2}),\text{weak-SOC}}^{\text{MSG},D^d_1}=23$; \newline
	$n_{(-\frac{1}{2},\frac{1}{2},\frac{1}{2}),\text{weak-SOC}}^{\text{MSG},D^d_1}=23$; \newline
	$n_{(-\frac{1}{2},0,0),\text{weak-SOC}}^{\text{MSG},D^d_1}=11$, $n_{(-\frac{1}{2},0,0),\text{weak-SOC}}^{\text{MSG},D^d_2}=12$; \newline
	$n_{(0,\frac{1}{2},0),\text{weak-SOC}}^{\text{MSG},D^d_1}=11$, $n_{(0,\frac{1}{2},0),\text{weak-SOC}}^{\text{MSG},D^d_2}=12$; \newline
	$n_{(-\frac{1}{2},0,\frac{1}{2}),\text{weak-SOC}}^{\text{MSG},D^d_1}=23$; \newline
	$n_{(0,\frac{1}{2},-\frac{1}{2}),\text{weak-SOC}}^{\text{MSG},D^d_1}=23$ \\
	\noalign{\vskip 5pt}
	\hline
	\noalign{\vskip 5pt}
	
	$(0, 1, 0)$ &
	$n_{(0,0,0),\text{weak-SOC}}^{\text{MSG},D^d_1}=14$, $n_{(0,0,0),\text{weak-SOC}}^{\text{MSG},D^d_2}=9$; \newline
	$n_{(-\frac{1}{2},\frac{1}{2},0),\text{weak-SOC}}^{\text{MSG},D^d_1}=11$, $n_{(-\frac{1}{2},\frac{1}{2},0),\text{weak-SOC}}^{\text{MSG},D^d_2}=12$; \newline
	$n_{(0,0,\frac{1}{2}),\text{weak-SOC}}^{\text{MSG},D^d_1}=11$, $n_{(0,0,\frac{1}{2}),\text{weak-SOC}}^{\text{MSG},D^d_2}=12$; \newline
	$n_{(-\frac{1}{2},\frac{1}{2},\frac{1}{2}),\text{weak-SOC}}^{\text{MSG},D^d_1}=11$, $n_{(-\frac{1}{2},\frac{1}{2},\frac{1}{2}),\text{weak-SOC}}^{\text{MSG},D^d_2}=12$; \newline
	$n_{(-\frac{1}{2},0,0),\text{weak-SOC}}^{\text{MSG},D^d_1}=11$, $n_{(-\frac{1}{2},0,0),\text{weak-SOC}}^{\text{MSG},D^d_2}=12$; \newline
	$n_{(-\frac{1}{2},0,\frac{1}{2}),\text{weak-SOC}}^{\text{MSG},D^d_1}=23$ \\
	\noalign{\vskip 5pt}
	\hline
	\noalign{\vskip 5pt}
	
	$(n_x, 0, n_z)$ &
	$n_{(0,0,0),\text{weak-SOC}}^{\text{MSG},D^d_1}=14$, $n_{(0,0,0),\text{weak-SOC}}^{\text{MSG},D^d_2}=9$; \newline
	$n_{(-\frac{1}{2},\frac{1}{2},0),\text{weak-SOC}}^{\text{MSG},D^d_1}=11$, $n_{(-\frac{1}{2},\frac{1}{2},0),\text{weak-SOC}}^{\text{MSG},D^d_2}=12$; \newline
	$n_{(0,0,\frac{1}{2}),\text{weak-SOC}}^{\text{MSG},D^d_1}=23$; \newline
	$n_{(-\frac{1}{2},\frac{1}{2},\frac{1}{2}),\text{weak-SOC}}^{\text{MSG},D^d_1}=23$; \newline
	$n_{(-\frac{1}{2},0,0),\text{weak-SOC}}^{\text{MSG},D^d_1}=11$, $n_{(-\frac{1}{2},0,0),\text{weak-SOC}}^{\text{MSG},D^d_2}=12$; \newline
	$n_{(-\frac{1}{2},0,\frac{1}{2}),\text{weak-SOC}}^{\text{MSG},D^d_1}=23$ \\
	\noalign{\vskip 5pt}
	\hline
	\noalign{\vskip 5pt}
	
	$(1, 0, 0)$ &
	$n_{(0,0,0),\text{weak-SOC}}^{\text{MSG},D^d_1}=14$, $n_{(0,0,0),\text{weak-SOC}}^{\text{MSG},D^d_2}=9$; \newline
	$n_{(-\frac{1}{2},\frac{1}{2},0),\text{weak-SOC}}^{\text{MSG},D^d_1}=11$, $n_{(-\frac{1}{2},\frac{1}{2},0),\text{weak-SOC}}^{\text{MSG},D^d_2}=12$; \newline
	$n_{(0,0,\frac{1}{2}),\text{weak-SOC}}^{\text{MSG},D^d_1}=12$, $n_{(0,0,\frac{1}{2}),\text{weak-SOC}}^{\text{MSG},D^d_2}=11$; \newline
	$n_{(-\frac{1}{2},\frac{1}{2},\frac{1}{2}),\text{weak-SOC}}^{\text{MSG},D^d_1}=11$, $n_{(-\frac{1}{2},\frac{1}{2},\frac{1}{2}),\text{weak-SOC}}^{\text{MSG},D^d_2}=12$; \newline
	$n_{(-\frac{1}{2},0,0),\text{weak-SOC}}^{\text{MSG},D^d_1}=11$, $n_{(-\frac{1}{2},0,0),\text{weak-SOC}}^{\text{MSG},D^d_2}=12$; \newline
	$n_{(-\frac{1}{2},0,\frac{1}{2}),\text{weak-SOC}}^{\text{MSG},D^d_1}=23$ \\
	\noalign{\vskip 5pt}
	\hline
	\noalign{\vskip 5pt}
	
	$(0, n_y, n_z)$ &
	$n_{(0,0,0),\text{weak-SOC}}^{\text{MSG},D^d_1}=14$, $n_{(0,0,0),\text{weak-SOC}}^{\text{MSG},D^d_2}=9$; \newline
	$n_{(-\frac{1}{2},\frac{1}{2},0),\text{weak-SOC}}^{\text{MSG},D^d_1}=11$, $n_{(-\frac{1}{2},\frac{1}{2},0),\text{weak-SOC}}^{\text{MSG},D^d_2}=12$; \newline
	$n_{(0,0,\frac{1}{2}),\text{weak-SOC}}^{\text{MSG},D^d_1}=23$; \newline
	$n_{(-\frac{1}{2},\frac{1}{2},\frac{1}{2}),\text{weak-SOC}}^{\text{MSG},D^d_1}=23$; \newline
	$n_{(-\frac{1}{2},0,0),\text{weak-SOC}}^{\text{MSG},D^d_1}=11$, $n_{(-\frac{1}{2},0,0),\text{weak-SOC}}^{\text{MSG},D^d_2}=12$; \newline
	$n_{(-\frac{1}{2},0,\frac{1}{2}),\text{weak-SOC}}^{\text{MSG},D^d_1}=23$ \\
	\noalign{\vskip 5pt}
	\hline
	\noalign{\vskip 5pt}
	
	$(0, 0, 1)$ &
	$n_{(0,0,0),\text{weak-SOC}}^{\text{MSG},D^d_1}=14$, $n_{(0,0,0),\text{weak-SOC}}^{\text{MSG},D^d_2}=9$; \newline
	$n_{(-\frac{1}{2},\frac{1}{2},0),\text{weak-SOC}}^{\text{MSG},D^d_1}=11$, $n_{(-\frac{1}{2},\frac{1}{2},0),\text{weak-SOC}}^{\text{MSG},D^d_2}=12$; \newline
	$n_{(0,0,\frac{1}{2}),\text{weak-SOC}}^{\text{MSG},D^d_1}=11$, $n_{(0,0,\frac{1}{2}),\text{weak-SOC}}^{\text{MSG},D^d_2}=12$; \newline
	$n_{(-\frac{1}{2},\frac{1}{2},\frac{1}{2}),\text{weak-SOC}}^{\text{MSG},D^d_1}=11$, $n_{(-\frac{1}{2},\frac{1}{2},\frac{1}{2}),\text{weak-SOC}}^{\text{MSG},D^d_2}=12$; \newline
	$n_{(-\frac{1}{2},0,0),\text{weak-SOC}}^{\text{MSG},D^d_1}=11$, $n_{(-\frac{1}{2},0,0),\text{weak-SOC}}^{\text{MSG},D^d_2}=12$; \newline
	$n_{(-\frac{1}{2},0,\frac{1}{2}),\text{weak-SOC}}^{\text{MSG},D^d_1}=11$, $n_{(-\frac{1}{2},0,\frac{1}{2}),\text{weak-SOC}}^{\text{MSG},D^d_2}=12$ \\
	\noalign{\vskip 5pt}
	\hline
	\noalign{\vskip 5pt}
	
	$(n_x, n_y, 0)$ &
	$n_{(0,0,0),\text{weak-SOC}}^{\text{MSG},D^d_1}=14$, $n_{(0,0,0),\text{weak-SOC}}^{\text{MSG},D^d_2}=9$; \newline
	$n_{(-\frac{1}{2},\frac{1}{2},0),\text{weak-SOC}}^{\text{MSG},D^d_1}=11$, $n_{(-\frac{1}{2},\frac{1}{2},0),\text{weak-SOC}}^{\text{MSG},D^d_2}=12$; \newline
	$n_{(0,0,\frac{1}{2}),\text{weak-SOC}}^{\text{MSG},D^d_1}=23$; \newline
	$n_{(-\frac{1}{2},\frac{1}{2},\frac{1}{2}),\text{weak-SOC}}^{\text{MSG},D^d_1}=23$; \newline
	$n_{(-\frac{1}{2},0,0),\text{weak-SOC}}^{\text{MSG},D^d_1}=11$, $n_{(-\frac{1}{2},0,0),\text{weak-SOC}}^{\text{MSG},D^d_2}=12$; \newline
	$n_{(0,\frac{1}{2},0),\text{weak-SOC}}^{\text{MSG},D^d_1}=11$, $n_{(0,\frac{1}{2},0),\text{weak-SOC}}^{\text{MSG},D^d_2}=12$; \newline
	$n_{(-\frac{1}{2},0,\frac{1}{2}),\text{weak-SOC}}^{\text{MSG},D^d_1}=23$; \newline
	$n_{(0,\frac{1}{2},-\frac{1}{2}),\text{weak-SOC}}^{\text{MSG},D^d_1}=23$ \\
	\noalign{\vskip 5pt}
	\hline
	\hline
	\label{c2}
\end{longtable*}

\begin{table*}[htbp]
	\centering
	\caption{The table presents the coordinates of the HSPs under the SSG of Mn$_2$AlB$_2$, together with the characters of the (co-)irreps of their little groups. The coordinates of K2 and K4 used here are equivalent to those in Tables~\ref{fig:ssg}, \ref{c1}, and \ref{c2}, differing only by the reciprocal lattice vector $(0,-1,0)$. Only unitary SSG operations with nonzero characters are shown. Here, $\phi \in [0, 2\pi)$ denotes the angle of rotation around the MM direction in $C_n(\phi)$. The SSG operations are listed as table columns, so each character is aligned with its corresponding operation. The (co-)irrep labels follow the enumeration obtained from our systematic calculation of the little groups at momentum \(\bm{k}\) and their (co-)irreps for all collinear SSGs~\cite{ssgsi}. Here we show only the (co-)irreps relevant to the electronic system; therefore, the labels shown here do not start from \(D_1\). In the full little groups, the omitted unitary operations are \((C_{2\perp} || -y, -x, \frac{1}{2}+z)\), \((C_{2\perp} || y, x, \frac{1}{2}+z)\), \((C_{2\perp} || x, y, \frac{1}{2}-z)\), \((C_{2\perp} || -x, -y, \frac{1}{2}-z)\), \((C_{2\perp} || y, x, \frac{1}{2}-z)\), \((C_{2\perp} || -y, -x, \frac{1}{2}-z)\), \((C_{2\perp} || -x, -y, \frac{1}{2}+z)\), and \((C_{2\perp} || x, y, \frac{1}{2}+z)\) for HSPs K1--K4, and \((C_{2\perp} || x, y, \frac{1}{2}-z)\), \((C_{2\perp} || -x, -y, \frac{1}{2}-z)\), \((C_{2\perp} || -x, -y, \frac{1}{2}+z)\), and \((C_{2\perp} || x, y, \frac{1}{2}+z)\) for HSPs K5 and K6. Since these operations have zero characters for the (co-)irreps listed here, they and their corresponding zero character entries are omitted.}
	\setlength{\tabcolsep}{3pt}
	\renewcommand{\arraystretch}{1.25}
	\textbf{HSPs K1--K4}\\[2pt]
	\resizebox{1\textwidth}{!}{%
		\begin{tabular}{cccccccccc}
			\hline
			HSP & (Co-)irrep & \multicolumn{8}{c}{SSG operation} \\
			\cline{3-10}
			& & $(E || x, y, z)$ & $(E || -x, -y, z)$ & $(E || -y, -x, -z)$ & $(E || y, x, -z)$ & $(E || -x, -y, -z)$ & $(E || x, y, -z)$ & $(E || y, x, z)$ & $(E || -y, -x, z)$ \\
			\hline
			K1:$(0,0,0)$ & $D_{33}$ & $2 \cos \left(\frac{\phi }{2}\right)$ & $2 \cos \left(\frac{\phi }{2}\right)$ & $2 \cos \left(\frac{\phi }{2}\right)$ & $2 \cos \left(\frac{\phi }{2}\right)$ & $2 \cos \left(\frac{\phi }{2}\right)$ & $2 \cos \left(\frac{\phi }{2}\right)$ & $2 \cos \left(\frac{\phi }{2}\right)$ & $2 \cos \left(\frac{\phi }{2}\right)$ \\
			& $D_{34}$ & $2 \cos \left(\frac{\phi }{2}\right)$ & $-2 \cos \left(\frac{\phi }{2}\right)$ & $-2 \cos \left(\frac{\phi }{2}\right)$ & $2 \cos \left(\frac{\phi }{2}\right)$ & $2 \cos \left(\frac{\phi }{2}\right)$ & $-2 \cos \left(\frac{\phi }{2}\right)$ & $-2 \cos \left(\frac{\phi }{2}\right)$ & $2 \cos \left(\frac{\phi }{2}\right)$ \\
			& $D_{35}$ & $2 \cos \left(\frac{\phi }{2}\right)$ & $2 \cos \left(\frac{\phi }{2}\right)$ & $-2 \cos \left(\frac{\phi }{2}\right)$ & $-2 \cos \left(\frac{\phi }{2}\right)$ & $2 \cos \left(\frac{\phi }{2}\right)$ & $2 \cos \left(\frac{\phi }{2}\right)$ & $-2 \cos \left(\frac{\phi }{2}\right)$ & $-2 \cos \left(\frac{\phi }{2}\right)$ \\
			& $D_{36}$ & $2 \cos \left(\frac{\phi }{2}\right)$ & $-2 \cos \left(\frac{\phi }{2}\right)$ & $2 \cos \left(\frac{\phi }{2}\right)$ & $-2 \cos \left(\frac{\phi }{2}\right)$ & $2 \cos \left(\frac{\phi }{2}\right)$ & $-2 \cos \left(\frac{\phi }{2}\right)$ & $2 \cos \left(\frac{\phi }{2}\right)$ & $-2 \cos \left(\frac{\phi }{2}\right)$ \\
			& $D_{37}$ & $2 \cos \left(\frac{\phi }{2}\right)$ & $2 \cos \left(\frac{\phi }{2}\right)$ & $2 \cos \left(\frac{\phi }{2}\right)$ & $2 \cos \left(\frac{\phi }{2}\right)$ & $-2 \cos \left(\frac{\phi }{2}\right)$ & $-2 \cos \left(\frac{\phi }{2}\right)$ & $-2 \cos \left(\frac{\phi }{2}\right)$ & $-2 \cos \left(\frac{\phi }{2}\right)$ \\
			& $D_{38}$ & $2 \cos \left(\frac{\phi }{2}\right)$ & $-2 \cos \left(\frac{\phi }{2}\right)$ & $-2 \cos \left(\frac{\phi }{2}\right)$ & $2 \cos \left(\frac{\phi }{2}\right)$ & $-2 \cos \left(\frac{\phi }{2}\right)$ & $2 \cos \left(\frac{\phi }{2}\right)$ & $2 \cos \left(\frac{\phi }{2}\right)$ & $-2 \cos \left(\frac{\phi }{2}\right)$ \\
			& $D_{39}$ & $2 \cos \left(\frac{\phi }{2}\right)$ & $2 \cos \left(\frac{\phi }{2}\right)$ & $-2 \cos \left(\frac{\phi }{2}\right)$ & $-2 \cos \left(\frac{\phi }{2}\right)$ & $-2 \cos \left(\frac{\phi }{2}\right)$ & $-2 \cos \left(\frac{\phi }{2}\right)$ & $2 \cos \left(\frac{\phi }{2}\right)$ & $2 \cos \left(\frac{\phi }{2}\right)$ \\
			& $D_{40}$ & $2 \cos \left(\frac{\phi }{2}\right)$ & $-2 \cos \left(\frac{\phi }{2}\right)$ & $2 \cos \left(\frac{\phi }{2}\right)$ & $-2 \cos \left(\frac{\phi }{2}\right)$ & $-2 \cos \left(\frac{\phi }{2}\right)$ & $2 \cos \left(\frac{\phi }{2}\right)$ & $-2 \cos \left(\frac{\phi }{2}\right)$ & $2 \cos \left(\frac{\phi }{2}\right)$ \\
			\hline
			K2:$(-\frac{1}{2},-\frac{1}{2},0)$ & $D_{33}$ & $2 \cos \left(\frac{\phi }{2}\right)$ & $2 \cos \left(\frac{\phi }{2}\right)$ & $2 \cos \left(\frac{\phi }{2}\right)$ & $2 \cos \left(\frac{\phi }{2}\right)$ & $2 \cos \left(\frac{\phi }{2}\right)$ & $2 \cos \left(\frac{\phi }{2}\right)$ & $2 \cos \left(\frac{\phi }{2}\right)$ & $2 \cos \left(\frac{\phi }{2}\right)$ \\
			& $D_{34}$ & $2 \cos \left(\frac{\phi }{2}\right)$ & $-2 \cos \left(\frac{\phi }{2}\right)$ & $-2 \cos \left(\frac{\phi }{2}\right)$ & $2 \cos \left(\frac{\phi }{2}\right)$ & $2 \cos \left(\frac{\phi }{2}\right)$ & $-2 \cos \left(\frac{\phi }{2}\right)$ & $-2 \cos \left(\frac{\phi }{2}\right)$ & $2 \cos \left(\frac{\phi }{2}\right)$ \\
			& $D_{35}$ & $2 \cos \left(\frac{\phi }{2}\right)$ & $2 \cos \left(\frac{\phi }{2}\right)$ & $-2 \cos \left(\frac{\phi }{2}\right)$ & $-2 \cos \left(\frac{\phi }{2}\right)$ & $2 \cos \left(\frac{\phi }{2}\right)$ & $2 \cos \left(\frac{\phi }{2}\right)$ & $-2 \cos \left(\frac{\phi }{2}\right)$ & $-2 \cos \left(\frac{\phi }{2}\right)$ \\
			& $D_{36}$ & $2 \cos \left(\frac{\phi }{2}\right)$ & $-2 \cos \left(\frac{\phi }{2}\right)$ & $2 \cos \left(\frac{\phi }{2}\right)$ & $-2 \cos \left(\frac{\phi }{2}\right)$ & $2 \cos \left(\frac{\phi }{2}\right)$ & $-2 \cos \left(\frac{\phi }{2}\right)$ & $2 \cos \left(\frac{\phi }{2}\right)$ & $-2 \cos \left(\frac{\phi }{2}\right)$ \\
			& $D_{37}$ & $2 \cos \left(\frac{\phi }{2}\right)$ & $2 \cos \left(\frac{\phi }{2}\right)$ & $2 \cos \left(\frac{\phi }{2}\right)$ & $2 \cos \left(\frac{\phi }{2}\right)$ & $-2 \cos \left(\frac{\phi }{2}\right)$ & $-2 \cos \left(\frac{\phi }{2}\right)$ & $-2 \cos \left(\frac{\phi }{2}\right)$ & $-2 \cos \left(\frac{\phi }{2}\right)$ \\
			& $D_{38}$ & $2 \cos \left(\frac{\phi }{2}\right)$ & $-2 \cos \left(\frac{\phi }{2}\right)$ & $-2 \cos \left(\frac{\phi }{2}\right)$ & $2 \cos \left(\frac{\phi }{2}\right)$ & $-2 \cos \left(\frac{\phi }{2}\right)$ & $2 \cos \left(\frac{\phi }{2}\right)$ & $2 \cos \left(\frac{\phi }{2}\right)$ & $-2 \cos \left(\frac{\phi }{2}\right)$ \\
			& $D_{39}$ & $2 \cos \left(\frac{\phi }{2}\right)$ & $2 \cos \left(\frac{\phi }{2}\right)$ & $-2 \cos \left(\frac{\phi }{2}\right)$ & $-2 \cos \left(\frac{\phi }{2}\right)$ & $-2 \cos \left(\frac{\phi }{2}\right)$ & $-2 \cos \left(\frac{\phi }{2}\right)$ & $2 \cos \left(\frac{\phi }{2}\right)$ & $2 \cos \left(\frac{\phi }{2}\right)$ \\
			& $D_{40}$ & $2 \cos \left(\frac{\phi }{2}\right)$ & $-2 \cos \left(\frac{\phi }{2}\right)$ & $2 \cos \left(\frac{\phi }{2}\right)$ & $-2 \cos \left(\frac{\phi }{2}\right)$ & $-2 \cos \left(\frac{\phi }{2}\right)$ & $2 \cos \left(\frac{\phi }{2}\right)$ & $-2 \cos \left(\frac{\phi }{2}\right)$ & $2 \cos \left(\frac{\phi }{2}\right)$ \\
			\hline
			K3:$(0,0,\frac{1}{2})$ & $D_{17}$ & $2 \cos \left(\frac{\phi }{2}\right)$ & $2 \cos \left(\frac{\phi }{2}\right)$ & $2 \mathrm{i} \sin \left(\frac{\phi }{2}\right)$ & $2 \mathrm{i} \sin \left(\frac{\phi }{2}\right)$ & $2 \mathrm{i} \sin \left(\frac{\phi }{2}\right)$ & $2 \mathrm{i} \sin \left(\frac{\phi }{2}\right)$ & $2 \cos \left(\frac{\phi }{2}\right)$ & $2 \cos \left(\frac{\phi }{2}\right)$ \\
			& $D_{18}$ & $2 \cos \left(\frac{\phi }{2}\right)$ & $-2 \cos \left(\frac{\phi }{2}\right)$ & $-2 \mathrm{i} \sin \left(\frac{\phi }{2}\right)$ & $2 \mathrm{i} \sin \left(\frac{\phi }{2}\right)$ & $2 \mathrm{i} \sin \left(\frac{\phi }{2}\right)$ & $-2 \mathrm{i} \sin \left(\frac{\phi }{2}\right)$ & $-2 \cos \left(\frac{\phi }{2}\right)$ & $2 \cos \left(\frac{\phi }{2}\right)$ \\
			& $D_{19}$ & $2 \cos \left(\frac{\phi }{2}\right)$ & $2 \cos \left(\frac{\phi }{2}\right)$ & $-2 \mathrm{i} \sin \left(\frac{\phi }{2}\right)$ & $-2 \mathrm{i} \sin \left(\frac{\phi }{2}\right)$ & $2 \mathrm{i} \sin \left(\frac{\phi }{2}\right)$ & $2 \mathrm{i} \sin \left(\frac{\phi }{2}\right)$ & $-2 \cos \left(\frac{\phi }{2}\right)$ & $-2 \cos \left(\frac{\phi }{2}\right)$ \\
			& $D_{20}$ & $2 \cos \left(\frac{\phi }{2}\right)$ & $-2 \cos \left(\frac{\phi }{2}\right)$ & $2 \mathrm{i} \sin \left(\frac{\phi }{2}\right)$ & $-2 \mathrm{i} \sin \left(\frac{\phi }{2}\right)$ & $2 \mathrm{i} \sin \left(\frac{\phi }{2}\right)$ & $-2 \mathrm{i} \sin \left(\frac{\phi }{2}\right)$ & $2 \cos \left(\frac{\phi }{2}\right)$ & $-2 \cos \left(\frac{\phi }{2}\right)$ \\
			& $D_{21}$ & $2 \cos \left(\frac{\phi }{2}\right)$ & $2 \cos \left(\frac{\phi }{2}\right)$ & $2 \mathrm{i} \sin \left(\frac{\phi }{2}\right)$ & $2 \mathrm{i} \sin \left(\frac{\phi }{2}\right)$ & $-2 \mathrm{i} \sin \left(\frac{\phi }{2}\right)$ & $-2 \mathrm{i} \sin \left(\frac{\phi }{2}\right)$ & $-2 \cos \left(\frac{\phi }{2}\right)$ & $-2 \cos \left(\frac{\phi }{2}\right)$ \\
			& $D_{22}$ & $2 \cos \left(\frac{\phi }{2}\right)$ & $-2 \cos \left(\frac{\phi }{2}\right)$ & $-2 \mathrm{i} \sin \left(\frac{\phi }{2}\right)$ & $2 \mathrm{i} \sin \left(\frac{\phi }{2}\right)$ & $-2 \mathrm{i} \sin \left(\frac{\phi }{2}\right)$ & $2 \mathrm{i} \sin \left(\frac{\phi }{2}\right)$ & $2 \cos \left(\frac{\phi }{2}\right)$ & $-2 \cos \left(\frac{\phi }{2}\right)$ \\
			& $D_{23}$ & $2 \cos \left(\frac{\phi }{2}\right)$ & $2 \cos \left(\frac{\phi }{2}\right)$ & $-2 \mathrm{i} \sin \left(\frac{\phi }{2}\right)$ & $-2 \mathrm{i} \sin \left(\frac{\phi }{2}\right)$ & $-2 \mathrm{i} \sin \left(\frac{\phi }{2}\right)$ & $-2 \mathrm{i} \sin \left(\frac{\phi }{2}\right)$ & $2 \cos \left(\frac{\phi }{2}\right)$ & $2 \cos \left(\frac{\phi }{2}\right)$ \\
			& $D_{24}$ & $2 \cos \left(\frac{\phi }{2}\right)$ & $-2 \cos \left(\frac{\phi }{2}\right)$ & $2 \mathrm{i} \sin \left(\frac{\phi }{2}\right)$ & $-2 \mathrm{i} \sin \left(\frac{\phi }{2}\right)$ & $-2 \mathrm{i} \sin \left(\frac{\phi }{2}\right)$ & $2 \mathrm{i} \sin \left(\frac{\phi }{2}\right)$ & $-2 \cos \left(\frac{\phi }{2}\right)$ & $2 \cos \left(\frac{\phi }{2}\right)$ \\
			\hline
			K4:$(-\frac{1}{2},-\frac{1}{2},\frac{1}{2})$ & $D_{17}$ & $2 \cos \left(\frac{\phi }{2}\right)$ & $2 \cos \left(\frac{\phi }{2}\right)$ & $2 \mathrm{i} \sin \left(\frac{\phi }{2}\right)$ & $2 \mathrm{i} \sin \left(\frac{\phi }{2}\right)$ & $2 \mathrm{i} \sin \left(\frac{\phi }{2}\right)$ & $2 \mathrm{i} \sin \left(\frac{\phi }{2}\right)$ & $2 \cos \left(\frac{\phi }{2}\right)$ & $2 \cos \left(\frac{\phi }{2}\right)$ \\
			& $D_{18}$ & $2 \cos \left(\frac{\phi }{2}\right)$ & $-2 \cos \left(\frac{\phi }{2}\right)$ & $-2 \mathrm{i} \sin \left(\frac{\phi }{2}\right)$ & $2 \mathrm{i} \sin \left(\frac{\phi }{2}\right)$ & $2 \mathrm{i} \sin \left(\frac{\phi }{2}\right)$ & $-2 \mathrm{i} \sin \left(\frac{\phi }{2}\right)$ & $-2 \cos \left(\frac{\phi }{2}\right)$ & $2 \cos \left(\frac{\phi }{2}\right)$ \\
			& $D_{19}$ & $2 \cos \left(\frac{\phi }{2}\right)$ & $2 \cos \left(\frac{\phi }{2}\right)$ & $-2 \mathrm{i} \sin \left(\frac{\phi }{2}\right)$ & $-2 \mathrm{i} \sin \left(\frac{\phi }{2}\right)$ & $2 \mathrm{i} \sin \left(\frac{\phi }{2}\right)$ & $2 \mathrm{i} \sin \left(\frac{\phi }{2}\right)$ & $-2 \cos \left(\frac{\phi }{2}\right)$ & $-2 \cos \left(\frac{\phi }{2}\right)$ \\
			& $D_{20}$ & $2 \cos \left(\frac{\phi }{2}\right)$ & $-2 \cos \left(\frac{\phi }{2}\right)$ & $2 \mathrm{i} \sin \left(\frac{\phi }{2}\right)$ & $-2 \mathrm{i} \sin \left(\frac{\phi }{2}\right)$ & $2 \mathrm{i} \sin \left(\frac{\phi }{2}\right)$ & $-2 \mathrm{i} \sin \left(\frac{\phi }{2}\right)$ & $2 \cos \left(\frac{\phi }{2}\right)$ & $-2 \cos \left(\frac{\phi }{2}\right)$ \\
			& $D_{21}$ & $2 \cos \left(\frac{\phi }{2}\right)$ & $2 \cos \left(\frac{\phi }{2}\right)$ & $2 \mathrm{i} \sin \left(\frac{\phi }{2}\right)$ & $2 \mathrm{i} \sin \left(\frac{\phi }{2}\right)$ & $-2 \mathrm{i} \sin \left(\frac{\phi }{2}\right)$ & $-2 \mathrm{i} \sin \left(\frac{\phi }{2}\right)$ & $-2 \cos \left(\frac{\phi }{2}\right)$ & $-2 \cos \left(\frac{\phi }{2}\right)$ \\
			& $D_{22}$ & $2 \cos \left(\frac{\phi }{2}\right)$ & $-2 \cos \left(\frac{\phi }{2}\right)$ & $-2 \mathrm{i} \sin \left(\frac{\phi }{2}\right)$ & $2 \mathrm{i} \sin \left(\frac{\phi }{2}\right)$ & $-2 \mathrm{i} \sin \left(\frac{\phi }{2}\right)$ & $2 \mathrm{i} \sin \left(\frac{\phi }{2}\right)$ & $2 \cos \left(\frac{\phi }{2}\right)$ & $-2 \cos \left(\frac{\phi }{2}\right)$ \\
			& $D_{23}$ & $2 \cos \left(\frac{\phi }{2}\right)$ & $2 \cos \left(\frac{\phi }{2}\right)$ & $-2 \mathrm{i} \sin \left(\frac{\phi }{2}\right)$ & $-2 \mathrm{i} \sin \left(\frac{\phi }{2}\right)$ & $-2 \mathrm{i} \sin \left(\frac{\phi }{2}\right)$ & $-2 \mathrm{i} \sin \left(\frac{\phi }{2}\right)$ & $2 \cos \left(\frac{\phi }{2}\right)$ & $2 \cos \left(\frac{\phi }{2}\right)$ \\
			& $D_{24}$ & $2 \cos \left(\frac{\phi }{2}\right)$ & $-2 \cos \left(\frac{\phi }{2}\right)$ & $2 \mathrm{i} \sin \left(\frac{\phi }{2}\right)$ & $-2 \mathrm{i} \sin \left(\frac{\phi }{2}\right)$ & $-2 \mathrm{i} \sin \left(\frac{\phi }{2}\right)$ & $2 \mathrm{i} \sin \left(\frac{\phi }{2}\right)$ & $-2 \cos \left(\frac{\phi }{2}\right)$ & $2 \cos \left(\frac{\phi }{2}\right)$ \\
			\hline
		\end{tabular}
	}
	\vspace{0.8em}
	
	\textbf{HSPs K5--K6}\\[2pt]
	\resizebox{0.6\textwidth}{!}{%
		\begin{tabular}{cccccc}
			\hline
			High-symmetry point & (Co-)irreducible representation & \multicolumn{4}{c}{SSG operation} \\
			\cline{3-6}
			& & $(E || x, y, z)$ & $(E || -x, -y, z)$ & $(E || -x, -y, -z)$ & $(E || x, y, -z)$ \\
			\hline
			K5:$(-\frac{1}{2},0,0)$ & $D_{17}$ & $2 \cos \left(\frac{\phi }{2}\right)$ & $2 \cos \left(\frac{\phi }{2}\right)$ & $2 \cos \left(\frac{\phi }{2}\right)$ & $2 \cos \left(\frac{\phi }{2}\right)$ \\
			& $D_{18}$ & $2 \cos \left(\frac{\phi }{2}\right)$ & $2 \cos \left(\frac{\phi }{2}\right)$ & $-2 \cos \left(\frac{\phi }{2}\right)$ & $-2 \cos \left(\frac{\phi }{2}\right)$ \\
			& $D_{19}$ & $2 \cos \left(\frac{\phi }{2}\right)$ & $-2 \cos \left(\frac{\phi }{2}\right)$ & $2 \cos \left(\frac{\phi }{2}\right)$ & $-2 \cos \left(\frac{\phi }{2}\right)$ \\
			& $D_{20}$ & $2 \cos \left(\frac{\phi }{2}\right)$ & $-2 \cos \left(\frac{\phi }{2}\right)$ & $-2 \cos \left(\frac{\phi }{2}\right)$ & $2 \cos \left(\frac{\phi }{2}\right)$ \\
			\hline
			K6:$(-\frac{1}{2},0,\frac{1}{2})$ & $D_{9}$ & $2 \cos \left(\frac{\phi }{2}\right)$ & $2 \cos \left(\frac{\phi }{2}\right)$ & $2 \mathrm{i} \sin \left(\frac{\phi }{2}\right)$ & $2 \mathrm{i} \sin \left(\frac{\phi }{2}\right)$ \\
			& $D_{10}$ & $2 \cos \left(\frac{\phi }{2}\right)$ & $2 \cos \left(\frac{\phi }{2}\right)$ & $-2 \mathrm{i} \sin \left(\frac{\phi }{2}\right)$ & $-2 \mathrm{i} \sin \left(\frac{\phi }{2}\right)$ \\
			& $D_{11}$ & $2 \cos \left(\frac{\phi }{2}\right)$ & $-2 \cos \left(\frac{\phi }{2}\right)$ & $2 \mathrm{i} \sin \left(\frac{\phi }{2}\right)$ & $-2 \mathrm{i} \sin \left(\frac{\phi }{2}\right)$ \\
			& $D_{12}$ & $2 \cos \left(\frac{\phi }{2}\right)$ & $-2 \cos \left(\frac{\phi }{2}\right)$ & $-2 \mathrm{i} \sin \left(\frac{\phi }{2}\right)$ & $2 \mathrm{i} \sin \left(\frac{\phi }{2}\right)$ \\
			\hline
		\end{tabular}
	}
	\label{h1}
\end{table*}

\FloatBarrier
\sectionnotoc{Maximal collinear SSG subgroups for LiCoO$_2$ and GaNi$_3$}\label{A7}

The paramagnetic phase of $\mathrm{LiCoO_2}$ (ICSD: 098270) crystallizes in SG $R\bar{3}m$ (No.~166), with the conventional lattice vectors given by
$\bm{a}_1=(0,-1.62466,4.68267)$,
$\bm{a}_2=(1.407,0.812332,4.68267)$,
and
$\bm{a}_3=(-1.407,0.812332,4.68267)$~\AA.
The fractional coordinates $(x,y,z)$ of all atoms in the conventional hexagonal cell with respect to the basis $(\bm{a}_1,\bm{a}_2,\bm{a}_3)$ are
\[
\begin{aligned}
	&\mathrm{Li}_1:(0,0,0.5),\quad
	\mathrm{Li}_2:(0.6667,0.3333,0.8333),\quad
	\mathrm{Li}_3:(0.3333,0.6667,0.1667),\\
	&\mathrm{Co}_1:(0,0,0),\quad
	\mathrm{Co}_2:(0.6667,0.3333,0.3333),\quad
	\mathrm{Co}_3:(0.3333,0.6667,0.6667),\\
	&\mathrm{O}_1:(0,0,0.2268),\quad
	\mathrm{O}_2:(0,0,0.7732),\quad
	\mathrm{O}_3:(0.6667,0.3333,0.5601),\\
	&\mathrm{O}_4:(0.6667,0.3333,0.1065),\quad
	\mathrm{O}_5:(0.3333,0.6667,0.8935),\quad
	\mathrm{O}_6:(0.3333,0.6667,0.4399).
\end{aligned}
\]

The SG operations of this structure are given in the Seitz notation $\{R|\bm{t}\}$, where $R$ and $\bm{t}$ denote the rotational and fractional translation parts of the SG operations, respectively. The notation acts on a fractional coordinate $\bm{r}=(x,y,z)$ in the conventional hexagonal cell as $\{R|\bm{t}\}:\bm{r}\rightarrow R\bm{r}+\bm{t}$. Following the notation in Ref.~\cite{BC}, $E$, $C_n$, $I$, $S_n$, and $\sigma$ denote the identity, $n$-fold rotation, inversion, improper rotation, and mirror reflection operations, respectively. The three centering translations of the conventional cell are $\bm{\tau}_1=(0,0,0),\bm{\tau}_2=\left(\frac{2}{3},\frac{1}{3},\frac{1}{3}\right),\bm{\tau}_3=\left(\frac{1}{3},\frac{2}{3},\frac{2}{3}\right)$.
Thus, the full SG operations are obtained as $\{R_i|\bm{\tau}_j\}$, where $i=1,\ldots,12$ and $j=1,2,3$. The rotational parts act on $(x,y,z)$ as
\[
\begin{alignedat}{3}
	\{E|\bm{0}\}&:(x,y,z)\rightarrow(x,y,z),\qquad&
	\{C_3^+|\bm{0}\}&:(x,y,z)\rightarrow(-y,x-y,z),\qquad&
	\{C_3^-|\bm{0}\}&:(x,y,z)\rightarrow(-x+y,-x,z),\\
	\{C'_{22}|\bm{0}\}&:(x,y,z)\rightarrow(y,x,-z),&
	\{C'_{23}|\bm{0}\}&:(x,y,z)\rightarrow(x-y,-y,-z),&
	\{C_{2z}|\bm{0}\}&:(x,y,z)\rightarrow(-x,-x+y,-z),\\
	\{I|\bm{0}\}&:(x,y,z)\rightarrow(-x,-y,-z),&
	\{S_6^-|\bm{0}\}&:(x,y,z)\rightarrow(y,-x+y,-z),&
	\{S_6^+|\bm{0}\}&:(x,y,z)\rightarrow(x-y,x,-z),\\
	\{\sigma_{d2}|\bm{0}\}&:(x,y,z)\rightarrow(-y,-x,z),&
	\{\sigma_{d3}|\bm{0}\}&:(x,y,z)\rightarrow(-x+y,y,z),&
	\{\sigma_z|\bm{0}\}&:(x,y,z)\rightarrow(x,x-y,z).
\end{alignedat}
\]

Based on these symmetry operations, the corresponding maximal collinear SSG subgroups of this SG can be constructed, as listed in Table~\ref{tab:LiCoO2_SSG}.
\begin{table}[htbp]
	\centering
	\caption{Maximal collinear SSG subgroups of SG No.~166 and nontrivial part of their corresponding symmetry operations, expressed in the basis of the conventional hexagonal cell of the paramagnetic phase. Here, $\bm{t}_1=\left(-\frac{1}{3},-\frac{2}{3},\frac{1}{3}\right)$ and $\bm{t}_2=(0,0,1)$.}
	\label{tab:LiCoO2_SSG}
	\renewcommand{\arraystretch}{1.2}
	\begin{tabular}{c|p{14cm}}
		\hline
		Maximal collinear SSG & \multicolumn{1}{c}{Nontrivial part of SSG operations}  \\
		\hline
		
		166.1.1.1.L\_1 &
		\(\begin{array}{llllll}
			\{E||E|\bm{0}\},&
			\{E||C_3^+|\bm{0}\},&
			\{E||C_3^-|\bm{0}\},&
			\{E||C'_{22}|\bm{0}\},&
			\{E||C'_{23}|\bm{0}\},&
			\{E||C_{2z}|\bm{0}\},\\[1pt]
			\{E||I|\bm{0}\},&
			\{E||S_6^-|\bm{0}\},&
			\{E||S_6^+|\bm{0}\},&
			\{E||\sigma_{d2}|\bm{0}\},&
			\{E||\sigma_{d3}|\bm{0}\},&
			\{E||\sigma_z|\bm{0}\}
		\end{array}\)
		\\
		\hline
		
		166.1.2.1.L\_1 &
		\(\begin{array}{llllll}
			\{E||E|\bm{0}\},&
			\{E||C_3^+|\bm{0}\},&
			\{E||C_3^-|\bm{0}\},&
			\{E||\sigma_{d2}|\bm{0}\},&
			\{E||\sigma_{d3}|\bm{0}\},&
			\{E||\sigma_z|\bm{0}\},\\[1pt]
			\{T||C'_{22}|\bm{0}\},&
			\{T||C'_{23}|\bm{0}\},&
			\{T||C_{2z}|\bm{0}\},&
			\{T||I|\bm{0}\},&
			\{T||S_6^-|\bm{0}\},&
			\{T||S_6^+|\bm{0}\}
		\end{array}\)
		\\
		\hline
		
		166.1.2.2.L\_1 &
		\(\begin{array}{llllll}
			\{E||E|\bm{0}\},&
			\{E||C_3^+|\bm{0}\},&
			\{E||C_3^-|\bm{0}\},&
			\{E||C'_{22}|\bm{0}\},&
			\{E||C'_{23}|\bm{0}\},&
			\{E||C_{2z}|\bm{0}\},\\[1pt]
			\{T||I|\bm{0}\},&
			\{T||S_6^-|\bm{0}\},&
			\{T||S_6^+|\bm{0}\},&
			\{T||\sigma_{d2}|\bm{0}\},&
			\{T||\sigma_{d3}|\bm{0}\},&
			\{T||\sigma_z|\bm{0}\}
		\end{array}\)
		\\
		\hline
		
		166.1.2.3.L\_1 &
		\(\begin{array}{llllll}
			\{E||E|\bm{0}\},&
			\{E||C_3^+|\bm{0}\},&
			\{E||C_3^-|\bm{0}\},&
			\{E||I|\bm{0}\},&
			\{E||S_6^-|\bm{0}\},&
			\{E||S_6^+|\bm{0}\},\\[1pt]
			\{T||C'_{22}|\bm{0}\},&
			\{T||C'_{23}|\bm{0}\},&
			\{T||C_{2z}|\bm{0}\},&
			\{T||\sigma_{d2}|\bm{0}\},&
			\{T||\sigma_{d3}|\bm{0}\},&
			\{T||\sigma_z|\bm{0}\}
		\end{array}\)
		\\
		\hline
		
		166.2.1.1.L\_1 &
		\(\begin{array}{llllll}
			\{E||E|\bm{0}\},&
			\{E||C_3^+|\bm{0}\},&
			\{E||C_3^-|\bm{0}\},&
			\{E||C'_{22}|\bm{0}\},&
			\{E||C'_{23}|\bm{0}\},&
			\{E||C_{2z}|\bm{0}\},\\[1pt]
			
			\{E||I|\bm{0}\},&
			\{E||S_6^-|\bm{0}\},&
			\{E||S_6^+|\bm{0}\},&
			\{E||\sigma_{d2}|\bm{0}\},&
			\{E||\sigma_{d3}|\bm{0}\},&
			\{E||\sigma_z|\bm{0}\},\\[1pt]
			
			\{T||E|\bm{t}_1\},&
			\{T||C_3^+|\bm{t}_1\},&
			\{T||C_3^-|\bm{t}_1\},&
			\{T||C'_{22}|\bm{t}_1\},&
			\{T||C'_{23}|\bm{t}_1\},&
			\{T||C_{2z}|\bm{t}_1\},\\[1pt]
			
			\{T||I|\bm{t}_1\},&
			\{T||S_6^-|\bm{t}_1\},&
			\{T||S_6^+|\bm{t}_1\},&
			\{T||\sigma_{d2}|\bm{t}_1\},&
			\{T||\sigma_{d3}|\bm{t}_1\},&
			\{T||\sigma_z|\bm{t}_1\}
		\end{array}\)
		\\
		\hline
		
		166.2.1.1.L\_2 &
		\(\begin{array}{llllll}
			\{E||E|\bm{0}\},&
			\{E||C_3^+|\bm{0}\},&
			\{E||C_3^-|\bm{0}\},&
			\{E||C'_{22}|\bm{t}_2\},&
			\{E||C'_{23}|\bm{t}_2\},&
			\{E||C_{2z}|\bm{t}_2\},\\[1pt]
			
			\{E||I|\bm{t}_2\},&
			\{E||S_6^-|\bm{t}_2\},&
			\{E||S_6^+|\bm{t}_2\},&
			\{E||\sigma_{d2}|\bm{0}\},&
			\{E||\sigma_{d3}|\bm{0}\},&
			\{E||\sigma_z|\bm{0}\},\\[1pt]
			
			\{T||E|\bm{t}_2\},&
			\{T||C_3^+|\bm{t}_2\},&
			\{T||C_3^-|\bm{t}_2\},&
			\{T||C'_{22}|\bm{0}\},&
			\{T||C'_{23}|\bm{0}\},&
			\{T||C_{2z}|\bm{0}\},\\[1pt]
			
			\{T||I|\bm{0}\},&
			\{T||S_6^-|\bm{0}\},&
			\{T||S_6^+|\bm{0}\},&
			\{T||\sigma_{d2}|\bm{t}_2\},&
			\{T||\sigma_{d3}|\bm{t}_2\},&
			\{T||\sigma_z|\bm{t}_2\}
		\end{array}\)
		\\
		\hline
		
		166.2.1.2.L\_1 &
		\(\begin{array}{llllll}
			\{E||E|\bm{0}\},&
			\{E||C_3^+|\bm{0}\},&
			\{E||C_3^-|\bm{0}\},&
			\{E||C'_{22}|\bm{0}\},&
			\{E||C'_{23}|\bm{0}\},&
			\{E||C_{2z}|\bm{0}\},\\[1pt]
			
			\{E||I|\bm{t}_2\},&
			\{E||S_6^-|\bm{t}_2\},&
			\{E||S_6^+|\bm{t}_2\},&
			\{E||\sigma_{d2}|\bm{t}_2\},&
			\{E||\sigma_{d3}|\bm{t}_2\},&
			\{E||\sigma_z|\bm{t}_2\},\\[1pt]
			
			\{T||E|\bm{t}_1\},&
			\{T||C_3^+|\bm{t}_1\},&
			\{T||C_3^-|\bm{t}_1\},&
			\{T||C'_{22}|\bm{t}_1\},&
			\{T||C'_{23}|\bm{t}_1\},&
			\{T||C_{2z}|\bm{t}_1\},\\[1pt]
			
			\{T||I|\bm{t}_1+\bm{t}_2\},&
			\{T||S_6^-|\bm{t}_1+\bm{t}_2\},&
			\{T||S_6^+|\bm{t}_1+\bm{t}_2\},&
			\{T||\sigma_{d2}|\bm{t}_1+\bm{t}_2\},&
			\{T||\sigma_{d3}|\bm{t}_1+\bm{t}_2\},&
			\{T||\sigma_z|\bm{t}_1+\bm{t}_2\}
		\end{array}\)
		\\
		\hline
		
		166.2.1.2.L\_2 &
		\(\begin{array}{llllll}
			\{E||E|\bm{0}\},&
			\{E||I|\bm{0}\},&
			\{E||S_6^-|\bm{0}\},&
			\{E||S_6^+|\bm{0}\},&
			\{E||C_3^-|\bm{0}\},&
			\{E||C_3^+|\bm{0}\},\\[1pt]
			
			\{E||C_{2z}|\bm{t}_2\},&
			\{E||C'_{22}|\bm{t}_2\},&
			\{E||C'_{23}|\bm{t}_2\},&
			\{E||\sigma_{d3}|-\bm{t}_2\},&
			\{E||\sigma_z|-\bm{t}_2\},&
			\{E||\sigma_{d2}|-\bm{t}_2\},\\[1pt]
			
			\{T||E|\bm{t}_2\},&
			\{T||I|\bm{t}_2\},&
			\{T||S_6^-|\bm{t}_2\},&
			\{T||S_6^+|\bm{t}_2\},&
			\{T||C_3^-|\bm{t}_2\},&
			\{T||C_3^+|\bm{t}_2\},\\[1pt]
			
			\{T||C_{2z}|\bm{0}\},&
			\{T||C'_{22}|\bm{0}\},&
			\{T||C'_{23}|\bm{0}\},&
			\{T||\sigma_{d3}|\bm{0}\},&
			\{T||\sigma_z|\bm{0}\},&
			\{T||\sigma_{d2}|\bm{0}\}
		\end{array}\)
		\\
		\hline
		
	\end{tabular}
\end{table}

Considering Co atoms as the only magnetic atoms, we examine which maximal collinear SSG subgroups of SG No.~166 allow nonzero MMs on the Co sites. Since $\mathrm{Co}_2$ and $\mathrm{Co}_3$ are generated from $\mathrm{Co}_1$ by the centering translations and are symmetry-equivalent to $\mathrm{Co}_1$, if the MM of $\mathrm{Co}_1$ is forced to vanish, the MMs of $\mathrm{Co}_2$ and $\mathrm{Co}_3$ also vanish:
\begin{equation}
	\bm{M}_{\mathrm{Co}_1}=\bm{0}
	\quad \Rightarrow \quad
	\bm{M}_{\mathrm{Co}_2}=\bm{M}_{\mathrm{Co}_3}=\bm{0}.
\end{equation}
The following analysis is therefore based on the representative Co site $\mathrm{Co}_1$.

For $166.1.2.1.\mathrm{L\_1}$, $166.1.2.2.\mathrm{L\_1}$ and $166.2.1.1.\mathrm{L\_2}$, the SSG operation $\{T||I|\bm{0}\}$ contains the real-space inversion operation $\{I|\bm{0}\}$. According to
\begin{equation}
	\{I|\bm{0}\}:(x,y,z)\rightarrow(-x,-y,-z),
\end{equation}
the representative Co site $\mathrm{Co}_1$ at $(0,0,0)$ is mapped onto itself:
\begin{equation}
	(0,0,0)\rightarrow(0,0,0).
\end{equation}
Meanwhile, the spin-space operation $T$ reverses the MM. Therefore, the MM at $\mathrm{Co}_1$ must satisfy
\begin{equation}
	\bm{M}_{\mathrm{Co}_1}=-\bm{M}_{\mathrm{Co}_1},
	\qquad \Rightarrow \qquad
	\bm{M}_{\mathrm{Co}_1}=\bm{0}.
\end{equation}
Consequently, the MMs of all Co atoms vanish for these two SSGs.

For $166.1.2.3.\mathrm{L\_1}$, the SSG operation $\{T||C'_{22}|\bm{0}\}$ contains the real-space operation $\{C'_{22}|\bm{0}\}$. According to
\begin{equation}
	\{C'_{22}|\bm{0}\}:(x,y,z)\rightarrow(y,x,-z),
\end{equation}
the representative Co site $\mathrm{Co}_1$ at $(0,0,0)$ is mapped onto itself:
\begin{equation}
	(0,0,0)\rightarrow(0,0,0).
\end{equation}
Meanwhile, the spin-space operation $T$ reverses the MM. Therefore, the MM at $\mathrm{Co}_1$ must satisfy
\begin{equation}
	\bm{M}_{\mathrm{Co}_1}=-\bm{M}_{\mathrm{Co}_1},
	\qquad \Rightarrow \qquad
	\bm{M}_{\mathrm{Co}_1}=\bm{0}.
\end{equation}
Consequently, the MMs of all Co atoms vanish for this SSG.

For $166.2.1.2.\mathrm{L\_1}$, the SSG operation $\{T||E|\bm{t}_1\}$, together with the centering translation $\bm{\tau}_3=(1/3,2/3,2/3)$, generates the operation $\{T||E|\bm{t}_2\}$, since
\begin{equation}
	\bm{t}_1+\bm{\tau}_3=\bm{t}_2 .
\end{equation}
The real-space part of this operation maps the Co site at $(0,0,0)$ to the Co site at $(0,0,1)$, which we label as $\mathrm{Co}_1'$:
\begin{equation}
	\{E|\bm{t}_2\}:\mathrm{Co}_1 \rightarrow \mathrm{Co}_1' .
\end{equation}
Because the spin-space operation is $T$, the MM is reversed, giving
\begin{equation}
	\bm{M}_{\mathrm{Co}_1'}=-\bm{M}_{\mathrm{Co}_1}.
\end{equation}
On the other hand, the SSG also contains the operation $\{E||E|\bm{t}_2\}$, whose real-space part gives the same mapping,
\begin{equation}
	\{E|\bm{t}_2\}:\mathrm{Co}_1 \rightarrow \mathrm{Co}_1' ,
\end{equation}
but whose spin-space part is $E$ and hence does not reverse the MM. Therefore,
\begin{equation}
	\bm{M}_{\mathrm{Co}_1'}=\bm{M}_{\mathrm{Co}_1}.
\end{equation}
Combining the two constraints gives
\begin{equation}
	\bm{M}_{\mathrm{Co}_1}=-\bm{M}_{\mathrm{Co}_1},
	\qquad \Rightarrow \qquad
	\bm{M}_{\mathrm{Co}_1}=\bm{0}.
\end{equation}
Consequently, the MMs of all Co atoms vanish for this SSG.

For $166.2.1.2.\mathrm{L\_2}$, the SSG operation $\{T||\sigma_z|\bm{0}\}$ contains the real-space operation $\{\sigma_z|\bm{0}\}$. According to
\begin{equation}
	\{\sigma_z|\bm{0}\}:(x,y,z)\rightarrow(x,x-y,z),
\end{equation}
the representative Co site $\mathrm{Co}_1$ at $(0,0,0)$ is mapped onto itself:
\begin{equation}
	(0,0,0)\rightarrow(0,0,0).
\end{equation}
Meanwhile, the spin-space operation $T$ reverses the MM. Therefore,
\begin{equation}
	\bm{M}_{\mathrm{Co}_1}=-\bm{M}_{\mathrm{Co}_1},
	\qquad \Rightarrow \qquad
	\bm{M}_{\mathrm{Co}_1}=\bm{0}.
\end{equation}
Consequently, the MMs of all Co atoms vanish for this SSG.

\begin{figure}[htbp]
	\centering
	\includegraphics[width=0.6\columnwidth]{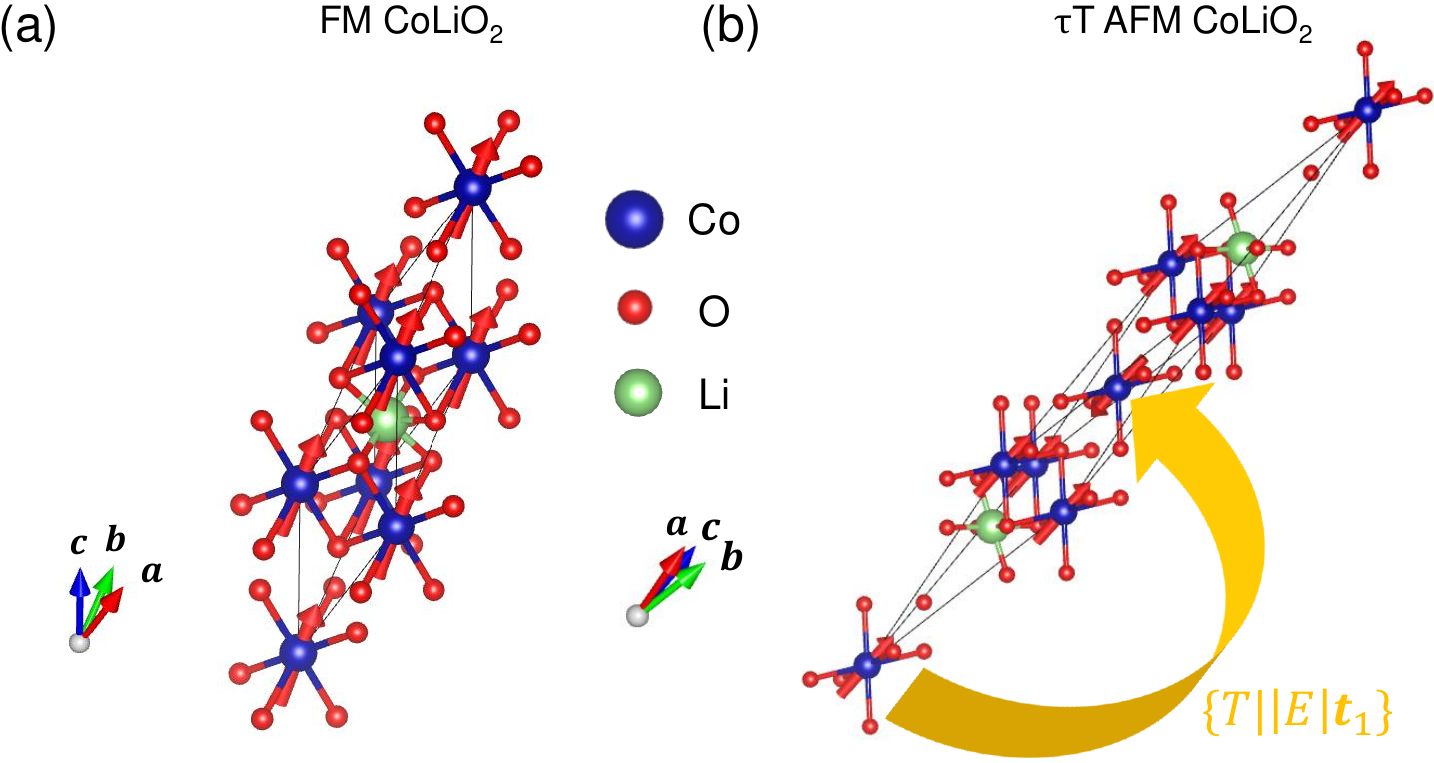}
	\caption{Constrained by the maximal collinear SSG subgroups of SG 166, LiCoO$_2$ owns two possible collinear magnetic orders as schematically depicted in (a) and (b), corresponding to FM and $\tau T$ AFM magnetic structures, respectively. Here the MMs at Co are chosen to be along the $(0,0,1)$ direction in Cartesian coordinates. In panel (b), the Co atoms carrying opposite MMs are related by the SSG operation $\lbrace T||E|\bm{t}_1\rbrace$.}
	\label{fig:LiCoO2_mag}
\end{figure}

In summary, among the maximal collinear SSG subgroups considered here, only \(166.1.1.1.\mathrm{L\_1}\) and \(166.2.1.1.\mathrm{L\_1}\) allow nonzero MMs at Co atoms. The former corresponds to an FM (see Fig.~\hyperref[fig:LiCoO2_mag]{\ref*{fig:LiCoO2_mag}(a)}), whereas the latter corresponds to a \(\tau T\) AFM (see Fig.~\hyperref[fig:LiCoO2_mag]{\ref*{fig:LiCoO2_mag}(b)}).

\FloatBarrier

The paramagnetic phase of $\mathrm{GaNi_3}$ (ICSD: 103856) crystallizes in SG $Pm\bar{3}m$ (No.~221), with the conventional lattice vectors given by $\bm{a}_1=(3.585,0,0)$, $\bm{a}_2=(0,3.585,0)$, and $\bm{a}_3=(0,0,3.585)$~\AA.
The fractional coordinates $(x,y,z)$ of all atoms in the conventional cubic cell with respect to the basis $(\bm{a}_1,\bm{a}_2,\bm{a}_3)$ are
\[
\mathrm{Ga}:(0,0,0),\qquad
\mathrm{Ni}_1:(0,0.5,0.5),\qquad
\mathrm{Ni}_2:(0.5,0,0.5),\qquad
\mathrm{Ni}_3:(0.5,0.5,0).
\]

The complete set of 48 SG operations of SG $Pm\bar{3}m$ (No.~221) is given by
\[
\begin{alignedat}{3}
	\{E|\bm{0}\}&:(x,y,z)\rightarrow(x,y,z),\qquad&
	\{C_{2z}|\bm{0}\}&:(x,y,z)\rightarrow(-x,-y,z),\qquad&
	\{C_{2y}|\bm{0}\}&:(x,y,z)\rightarrow(-x,y,-z),\\
	\{C_{2x}|\bm{0}\}&:(x,y,z)\rightarrow(x,-y,-z),&
	\{C_{31}^{+}|\bm{0}\}&:(x,y,z)\rightarrow(z,x,y),&
	\{C_{34}^{+}|\bm{0}\}&:(x,y,z)\rightarrow(z,-x,-y),\\
	\{C_{33}^{+}|\bm{0}\}&:(x,y,z)\rightarrow(-z,-x,y),&
	\{C_{32}^{+}|\bm{0}\}&:(x,y,z)\rightarrow(-z,x,-y),&
	\{C_{31}^{-}|\bm{0}\}&:(x,y,z)\rightarrow(y,z,x),\\
	\{C_{33}^{-}|\bm{0}\}&:(x,y,z)\rightarrow(-y,z,-x),&
	\{C_{32}^{-}|\bm{0}\}&:(x,y,z)\rightarrow(y,-z,-x),&
	\{C_{34}^{-}|\bm{0}\}&:(x,y,z)\rightarrow(-y,-z,x),\\
	\{C_{2a}|\bm{0}\}&:(x,y,z)\rightarrow(y,x,-z),&
	\{C_{2b}|\bm{0}\}&:(x,y,z)\rightarrow(-y,-x,-z),&
	\{C_{4z}^{-}|\bm{0}\}&:(x,y,z)\rightarrow(y,-x,z),\\
	\{C_{4z}^{+}|\bm{0}\}&:(x,y,z)\rightarrow(-y,x,z),&
	\{C_{4x}^{-}|\bm{0}\}&:(x,y,z)\rightarrow(x,z,-y),&
	\{C_{2d}|\bm{0}\}&:(x,y,z)\rightarrow(-x,z,y),\\
	\{C_{2f}|\bm{0}\}&:(x,y,z)\rightarrow(-x,-z,-y),&
	\{C_{4x}^{+}|\bm{0}\}&:(x,y,z)\rightarrow(x,-z,y),&
	\{C_{4y}^{+}|\bm{0}\}&:(x,y,z)\rightarrow(z,y,-x),\\
	\{C_{2c}|\bm{0}\}&:(x,y,z)\rightarrow(z,-y,x),&
	\{C_{4y}^{-}|\bm{0}\}&:(x,y,z)\rightarrow(-z,y,x),&
	\{C_{2e}|\bm{0}\}&:(x,y,z)\rightarrow(-z,-y,-x),\\
	\{I|\bm{0}\}&:(x,y,z)\rightarrow(-x,-y,-z),&
	\{\sigma_z|\bm{0}\}&:(x,y,z)\rightarrow(x,y,-z),&
	\{\sigma_y|\bm{0}\}&:(x,y,z)\rightarrow(x,-y,z),\\
	\{\sigma_x|\bm{0}\}&:(x,y,z)\rightarrow(-x,y,z),&
	\{S_{61}^{-}|\bm{0}\}&:(x,y,z)\rightarrow(-z,-x,-y),&
	\{S_{64}^{-}|\bm{0}\}&:(x,y,z)\rightarrow(-z,x,y),\\
	\{S_{63}^{-}|\bm{0}\}&:(x,y,z)\rightarrow(z,x,-y),&
	\{S_{62}^{-}|\bm{0}\}&:(x,y,z)\rightarrow(z,-x,y),&
	\{S_{61}^{+}|\bm{0}\}&:(x,y,z)\rightarrow(-y,-z,-x),\\
	\{S_{63}^{+}|\bm{0}\}&:(x,y,z)\rightarrow(y,-z,x),&
	\{S_{62}^{+}|\bm{0}\}&:(x,y,z)\rightarrow(-y,z,x),&
	\{S_{64}^{+}|\bm{0}\}&:(x,y,z)\rightarrow(y,z,-x),\\
	\{\sigma_{da}|\bm{0}\}&:(x,y,z)\rightarrow(-y,-x,z),&
	\{\sigma_{db}|\bm{0}\}&:(x,y,z)\rightarrow(y,x,z),&
	\{S_{4z}^{+}|\bm{0}\}&:(x,y,z)\rightarrow(y,-x,-z),\\
	\{S_{4z}^{-}|\bm{0}\}&:(x,y,z)\rightarrow(-y,x,-z),&
	\{S_{4x}^{+}|\bm{0}\}&:(x,y,z)\rightarrow(-x,-z,y),&
	\{\sigma_{dd}|\bm{0}\}&:(x,y,z)\rightarrow(x,-z,-y),\\
	\{\sigma_{df}|\bm{0}\}&:(x,y,z)\rightarrow(x,z,y),&
	\{S_{4x}^{-}|\bm{0}\}&:(x,y,z)\rightarrow(-x,z,-y),&
	\{S_{4y}^{-}|\bm{0}\}&:(x,y,z)\rightarrow(-z,-y,x),\\
	\{\sigma_{dc}|\bm{0}\}&:(x,y,z)\rightarrow(-z,y,-x),&
	\{S_{4y}^{+}|\bm{0}\}&:(x,y,z)\rightarrow(z,-y,-x),&
	\{\sigma_{de}|\bm{0}\}&:(x,y,z)\rightarrow(z,y,x).
\end{alignedat}
\]
Here, $E$, $C_n$, $I$, $S_n$, and $\sigma$ denote the identity, $n$-fold rotation, inversion, improper rotation, and mirror reflection operations, respectively.

\begingroup
\scriptsize
\renewcommand{\arraystretch}{1.2}
\setlength{\tabcolsep}{3pt}
\setlength{\LTcapwidth}{\textwidth}
\begin{longtable}{p{0.18\textwidth}|p{0.78\textwidth}}
	\caption{Maximal collinear SSG subgroups of SG No.~221 and the nontrivial parts of their corresponding symmetry operations, expressed in the basis of the conventional cubic cell of the paramagnetic phase. Here, $\bm{t}=(0,0,1)$.}
	\label{tab:GaNi3_SSG}\\
	\hline
	\multicolumn{1}{c|}{Maximal collinear SSG} & \multicolumn{1}{c}{Nontrivial part of SSG operations} \\
	\hline
	\endfirsthead
	\hline
	\multicolumn{1}{c|}{Maximal collinear SSG} & \multicolumn{1}{c}{Nontrivial part of SSG operations} \\
	\hline
	\endhead
	\hline
	\multicolumn{2}{r}{\emph{Continued on next page}} \\
	\endfoot
	\hline
	\endlastfoot
	\(221.1.1.1.\mathrm{L\_1}\) & \(\begin{array}{llllllll}
		\{E||E|\bm{0}\},&
		\{E||C_{2z}|\bm{0}\},&
		\{E||C_{2y}|\bm{0}\},&
		\{E||C_{2x}|\bm{0}\},&
		\{E||C_{31}^{+}|\bm{0}\},&
		\{E||C_{34}^{+}|\bm{0}\},&
		\{E||C_{33}^{+}|\bm{0}\},&
		\{E||C_{32}^{+}|\bm{0}\},
	\end{array}\) \\
	& \(\begin{array}{llllllll}
		\{E||C_{31}^{-}|\bm{0}\},&
		\{E||C_{33}^{-}|\bm{0}\},&
		\{E||C_{32}^{-}|\bm{0}\},&
		\{E||C_{34}^{-}|\bm{0}\},&
		\{E||C_{2a}|\bm{0}\},&
		\{E||C_{2b}|\bm{0}\},&
		\{E||C_{4z}^{-}|\bm{0}\},&
		\{E||C_{4z}^{+}|\bm{0}\},
	\end{array}\) \\
	& \(\begin{array}{llllllll}
		\{E||C_{4x}^{-}|\bm{0}\},&
		\{E||C_{2d}|\bm{0}\},&
		\{E||C_{2f}|\bm{0}\},&
		\{E||C_{4x}^{+}|\bm{0}\},&
		\{E||C_{4y}^{+}|\bm{0}\},&
		\{E||C_{2c}|\bm{0}\},&
		\{E||C_{4y}^{-}|\bm{0}\},&
		\{E||C_{2e}|\bm{0}\},
	\end{array}\) \\
	& \(\begin{array}{llllllll}
		\{E||I|\bm{0}\},&
		\{E||\sigma_z|\bm{0}\},&
		\{E||\sigma_y|\bm{0}\},&
		\{E||\sigma_x|\bm{0}\},&
		\{E||S_{61}^{-}|\bm{0}\},&
		\{E||S_{64}^{-}|\bm{0}\},&
		\{E||S_{63}^{-}|\bm{0}\},&
		\{E||S_{62}^{-}|\bm{0}\},
	\end{array}\) \\
	& \(\begin{array}{llllllll}
		\{E||S_{61}^{+}|\bm{0}\},&
		\{E||S_{63}^{+}|\bm{0}\},&
		\{E||S_{62}^{+}|\bm{0}\},&
		\{E||S_{64}^{+}|\bm{0}\},&
		\{E||\sigma_{da}|\bm{0}\},&
		\{E||\sigma_{db}|\bm{0}\},&
		\{E||S_{4z}^{+}|\bm{0}\},&
		\{E||S_{4z}^{-}|\bm{0}\},
	\end{array}\) \\
	& \(\begin{array}{llllllll}
		\{E||S_{4x}^{+}|\bm{0}\},&
		\{E||\sigma_{dd}|\bm{0}\},&
		\{E||\sigma_{df}|\bm{0}\},&
		\{E||S_{4x}^{-}|\bm{0}\},&
		\{E||S_{4y}^{-}|\bm{0}\},&
		\{E||\sigma_{dc}|\bm{0}\},&
		\{E||S_{4y}^{+}|\bm{0}\},&
		\{E||\sigma_{de}|\bm{0}\}
	\end{array}\) \\
	\hline
	\(221.1.2.1.\mathrm{L\_1}\) & \(\begin{array}{llllllll}
		\{E||E|\bm{0}\},&
		\{E||C_{2z}|\bm{0}\},&
		\{E||C_{2y}|\bm{0}\},&
		\{E||C_{2x}|\bm{0}\},&
		\{E||C_{31}^{+}|\bm{0}\},&
		\{E||C_{34}^{+}|\bm{0}\},&
		\{E||C_{33}^{+}|\bm{0}\},&
		\{E||C_{32}^{+}|\bm{0}\},
	\end{array}\) \\
	& \(\begin{array}{llllllll}
		\{E||C_{31}^{-}|\bm{0}\},&
		\{E||C_{33}^{-}|\bm{0}\},&
		\{E||C_{32}^{-}|\bm{0}\},&
		\{E||C_{34}^{-}|\bm{0}\},&
		\{E||\sigma_{db}|\bm{0}\},&
		\{E||\sigma_{da}|\bm{0}\},&
		\{E||S_{4z}^{-}|\bm{0}\},&
		\{E||S_{4z}^{+}|\bm{0}\},
	\end{array}\) \\
	& \(\begin{array}{llllllll}
		\{E||\sigma_{df}|\bm{0}\},&
		\{E||S_{4x}^{-}|\bm{0}\},&
		\{E||S_{4x}^{+}|\bm{0}\},&
		\{E||\sigma_{dd}|\bm{0}\},&
		\{E||\sigma_{de}|\bm{0}\},&
		\{E||S_{4y}^{+}|\bm{0}\},&
		\{E||\sigma_{dc}|\bm{0}\},&
		\{E||S_{4y}^{-}|\bm{0}\},
	\end{array}\) \\
	& \(\begin{array}{llllllll}
		\{T||C_{2a}|\bm{0}\},&
		\{T||C_{2b}|\bm{0}\},&
		\{T||C_{4z}^{-}|\bm{0}\},&
		\{T||C_{4z}^{+}|\bm{0}\},&
		\{T||C_{4x}^{-}|\bm{0}\},&
		\{T||C_{2d}|\bm{0}\},&
		\{T||C_{2f}|\bm{0}\},&
		\{T||C_{4x}^{+}|\bm{0}\},
	\end{array}\) \\
	& \(\begin{array}{llllllll}
		\{T||C_{4y}^{+}|\bm{0}\},&
		\{T||C_{2c}|\bm{0}\},&
		\{T||C_{4y}^{-}|\bm{0}\},&
		\{T||C_{2e}|\bm{0}\},&
		\{T||\sigma_z|\bm{0}\},&
		\{T||I|\bm{0}\},&
		\{T||\sigma_x|\bm{0}\},&
		\{T||\sigma_y|\bm{0}\},
	\end{array}\) \\
	& \(\begin{array}{llllllll}
		\{T||S_{63}^{-}|\bm{0}\},&
		\{T||S_{62}^{-}|\bm{0}\},&
		\{T||S_{61}^{-}|\bm{0}\},&
		\{T||S_{64}^{-}|\bm{0}\},&
		\{T||S_{64}^{+}|\bm{0}\},&
		\{T||S_{62}^{+}|\bm{0}\},&
		\{T||S_{63}^{+}|\bm{0}\},&
		\{T||S_{61}^{+}|\bm{0}\}
	\end{array}\) \\
	\hline
	\(221.1.2.2.\mathrm{L\_1}\) & \(\begin{array}{llllllll}
		\{E||E|\bm{0}\},&
		\{E||C_{2z}|\bm{0}\},&
		\{E||C_{2y}|\bm{0}\},&
		\{E||C_{2x}|\bm{0}\},&
		\{E||C_{31}^{+}|\bm{0}\},&
		\{E||C_{34}^{+}|\bm{0}\},&
		\{E||C_{33}^{+}|\bm{0}\},&
		\{E||C_{32}^{+}|\bm{0}\},
	\end{array}\) \\
	& \(\begin{array}{llllllll}
		\{E||C_{31}^{-}|\bm{0}\},&
		\{E||C_{33}^{-}|\bm{0}\},&
		\{E||C_{32}^{-}|\bm{0}\},&
		\{E||C_{34}^{-}|\bm{0}\},&
		\{E||C_{2a}|\bm{0}\},&
		\{E||C_{2b}|\bm{0}\},&
		\{E||C_{4z}^{-}|\bm{0}\},&
		\{E||C_{4z}^{+}|\bm{0}\},
	\end{array}\) \\
	& \(\begin{array}{llllllll}
		\{E||C_{4x}^{-}|\bm{0}\},&
		\{E||C_{2d}|\bm{0}\},&
		\{E||C_{2f}|\bm{0}\},&
		\{E||C_{4x}^{+}|\bm{0}\},&
		\{E||C_{4y}^{+}|\bm{0}\},&
		\{E||C_{2c}|\bm{0}\},&
		\{E||C_{4y}^{-}|\bm{0}\},&
		\{E||C_{2e}|\bm{0}\},
	\end{array}\) \\
	& \(\begin{array}{llllllll}
		\{T||I|\bm{0}\},&
		\{T||\sigma_z|\bm{0}\},&
		\{T||\sigma_y|\bm{0}\},&
		\{T||\sigma_x|\bm{0}\},&
		\{T||S_{61}^{-}|\bm{0}\},&
		\{T||S_{64}^{-}|\bm{0}\},&
		\{T||S_{63}^{-}|\bm{0}\},&
		\{T||S_{62}^{-}|\bm{0}\},
	\end{array}\) \\
	& \(\begin{array}{llllllll}
		\{T||S_{61}^{+}|\bm{0}\},&
		\{T||S_{63}^{+}|\bm{0}\},&
		\{T||S_{62}^{+}|\bm{0}\},&
		\{T||S_{64}^{+}|\bm{0}\},&
		\{T||\sigma_{da}|\bm{0}\},&
		\{T||\sigma_{db}|\bm{0}\},&
		\{T||S_{4z}^{+}|\bm{0}\},&
		\{T||S_{4z}^{-}|\bm{0}\},
	\end{array}\) \\
	& \(\begin{array}{llllllll}
		\{T||S_{4x}^{+}|\bm{0}\},&
		\{T||\sigma_{dd}|\bm{0}\},&
		\{T||\sigma_{df}|\bm{0}\},&
		\{T||S_{4x}^{-}|\bm{0}\},&
		\{T||S_{4y}^{-}|\bm{0}\},&
		\{T||\sigma_{dc}|\bm{0}\},&
		\{T||S_{4y}^{+}|\bm{0}\},&
		\{T||\sigma_{de}|\bm{0}\}
	\end{array}\) \\
	\hline
	\(221.1.2.3.\mathrm{L\_1}\) & \(\begin{array}{llllllll}
		\{E||E|\bm{0}\},&
		\{E||C_{2z}|\bm{0}\},&
		\{E||C_{2y}|\bm{0}\},&
		\{E||C_{2x}|\bm{0}\},&
		\{E||C_{31}^{+}|\bm{0}\},&
		\{E||C_{34}^{+}|\bm{0}\},&
		\{E||C_{33}^{+}|\bm{0}\},&
		\{E||C_{32}^{+}|\bm{0}\},
	\end{array}\) \\
	& \(\begin{array}{llllllll}
		\{E||C_{31}^{-}|\bm{0}\},&
		\{E||C_{33}^{-}|\bm{0}\},&
		\{E||C_{32}^{-}|\bm{0}\},&
		\{E||C_{34}^{-}|\bm{0}\},&
		\{E||I|\bm{0}\},&
		\{E||\sigma_z|\bm{0}\},&
		\{E||\sigma_y|\bm{0}\},&
		\{E||\sigma_x|\bm{0}\},
	\end{array}\) \\
	& \(\begin{array}{llllllll}
		\{E||S_{61}^{-}|\bm{0}\},&
		\{E||S_{64}^{-}|\bm{0}\},&
		\{E||S_{63}^{-}|\bm{0}\},&
		\{E||S_{62}^{-}|\bm{0}\},&
		\{E||S_{61}^{+}|\bm{0}\},&
		\{E||S_{63}^{+}|\bm{0}\},&
		\{E||S_{62}^{+}|\bm{0}\},&
		\{E||S_{64}^{+}|\bm{0}\},
	\end{array}\) \\
	& \(\begin{array}{llllllll}
		\{T||C_{2a}|\bm{0}\},&
		\{T||C_{2b}|\bm{0}\},&
		\{T||C_{4z}^{-}|\bm{0}\},&
		\{T||C_{4z}^{+}|\bm{0}\},&
		\{T||C_{4x}^{-}|\bm{0}\},&
		\{T||C_{2d}|\bm{0}\},&
		\{T||C_{2f}|\bm{0}\},&
		\{T||C_{4x}^{+}|\bm{0}\},
	\end{array}\) \\
	& \(\begin{array}{llllllll}
		\{T||C_{4y}^{+}|\bm{0}\},&
		\{T||C_{2c}|\bm{0}\},&
		\{T||C_{4y}^{-}|\bm{0}\},&
		\{T||C_{2e}|\bm{0}\},&
		\{T||\sigma_{da}|\bm{0}\},&
		\{T||\sigma_{db}|\bm{0}\},&
		\{T||S_{4z}^{+}|\bm{0}\},&
		\{T||S_{4z}^{-}|\bm{0}\},
	\end{array}\) \\
	& \(\begin{array}{llllllll}
		\{T||S_{4x}^{+}|\bm{0}\},&
		\{T||\sigma_{dd}|\bm{0}\},&
		\{T||\sigma_{df}|\bm{0}\},&
		\{T||S_{4x}^{-}|\bm{0}\},&
		\{T||S_{4y}^{-}|\bm{0}\},&
		\{T||\sigma_{dc}|\bm{0}\},&
		\{T||S_{4y}^{+}|\bm{0}\},&
		\{T||\sigma_{de}|\bm{0}\}
	\end{array}\) \\
	\hline
	\(221.2.1.1.\mathrm{L\_1}\) & \(\begin{array}{llllllll}
		\{E||E|\bm{0}\},&
		\{E||C_{2z}|\bm{0}\},&
		\{E||C_{2y}|\bm{0}\},&
		\{E||C_{2x}|\bm{0}\},&
		\{E||C_{31}^{+}|\bm{0}\},&
		\{E||C_{34}^{+}|\bm{0}\},&
		\{E||C_{33}^{+}|\bm{0}\},&
		\{E||C_{32}^{+}|\bm{0}\},
	\end{array}\) \\
	& \(\begin{array}{llllllll}
		\{E||C_{31}^{-}|\bm{0}\},&
		\{E||C_{33}^{-}|\bm{0}\},&
		\{E||C_{32}^{-}|\bm{0}\},&
		\{E||C_{34}^{-}|\bm{0}\},&
		\{E||C_{2a}|\bm{0}\},&
		\{E||C_{2b}|\bm{0}\},&
		\{E||C_{4z}^{-}|\bm{0}\},&
		\{E||C_{4z}^{+}|\bm{0}\},
	\end{array}\) \\
	& \(\begin{array}{llllllll}
		\{E||C_{4x}^{-}|\bm{0}\},&
		\{E||C_{2d}|\bm{0}\},&
		\{E||C_{2f}|\bm{0}\},&
		\{E||C_{4x}^{+}|\bm{0}\},&
		\{E||C_{4y}^{+}|\bm{0}\},&
		\{E||C_{2c}|\bm{0}\},&
		\{E||C_{4y}^{-}|\bm{0}\},&
		\{E||C_{2e}|\bm{0}\},
	\end{array}\) \\
	& \(\begin{array}{llllllll}
		\{E||I|\bm{0}\},&
		\{E||\sigma_z|\bm{0}\},&
		\{E||\sigma_y|\bm{0}\},&
		\{E||\sigma_x|\bm{0}\},&
		\{E||S_{61}^{-}|\bm{0}\},&
		\{E||S_{64}^{-}|\bm{0}\},&
		\{E||S_{63}^{-}|\bm{0}\},&
		\{E||S_{62}^{-}|\bm{0}\},
	\end{array}\) \\
	& \(\begin{array}{llllllll}
		\{E||S_{61}^{+}|\bm{0}\},&
		\{E||S_{63}^{+}|\bm{0}\},&
		\{E||S_{62}^{+}|\bm{0}\},&
		\{E||S_{64}^{+}|\bm{0}\},&
		\{E||\sigma_{da}|\bm{0}\},&
		\{E||\sigma_{db}|\bm{0}\},&
		\{E||S_{4z}^{+}|\bm{0}\},&
		\{E||S_{4z}^{-}|\bm{0}\},
	\end{array}\) \\
	& \(\begin{array}{llllllll}
		\{E||S_{4x}^{+}|\bm{0}\},&
		\{E||\sigma_{dd}|\bm{0}\},&
		\{E||\sigma_{df}|\bm{0}\},&
		\{E||S_{4x}^{-}|\bm{0}\},&
		\{E||S_{4y}^{-}|\bm{0}\},&
		\{E||\sigma_{dc}|\bm{0}\},&
		\{E||S_{4y}^{+}|\bm{0}\},&
		\{E||\sigma_{de}|\bm{0}\},
	\end{array}\) \\
	& \(\begin{array}{llllllll}
		\{T||E|\bm{t}\},&
		\{T||C_{2z}|\bm{t}\},&
		\{T||C_{2y}|\bm{t}\},&
		\{T||C_{2x}|\bm{t}\},&
		\{T||C_{31}^{+}|\bm{t}\},&
		\{T||C_{34}^{+}|\bm{t}\},&
		\{T||C_{33}^{+}|\bm{t}\},&
		\{T||C_{32}^{+}|\bm{t}\},
	\end{array}\) \\
	& \(\begin{array}{llllllll}
		\{T||C_{31}^{-}|\bm{t}\},&
		\{T||C_{33}^{-}|\bm{t}\},&
		\{T||C_{32}^{-}|\bm{t}\},&
		\{T||C_{34}^{-}|\bm{t}\},&
		\{T||C_{2a}|\bm{t}\},&
		\{T||C_{2b}|\bm{t}\},&
		\{T||C_{4z}^{-}|\bm{t}\},&
		\{T||C_{4z}^{+}|\bm{t}\},
	\end{array}\) \\
	& \(\begin{array}{llllllll}
		\{T||C_{4x}^{-}|\bm{t}\},&
		\{T||C_{2d}|\bm{t}\},&
		\{T||C_{2f}|\bm{t}\},&
		\{T||C_{4x}^{+}|\bm{t}\},&
		\{T||C_{4y}^{+}|\bm{t}\},&
		\{T||C_{2c}|\bm{t}\},&
		\{T||C_{4y}^{-}|\bm{t}\},&
		\{T||C_{2e}|\bm{t}\},
	\end{array}\) \\
	& \(\begin{array}{llllllll}
		\{T||I|\bm{t}\},&
		\{T||\sigma_z|\bm{t}\},&
		\{T||\sigma_y|\bm{t}\},&
		\{T||\sigma_x|\bm{t}\},&
		\{T||S_{61}^{-}|\bm{t}\},&
		\{T||S_{64}^{-}|\bm{t}\},&
		\{T||S_{63}^{-}|\bm{t}\},&
		\{T||S_{62}^{-}|\bm{t}\},
	\end{array}\) \\
	& \(\begin{array}{llllllll}
		\{T||S_{61}^{+}|\bm{t}\},&
		\{T||S_{63}^{+}|\bm{t}\},&
		\{T||S_{62}^{+}|\bm{t}\},&
		\{T||S_{64}^{+}|\bm{t}\},&
		\{T||\sigma_{da}|\bm{t}\},&
		\{T||\sigma_{db}|\bm{t}\},&
		\{T||S_{4z}^{+}|\bm{t}\},&
		\{T||S_{4z}^{-}|\bm{t}\},
	\end{array}\) \\
	& \(\begin{array}{llllllll}
		\{T||S_{4x}^{+}|\bm{t}\},&
		\{T||\sigma_{dd}|\bm{t}\},&
		\{T||\sigma_{df}|\bm{t}\},&
		\{T||S_{4x}^{-}|\bm{t}\},&
		\{T||S_{4y}^{-}|\bm{t}\},&
		\{T||\sigma_{dc}|\bm{t}\},&
		\{T||S_{4y}^{+}|\bm{t}\},&
		\{T||\sigma_{de}|\bm{t}\}
	\end{array}\) \\
	\hline
	\(221.2.1.1.\mathrm{L\_2}\) & \(\begin{array}{llllllll}
		\{E||E|\bm{0}\},&
		\{E||I|\bm{t}\},&
		\{E||\sigma_x|\bm{t}\},&
		\{E||\sigma_z|\bm{t}\},&
		\{E||\sigma_y|\bm{t}\},&
		\{E||C_{2x}|\bm{0}\},&
		\{E||C_{2z}|\bm{0}\},&
		\{E||C_{2y}|\bm{0}\},
	\end{array}\) \\
	& \(\begin{array}{llllllll}
		\{E||C_{34}^{+}|\bm{0}\},&
		\{E||C_{31}^{+}|\bm{0}\},&
		\{E||C_{32}^{+}|\bm{0}\},&
		\{E||C_{33}^{+}|\bm{0}\},&
		\{E||C_{34}^{-}|\bm{0}\},&
		\{E||C_{31}^{-}|\bm{0}\},&
		\{E||C_{32}^{-}|\bm{0}\},&
		\{E||C_{33}^{-}|\bm{0}\},
	\end{array}\) \\
	& \(\begin{array}{llllllll}
		\{E||S_{64}^{-}|\bm{t}\},&
		\{E||S_{61}^{-}|\bm{t}\},&
		\{E||S_{62}^{-}|\bm{t}\},&
		\{E||S_{63}^{-}|\bm{t}\},&
		\{E||S_{64}^{+}|\bm{t}\},&
		\{E||S_{61}^{+}|\bm{t}\},&
		\{E||S_{62}^{+}|\bm{t}\},&
		\{E||S_{63}^{+}|\bm{t}\},
	\end{array}\) \\
	& \(\begin{array}{llllllll}
		\{E||C_{4x}^{-}|\bm{t}\},&
		\{E||C_{4z}^{-}|\bm{t}\},&
		\{E||C_{4y}^{+}|\bm{t}\},&
		\{E||C_{4x}^{+}|\bm{t}\},&
		\{E||C_{4z}^{+}|\bm{t}\},&
		\{E||C_{4y}^{-}|\bm{t}\},&
		\{E||S_{4x}^{+}|\bm{0}\},&
		\{E||S_{4z}^{+}|\bm{0}\},
	\end{array}\) \\
	& \(\begin{array}{llllllll}
		\{E||S_{4y}^{-}|\bm{0}\},&
		\{E||S_{4x}^{-}|\bm{0}\},&
		\{E||S_{4z}^{-}|\bm{0}\},&
		\{E||S_{4y}^{+}|\bm{0}\},&
		\{E||C_{2c}|\bm{t}\},&
		\{E||C_{2e}|\bm{t}\},&
		\{E||C_{2b}|\bm{t}\},&
		\{E||C_{2f}|\bm{t}\},
	\end{array}\) \\
	& \(\begin{array}{llllllll}
		\{E||C_{2a}|\bm{t}\},&
		\{E||C_{2d}|\bm{t}\},&
		\{E||\sigma_{dc}|\bm{0}\},&
		\{E||\sigma_{de}|\bm{0}\},&
		\{E||\sigma_{db}|\bm{0}\},&
		\{E||\sigma_{df}|\bm{0}\},&
		\{E||\sigma_{da}|\bm{0}\},&
		\{E||\sigma_{dd}|\bm{0}\},
	\end{array}\) \\
	& \(\begin{array}{llllllll}
		\{T||E|\bm{t}\},&
		\{T||I|\bm{0}\},&
		\{T||\sigma_x|\bm{0}\},&
		\{T||\sigma_z|\bm{0}\},&
		\{T||\sigma_y|\bm{0}\},&
		\{T||C_{2x}|\bm{t}\},&
		\{T||C_{2z}|\bm{t}\},&
		\{T||C_{2y}|\bm{t}\},
	\end{array}\) \\
	& \(\begin{array}{llllllll}
		\{T||C_{34}^{+}|\bm{t}\},&
		\{T||C_{31}^{+}|\bm{t}\},&
		\{T||C_{32}^{+}|\bm{t}\},&
		\{T||C_{33}^{+}|\bm{t}\},&
		\{T||C_{34}^{-}|\bm{t}\},&
		\{T||C_{31}^{-}|\bm{t}\},&
		\{T||C_{32}^{-}|\bm{t}\},&
		\{T||C_{33}^{-}|\bm{t}\},
	\end{array}\) \\
	& \(\begin{array}{llllllll}
		\{T||S_{64}^{-}|\bm{0}\},&
		\{T||S_{61}^{-}|\bm{0}\},&
		\{T||S_{62}^{-}|\bm{0}\},&
		\{T||S_{63}^{-}|\bm{0}\},&
		\{T||S_{64}^{+}|\bm{0}\},&
		\{T||S_{61}^{+}|\bm{0}\},&
		\{T||S_{62}^{+}|\bm{0}\},&
		\{T||S_{63}^{+}|\bm{0}\},
	\end{array}\) \\
	& \(\begin{array}{llllllll}
		\{T||C_{4x}^{-}|\bm{0}\},&
		\{T||C_{4z}^{-}|\bm{0}\},&
		\{T||C_{4y}^{+}|\bm{0}\},&
		\{T||C_{4x}^{+}|\bm{0}\},&
		\{T||C_{4z}^{+}|\bm{0}\},&
		\{T||C_{4y}^{-}|\bm{0}\},&
		\{T||S_{4x}^{+}|\bm{t}\},&
		\{T||S_{4z}^{+}|\bm{t}\},
	\end{array}\) \\
	& \(\begin{array}{llllllll}
		\{T||S_{4y}^{-}|\bm{t}\},&
		\{T||S_{4x}^{-}|\bm{t}\},&
		\{T||S_{4z}^{-}|\bm{t}\},&
		\{T||S_{4y}^{+}|\bm{t}\},&
		\{T||C_{2c}|\bm{0}\},&
		\{T||C_{2e}|\bm{0}\},&
		\{T||C_{2b}|\bm{0}\},&
		\{T||C_{2f}|\bm{0}\},
	\end{array}\) \\
	& \(\begin{array}{llllllll}
		\{T||C_{2a}|\bm{0}\},&
		\{T||C_{2d}|\bm{0}\},&
		\{T||\sigma_{dc}|\bm{t}\},&
		\{T||\sigma_{de}|\bm{t}\},&
		\{T||\sigma_{db}|\bm{t}\},&
		\{T||\sigma_{df}|\bm{t}\},&
		\{T||\sigma_{da}|\bm{t}\},&
		\{T||\sigma_{dd}|\bm{t}\}
	\end{array}\) \\
	\hline
	\(221.2.1.2.\mathrm{L\_1}\) & \(\begin{array}{llllllll}
		\{E||E|\bm{0}\},&
		\{E||C_{2z}|\bm{0}\},&
		\{E||C_{2y}|\bm{0}\},&
		\{E||C_{2x}|\bm{0}\},&
		\{E||C_{31}^{+}|\bm{0}\},&
		\{E||C_{34}^{+}|\bm{0}\},&
		\{E||C_{33}^{+}|\bm{0}\},&
		\{E||C_{32}^{+}|\bm{0}\},
	\end{array}\) \\
	& \(\begin{array}{llllllll}
		\{E||C_{31}^{-}|\bm{0}\},&
		\{E||C_{33}^{-}|\bm{0}\},&
		\{E||C_{32}^{-}|\bm{0}\},&
		\{E||C_{34}^{-}|\bm{0}\},&
		\{E||C_{2a}|\bm{0}\},&
		\{E||C_{2b}|\bm{0}\},&
		\{E||C_{4z}^{-}|\bm{0}\},&
		\{E||C_{4z}^{+}|\bm{0}\},
	\end{array}\) \\
	& \(\begin{array}{llllllll}
		\{E||C_{4x}^{-}|\bm{0}\},&
		\{E||C_{2d}|\bm{0}\},&
		\{E||C_{2f}|\bm{0}\},&
		\{E||C_{4x}^{+}|\bm{0}\},&
		\{E||C_{4y}^{+}|\bm{0}\},&
		\{E||C_{2c}|\bm{0}\},&
		\{E||C_{4y}^{-}|\bm{0}\},&
		\{E||C_{2e}|\bm{0}\},
	\end{array}\) \\
	& \(\begin{array}{llllllll}
		\{E||I|\bm{t}\},&
		\{E||\sigma_z|\bm{t}\},&
		\{E||\sigma_y|\bm{t}\},&
		\{E||\sigma_x|\bm{t}\},&
		\{E||S_{61}^{-}|\bm{t}\},&
		\{E||S_{64}^{-}|\bm{t}\},&
		\{E||S_{63}^{-}|\bm{t}\},&
		\{E||S_{62}^{-}|\bm{t}\},
	\end{array}\) \\
	& \(\begin{array}{llllllll}
		\{E||S_{61}^{+}|\bm{t}\},&
		\{E||S_{63}^{+}|\bm{t}\},&
		\{E||S_{62}^{+}|\bm{t}\},&
		\{E||S_{64}^{+}|\bm{t}\},&
		\{E||\sigma_{da}|\bm{t}\},&
		\{E||\sigma_{db}|\bm{t}\},&
		\{E||S_{4z}^{+}|\bm{t}\},&
		\{E||S_{4z}^{-}|\bm{t}\},
	\end{array}\) \\
	& \(\begin{array}{llllllll}
		\{E||S_{4x}^{+}|\bm{t}\},&
		\{E||\sigma_{dd}|\bm{t}\},&
		\{E||\sigma_{df}|\bm{t}\},&
		\{E||S_{4x}^{-}|\bm{t}\},&
		\{E||S_{4y}^{-}|\bm{t}\},&
		\{E||\sigma_{dc}|\bm{t}\},&
		\{E||S_{4y}^{+}|\bm{t}\},&
		\{E||\sigma_{de}|\bm{t}\},
	\end{array}\) \\
	& \(\begin{array}{llllllll}
		\{T||E|\bm{t}\},&
		\{T||C_{2z}|\bm{t}\},&
		\{T||C_{2y}|\bm{t}\},&
		\{T||C_{2x}|\bm{t}\},&
		\{T||C_{31}^{+}|\bm{t}\},&
		\{T||C_{34}^{+}|\bm{t}\},&
		\{T||C_{33}^{+}|\bm{t}\},&
		\{T||C_{32}^{+}|\bm{t}\},
	\end{array}\) \\
	& \(\begin{array}{llllllll}
		\{T||C_{31}^{-}|\bm{t}\},&
		\{T||C_{33}^{-}|\bm{t}\},&
		\{T||C_{32}^{-}|\bm{t}\},&
		\{T||C_{34}^{-}|\bm{t}\},&
		\{T||C_{2a}|\bm{t}\},&
		\{T||C_{2b}|\bm{t}\},&
		\{T||C_{4z}^{-}|\bm{t}\},&
		\{T||C_{4z}^{+}|\bm{t}\},
	\end{array}\) \\
	& \(\begin{array}{llllllll}
		\{T||C_{4x}^{-}|\bm{t}\},&
		\{T||C_{2d}|\bm{t}\},&
		\{T||C_{2f}|\bm{t}\},&
		\{T||C_{4x}^{+}|\bm{t}\},&
		\{T||C_{4y}^{+}|\bm{t}\},&
		\{T||C_{2c}|\bm{t}\},&
		\{T||C_{4y}^{-}|\bm{t}\},&
		\{T||C_{2e}|\bm{t}\},
	\end{array}\) \\
	& \(\begin{array}{llllllll}
		\{T||I|\bm{0}\},&
		\{T||\sigma_z|\bm{0}\},&
		\{T||\sigma_y|\bm{0}\},&
		\{T||\sigma_x|\bm{0}\},&
		\{T||S_{61}^{-}|\bm{0}\},&
		\{T||S_{64}^{-}|\bm{0}\},&
		\{T||S_{63}^{-}|\bm{0}\},&
		\{T||S_{62}^{-}|\bm{0}\},
	\end{array}\) \\
	& \(\begin{array}{llllllll}
		\{T||S_{61}^{+}|\bm{0}\},&
		\{T||S_{63}^{+}|\bm{0}\},&
		\{T||S_{62}^{+}|\bm{0}\},&
		\{T||S_{64}^{+}|\bm{0}\},&
		\{T||\sigma_{da}|\bm{0}\},&
		\{T||\sigma_{db}|\bm{0}\},&
		\{T||S_{4z}^{+}|\bm{0}\},&
		\{T||S_{4z}^{-}|\bm{0}\},
	\end{array}\) \\
	& \(\begin{array}{llllllll}
		\{T||S_{4x}^{+}|\bm{0}\},&
		\{T||\sigma_{dd}|\bm{0}\},&
		\{T||\sigma_{df}|\bm{0}\},&
		\{T||S_{4x}^{-}|\bm{0}\},&
		\{T||S_{4y}^{-}|\bm{0}\},&
		\{T||\sigma_{dc}|\bm{0}\},&
		\{T||S_{4y}^{+}|\bm{0}\},&
		\{T||\sigma_{de}|\bm{0}\}
	\end{array}\) \\
	\hline
	\(221.2.1.2.\mathrm{L\_2}\) & \(\begin{array}{llllllll}
		\{E||E|\bm{0}\},&
		\{E||I|\bm{0}\},&
		\{E||\sigma_x|\bm{0}\},&
		\{E||\sigma_z|\bm{0}\},&
		\{E||\sigma_y|\bm{0}\},&
		\{E||C_{2x}|\bm{0}\},&
		\{E||C_{2z}|\bm{0}\},&
		\{E||C_{2y}|\bm{0}\},
	\end{array}\) \\
	& \(\begin{array}{llllllll}
		\{E||C_{34}^{+}|\bm{0}\},&
		\{E||C_{31}^{+}|\bm{0}\},&
		\{E||C_{32}^{+}|\bm{0}\},&
		\{E||C_{33}^{+}|\bm{0}\},&
		\{E||C_{34}^{-}|\bm{0}\},&
		\{E||C_{31}^{-}|\bm{0}\},&
		\{E||C_{32}^{-}|\bm{0}\},&
		\{E||C_{33}^{-}|\bm{0}\},
	\end{array}\) \\
	& \(\begin{array}{llllllll}
		\{E||S_{64}^{-}|\bm{0}\},&
		\{E||S_{61}^{-}|\bm{0}\},&
		\{E||S_{62}^{-}|\bm{0}\},&
		\{E||S_{63}^{-}|\bm{0}\},&
		\{E||S_{64}^{+}|\bm{0}\},&
		\{E||S_{61}^{+}|\bm{0}\},&
		\{E||S_{62}^{+}|\bm{0}\},&
		\{E||S_{63}^{+}|\bm{0}\},
	\end{array}\) \\
	& \(\begin{array}{llllllll}
		\{E||C_{4x}^{-}|\bm{t}\},&
		\{E||C_{4z}^{-}|\bm{t}\},&
		\{E||C_{4y}^{+}|\bm{t}\},&
		\{E||C_{4x}^{+}|\bm{t}\},&
		\{E||C_{4z}^{+}|\bm{t}\},&
		\{E||C_{4y}^{-}|\bm{t}\},&
		\{E||S_{4x}^{+}|\bm{t}\},&
		\{E||S_{4z}^{+}|\bm{t}\},
	\end{array}\) \\
	& \(\begin{array}{llllllll}
		\{E||S_{4y}^{-}|\bm{t}\},&
		\{E||S_{4x}^{-}|\bm{t}\},&
		\{E||S_{4z}^{-}|\bm{t}\},&
		\{E||S_{4y}^{+}|\bm{t}\},&
		\{E||C_{2c}|\bm{t}\},&
		\{E||C_{2e}|\bm{t}\},&
		\{E||C_{2b}|\bm{t}\},&
		\{E||C_{2f}|\bm{t}\},
	\end{array}\) \\
	& \(\begin{array}{llllllll}
		\{E||C_{2a}|\bm{t}\},&
		\{E||C_{2d}|\bm{t}\},&
		\{E||\sigma_{dc}|\bm{t}\},&
		\{E||\sigma_{de}|\bm{t}\},&
		\{E||\sigma_{db}|\bm{t}\},&
		\{E||\sigma_{df}|\bm{t}\},&
		\{E||\sigma_{da}|\bm{t}\},&
		\{E||\sigma_{dd}|\bm{t}\},
	\end{array}\) \\
	& \(\begin{array}{llllllll}
		\{T||E|\bm{t}\},&
		\{T||I|\bm{t}\},&
		\{T||\sigma_x|\bm{t}\},&
		\{T||\sigma_z|\bm{t}\},&
		\{T||\sigma_y|\bm{t}\},&
		\{T||C_{2x}|\bm{t}\},&
		\{T||C_{2z}|\bm{t}\},&
		\{T||C_{2y}|\bm{t}\},
	\end{array}\) \\
	& \(\begin{array}{llllllll}
		\{T||C_{34}^{+}|\bm{t}\},&
		\{T||C_{31}^{+}|\bm{t}\},&
		\{T||C_{32}^{+}|\bm{t}\},&
		\{T||C_{33}^{+}|\bm{t}\},&
		\{T||C_{34}^{-}|\bm{t}\},&
		\{T||C_{31}^{-}|\bm{t}\},&
		\{T||C_{32}^{-}|\bm{t}\},&
		\{T||C_{33}^{-}|\bm{t}\},
	\end{array}\) \\
	& \(\begin{array}{llllllll}
		\{T||S_{64}^{-}|\bm{t}\},&
		\{T||S_{61}^{-}|\bm{t}\},&
		\{T||S_{62}^{-}|\bm{t}\},&
		\{T||S_{63}^{-}|\bm{t}\},&
		\{T||S_{64}^{+}|\bm{t}\},&
		\{T||S_{61}^{+}|\bm{t}\},&
		\{T||S_{62}^{+}|\bm{t}\},&
		\{T||S_{63}^{+}|\bm{t}\},
	\end{array}\) \\
	& \(\begin{array}{llllllll}
		\{T||C_{4x}^{-}|\bm{0}\},&
		\{T||C_{4z}^{-}|\bm{0}\},&
		\{T||C_{4y}^{+}|\bm{0}\},&
		\{T||C_{4x}^{+}|\bm{0}\},&
		\{T||C_{4z}^{+}|\bm{0}\},&
		\{T||C_{4y}^{-}|\bm{0}\},&
		\{T||S_{4x}^{+}|\bm{0}\},&
		\{T||S_{4z}^{+}|\bm{0}\},
	\end{array}\) \\
	& \(\begin{array}{llllllll}
		\{T||S_{4y}^{-}|\bm{0}\},&
		\{T||S_{4x}^{-}|\bm{0}\},&
		\{T||S_{4z}^{-}|\bm{0}\},&
		\{T||S_{4y}^{+}|\bm{0}\},&
		\{T||C_{2c}|\bm{0}\},&
		\{T||C_{2e}|\bm{0}\},&
		\{T||C_{2b}|\bm{0}\},&
		\{T||C_{2f}|\bm{0}\},
	\end{array}\) \\
	& \(\begin{array}{llllllll}
		\{T||C_{2a}|\bm{0}\},&
		\{T||C_{2d}|\bm{0}\},&
		\{T||\sigma_{dc}|\bm{0}\},&
		\{T||\sigma_{de}|\bm{0}\},&
		\{T||\sigma_{db}|\bm{0}\},&
		\{T||\sigma_{df}|\bm{0}\},&
		\{T||\sigma_{da}|\bm{0}\},&
		\{T||\sigma_{dd}|\bm{0}\}
	\end{array}\) \\
	\hline
\end{longtable}
\endgroup

Considering Ni atoms as the only magnetic atoms, we examine which maximal collinear SSG subgroups of SG No.~221 allow nonzero MMs on the Ni sites. The three Ni atoms are related by the SSG operations $\{E||C_{31}^{+}|\bm{0}\}$ and $\{E||C_{31}^{-}|\bm{0}\}$, which are common to all the maximal collinear SSG subgroups of SG No.~221 considered here. Indeed, according to
\begin{equation}
	\{C_{31}^{+}|\bm{0}\}:(x,y,z)\rightarrow(z,x,y),
	\qquad
	\{C_{31}^{-}|\bm{0}\}:(x,y,z)\rightarrow(y,z,x),
\end{equation}
the representative Ni site $\mathrm{Ni}_1=(0,1/2,1/2)$ is mapped to the other two Ni sites:
\begin{equation}
	(0,1/2,1/2)\rightarrow(1/2,0,1/2),
	\qquad
	(0,1/2,1/2)\rightarrow(1/2,1/2,0).
\end{equation}
Thus, if the MM of $\mathrm{Ni}_1$ is forced to vanish, the MMs of all Ni atoms also vanish:
\begin{equation}
	\bm{M}_{\mathrm{Ni}_1}=\bm{0}
	\quad \Rightarrow \quad
	\bm{M}_{\mathrm{Ni}_2}=\bm{M}_{\mathrm{Ni}_3}=\bm{0}.
\end{equation}
The following analysis is therefore based on the representative Ni site $\mathrm{Ni}_1$.

For $221.1.2.1.\mathrm{L\_1}$, the SSG operation $\{T||\sigma_x|\bm{0}\}$ contains the real-space operation $\{\sigma_x|\bm{0}\}$. According to
\begin{equation}
	\{\sigma_x|\bm{0}\}:(x,y,z)\rightarrow(-x,y,z),
\end{equation}
we have
\begin{equation}
	(0,1/2,1/2)\rightarrow(0,1/2,1/2).
\end{equation}
Thus, $\mathrm{Ni}_1$ is mapped onto itself. Since the spin-space operation $T$ reverses the MM, the MM at $\mathrm{Ni}_1$ satisfies
\begin{equation}
	\bm{M}_{\mathrm{Ni}_1}=-\bm{M}_{\mathrm{Ni}_1},
	\qquad \Rightarrow \qquad
	\bm{M}_{\mathrm{Ni}_1}=\bm{0}.
\end{equation}

For $221.2.1.1.\mathrm{L\_1}$, the SSG operation $\{T||C_{2y}|\bm{t}\}$ contains the real-space operation $\{C_{2y}|\bm{t}\}$. According to
\begin{equation}
	\{C_{2y}|\bm{0}\}:(x,y,z)\rightarrow(-x,y,-z),
\end{equation}
we have
\begin{equation}
	\{C_{2y}|\bm{t}\}:(0,1/2,1/2)\rightarrow \bm{t}+(0,1/2,-1/2)
	= (0,1/2,1/2).
\end{equation}
Thus, $\mathrm{Ni}_1$ is mapped onto itself modulo an allowed translation. Since the spin-space operation $T$ reverses the MM,
\begin{equation}
	\bm{M}_{\mathrm{Ni}_1}=-\bm{M}_{\mathrm{Ni}_1},
	\qquad \Rightarrow \qquad
	\bm{M}_{\mathrm{Ni}_1}=\bm{0}.
\end{equation}

For $221.2.1.2.\mathrm{L\_2}$, the SSG operation $\{T||\sigma_z|\bm{t}\}$ contains the real-space operation $\{\sigma_z|\bm{t}\}$. According to
\begin{equation}
	\{\sigma_z|\bm{0}\}:(x,y,z)\rightarrow(x,y,-z),
\end{equation}
we have
\begin{equation}
	\{\sigma_z|\bm{t}\}:(0,1/2,1/2)\rightarrow \bm{t}+(0,1/2,-1/2)
	= (0,1/2,1/2).
\end{equation}
Thus, $\mathrm{Ni}_1$ is mapped onto itself modulo an allowed translation. Since the spin-space operation $T$ reverses the MM,
\begin{equation}
	\bm{M}_{\mathrm{Ni}_1}=-\bm{M}_{\mathrm{Ni}_1},
	\qquad \Rightarrow \qquad
	\bm{M}_{\mathrm{Ni}_1}=\bm{0}.
\end{equation}
Consequently, the MMs of all Ni atoms vanish for $221.1.2.1.\mathrm{L\_1}$, $221.2.1.1.\mathrm{L\_1}$, and $221.2.1.2.\mathrm{L\_2}$.

For $221.1.2.2.\mathrm{L\_1}$, $221.2.1.1.\mathrm{L\_2}$, and $221.2.1.2.\mathrm{L\_1}$, the SSG operation $\{T||I|\bm{0}\}$ contains the real-space inversion operation $\{I|\bm{0}\}$. According to
\begin{equation}
	\{I|\bm{0}\}:(x,y,z)\rightarrow(-x,-y,-z),
\end{equation}
we have
\begin{equation}
	(0,1/2,1/2)\rightarrow(0,-1/2,-1/2).
\end{equation}
We label the Ni atom at $(0,-1/2,-1/2)$ as $\mathrm{Ni}_1'$. This site is equivalent to $\mathrm{Ni}_1=(0,1/2,1/2)$ modulo a lattice translation for $221.1.2.2.\mathrm{L\_1}$, and modulo a magnetic-cell lattice translation for $221.2.1.1.\mathrm{L\_2}$ and $221.2.1.2.\mathrm{L\_1}$:
\begin{equation}
	\mathrm{Ni}_1' \equiv \mathrm{Ni}_1 .
\end{equation}
Since the spin-space operation $T$ reverses the MM, the operation $\{T||I|\bm{0}\}$ gives
\begin{equation}
	\bm{M}_{\mathrm{Ni}_1'}=-\bm{M}_{\mathrm{Ni}_1}.
\end{equation}
On the other hand, because $\mathrm{Ni}_1'$ is equivalent to $\mathrm{Ni}_1$ by the corresponding lattice or magnetic-cell lattice translation, we also have
\begin{equation}
	\bm{M}_{\mathrm{Ni}_1'}=\bm{M}_{\mathrm{Ni}_1}.
\end{equation}
Combining the two constraints gives
\begin{equation}
	\bm{M}_{\mathrm{Ni}_1}=-\bm{M}_{\mathrm{Ni}_1},
	\qquad \Rightarrow \qquad
	\bm{M}_{\mathrm{Ni}_1}=\bm{0}.
\end{equation}
Consequently, the MMs of all Ni atoms vanish for these three SSGs.

For $221.1.2.3.\mathrm{L\_1}$, the SSG operation $\{T||C_{4x}^{-}|\bm{0}\}$ contains the real-space operation $\{C_{4x}^{-}|\bm{0}\}$. According to
\begin{equation}
	\{C_{4x}^{-}|\bm{0}\}:(x,y,z)\rightarrow(x,z,-y),
\end{equation}
we have
\begin{equation}
	(0,1/2,1/2)\rightarrow(0,1/2,-1/2).
\end{equation}
The position $(0,1/2,-1/2)$ is equivalent to $\mathrm{Ni}_1=(0,1/2,1/2)$ modulo a lattice translation:
\begin{equation}
	(0,1/2,-1/2)\equiv(0,1/2,1/2).
\end{equation}
Since the spin-space operation $T$ reverses the MM, the MM of the Ni atom at $(0,1/2,-1/2)$, labeled as $\mathrm{Ni}_1'$, satisfies
\begin{equation}
	\bm{M}_{\mathrm{Ni}_1'}=-\bm{M}_{\mathrm{Ni}_1}.
\end{equation}
However, because $\mathrm{Ni}_1'$ is equivalent to $\mathrm{Ni}_1$ by a lattice translation, we also have
\begin{equation}
	\bm{M}_{\mathrm{Ni}_1'}=\bm{M}_{\mathrm{Ni}_1}.
\end{equation}
Combining the two constraints gives
\begin{equation}
	\bm{M}_{\mathrm{Ni}_1}=-\bm{M}_{\mathrm{Ni}_1},
	\qquad \Rightarrow \qquad
	\bm{M}_{\mathrm{Ni}_1}=\bm{0}.
\end{equation}
Consequently, the MMs of all Ni atoms vanish for this SSG.

\begin{figure}[htbp]
	\centering
	\includegraphics[width=0.5\columnwidth]{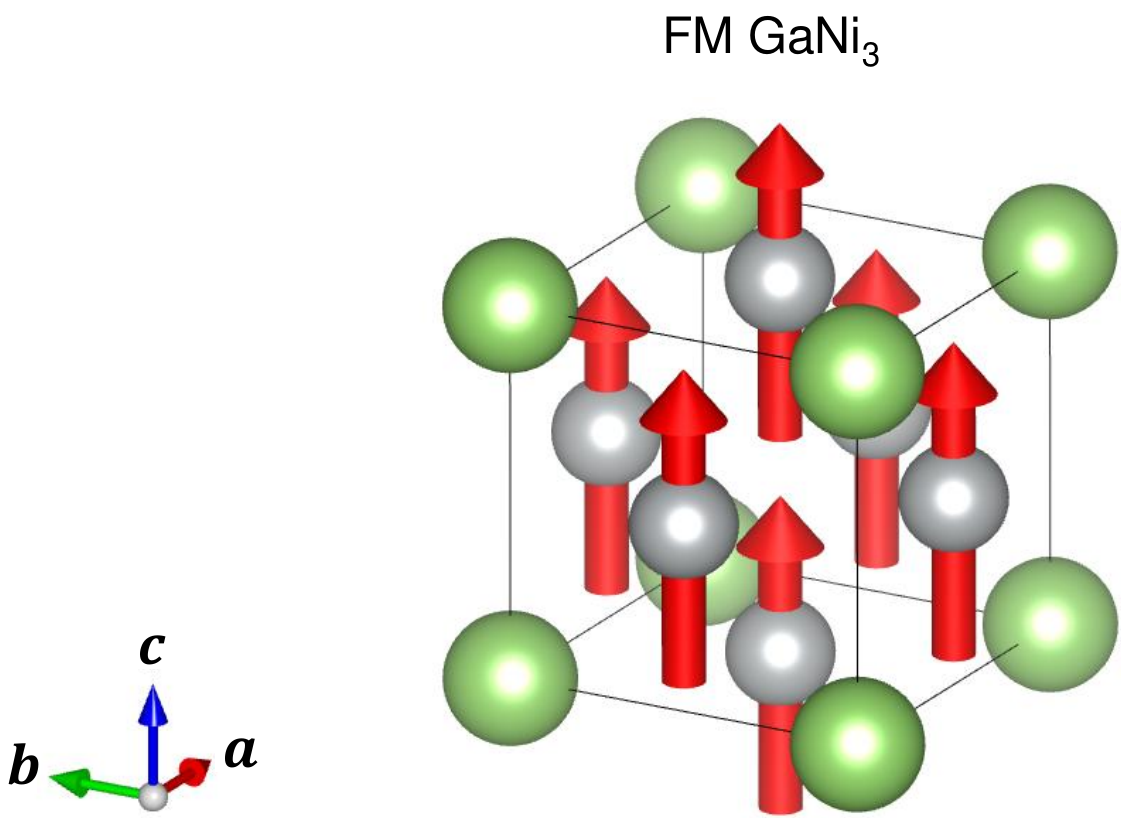}
	\caption{
		Constrained by the maximal collinear SSG subgroups of SG 221, the only possible candidate collinear magnetic order of GaNi$_3$ is FM order. Here the MMs at Ni are chosen to be all along the $c$ axis.
	}
	\label{fig:GaNi3_mag}
\end{figure}

In summary, among the maximal collinear SSG subgroups considered here, only \(221.1.1.1.\mathrm{L\_1}\) allows nonzero MMs at Ni atoms. This SSG corresponds to an FM (see Fig.~\ref{fig:GaNi3_mag}).

\end{appendices}

\end{document}